\documentclass[a4paper,11pt]{article}
\pdfoutput=1 
\usepackage{axodraw2}
\usepackage{jheppub} 

\usepackage[T1]{fontenc} 
\usepackage{subfigure}
\usepackage[normalem]{ulem}
\usepackage{epsfig,amsfonts,amsthm}
\usepackage[normalem]{ulem}
\usepackage{cancel}
\usepackage{amsmath,amssymb}
\usepackage{array}
\usepackage{amsmath}
\usepackage{amsfonts}
\usepackage{amssymb}
\usepackage{wrapfig}
\usepackage{graphicx}

\newcommand{\be}{\begin{equation}}
\newcommand{\ee}{\end{equation}}
\newcommand{\bea}{\begin{eqnarray}}
\newcommand{\eea}{\end{eqnarray}}

\newcommand{\doublet}[2]{ \left( \begin{array}{c}#1 \\ #2 \end{array}\right) }

\usepackage{color}

\def\<{\left\langle}
\def\>{\right\rangle}

\def\lsim{\mathrel{\rlap{\lower4pt\hbox{\hskip1pt$\sim$}}
    \raise1pt\hbox{$<$}}}         
\def\gsim{\mathrel{\rlap{\lower4pt\hbox{\hskip1pt$\sim$}}
    \raise1pt\hbox{$>$}}}         

\def\<{\left\langle}
\def\>{\right\rangle}

\newcommand{\bt}{\begin{tabular}}
\newcommand{\et}{\end{tabular}}

\title{\boldmath CP violation from charged Higgs bosons in the three Higgs doublet model}

\author[a]{Heather E.\ Logan,}
\author[b,c]{Stefano Moretti,}
\author[c,d]{Diana Rojas-Ciofalo,}
\author[c]{and Muyuan Song}

\affiliation[a]{Ottawa-Carleton Institute for Physics, Carleton University, Ottawa, Ontario K1S 5B6, Canada}
\affiliation[b]{Particle Physics Department, Rutherford Appleton Laboratory, Chilton, Didcot, Oxon OX11 0QX, United Kingdom}
\affiliation[c]{School of Physics and Astronomy, University of Southampton, Southampton, SO17 1BJ, United Kingdom}
\affiliation[d]{National Centre for Nuclear Research, Pasteura 7, 02-093 Warsaw, Poland}

\emailAdd{logan@physics.carleton.ca}
\emailAdd{S.Moretti@soton.ac.uk}
\emailAdd{Diana.Rojas-Ciofalo@ncbj.gov.pl}
\emailAdd{ms32g13@soton.ac.uk}

\abstract{We demonstrate a new type of cancellation of contributions to the electron and neutron electric dipole moments (EDMs) that occurs in three Higgs doublet models (3HDMs) when CP violation appears in the charged Higgs sector.  The cancellation becomes exact when the two physical charged Higgs bosons in the model are degenerate in mass.  Depending on the model parameters, degeneracies at the 10\% level are however sufficient to evade current bounds on the electron and neutron EDMs.  We demonstrate that viable parameter space remains with both charged Higgs bosons lighter than 500~GeV and large CP-violating phases while also satisfying theoretical constraints from perturbativity and experimental ones from $\bar{B} \to X_s \gamma$ and direct searches. 
}

\begin{document} 
\maketitle
\flushbottom

\section{Introduction}
\label{sec:intro}

In order to explain the observed baryon asymmetry of the universe \cite{Canetti2012}, a new source of charge-parity (CP) violation beyond the single complex phase in the Standard Model (SM) Cabibbo-Kobayashi-Maskawa (CKM) quark mixing matrix is needed \cite{Sakharov1991a}.  At the same time, new CP-violating physics with direct couplings to quarks or leptons is becoming increasingly tightly constrained by measurements that set stringent upper limits on the electric dipole moments (EDMs) of the neutron~\cite{Abel2020} and electron~\cite{Andreev2018a}.  These improved limits have led particle  theorists to consider models in which the additional CP violation is sequestered in a hidden or dark sector that does not couple directly to SM fermions \cite{CorderoCid2016, Azevedo2018, Carena2019, Okawa2019, Carena2020,Keus2020, CorderoCid2020}.

However, it does remain possible to evade the EDM limits through cancellations among contributions to the EDMs while maintaining large CP-violating phases.  We focus here on models with CP violation in an extended Higgs sector in which the couplings of the additional Higgs bosons to fermions are not suppressed.  Such a cancellation for the electron EDM (eEDM) was demonstrated in a CP-violating two Higgs doublet model (2HDM) in Ref.~\cite{Kanemura2020}, in which CP-violating contributions from neutral Higgs boson couplings cancel between two different Barr-Zee diagrams.  

If one considers  three Higgs doublet models (3HDMs) in the presence of CP violation in the charged Higgs sector, another type of cancellation takes place in both 
 the eEDM and neutron EDM (nEDM). Quite  similarly to the Glashow-Iliopoulos-Maiani (GIM) mechanism~\cite{Glashow1970} that suppresses flavor-changing neutral currents (FCNCs)  induced by loop diagrams involving a sum over fermions, the cancellation in the 3HDM involves a sum over the two charged Higgs bosons of the model and becomes exact when these are degenerate in mass.  However, even away from this limit and 
depending on the other model parameters, chiefly the charged Higgs boson Yukawa couplings, mass 
degeneracies of ${\cal O}(10\%)$  are still sufficient to evade current bounds on the electron and neutron EDMs.  Under these conditions, we show that significant parameter space regions remain in a 3HDM, including regions with one or both  charged Higgs bosons being lighter than $m_t$ (the top quark mass), while simultaneously satisfying both theoretical and experimental constraints. 

{In this paper we focus on CP-violating effects within the charged Higgs sector of the 3HDM in which natural flavour conservation (NFC)~\cite{Glashow1977,Paschos} ensures the absence of tree-level
FCNCs via Higgs interactions.  In this kind of models the Higgs sector CP violation arises from four physical CP-violating phases in the scalar potential.  These phases generically lead to CP violation in both the neutral and charged Higgs sectors.
Previous studies of CP violation in extended Higgs sectors have primarily focused on CP violation in the neutral scalar sector, indeed, this is the only type of Higgs sector CP violation that is possible in 2HDMs with NFC. (Removing the requirement for NFC does open the possibility of CP violation in the couplings of the single charged Higgs boson of the Aligned 2HDM, though~\cite{Jung2014}.)  The cancellation mechanism that
we study in this paper does not appear in the Aligned 2HDM because that model contains only a single physical charged Higgs boson. To highlight the novel cancellation mechanism for CP-odd observables like the nEDM and eEDM, in this work, we will limit ourselves to the case in which CP violation appears only in the charged Higgs sector.  We will show that it is possible to do this through a judicious choice of three of the four physical CP-violating phases in the scalar potential, leaving one physical phase free to control the CP violation in the charged Higgs sector. We leave a full analysis including CP violation in the neutral scalar sector
of the 3HDM to future work. }

This paper is organized as follows.  In Sec.~\ref{sec:model}, we describe the 3HDM and define our notation.  In Sec.~\ref{sec:edms}, we describe the calculation of the electron and neutron EDMs.  In Sec.~\ref{sec:hiding}, we explain the physics behind the aforementioned cancellation mechanism.  In Sec.~\ref{sec:numerics} we present numerical results showing the allowed CP-violating parameter space given the current EDM constraints along with constraints from $\bar{B} \to X_s \gamma$ and searches for charged Higgs bosons at colliders.  In Sec.~\ref{sec:conclusions}, we conclude.  Details of our implementation of the $\bar{B} \to X_s \gamma$ and EDM constraints are given in two appendices.

\section{Three Higgs doublet model}
\label{sec:model}

The model contains three scalar SU(2)$_L$ doublets, denoted $\Phi_1$, $\Phi_2$, $\Phi_3$, with 
\bea
 \Phi_i = \doublet{\phi^+_i}{(v_i + \phi^{0,r}_i + i\phi^{0,i}_i)/\sqrt{2}}.\
\eea
The vacuum expectation values (VEVs) $v_i$ of the three Higgs doublets can be chosen real through an independent rephasing of each doublet. They are constrained by the $W^\pm$ boson mass to satisfy $v = \sqrt{v^2_1 + v^2_2 + v^2_3} \approx 246$ GeV.

In order to avoid FCNCs through Higgs interactions, we will impose NFC~\cite{Glashow1977,Paschos} by allowing each type of fermion to couple to only a single Higgs doublet.  To this end, we require that the scalar potential be invariant under three $Z_2$ symmetries that each act on one of the $\Phi_i$.\footnote{Only two of these $Z_2$ symmetries need to be imposed by hand as the third follows accidentally.  The Type-I version of the model can be achieved by imposing only a single $Z_2$ symmetry, which opens the possibility of additional terms in the scalar potential. We do not consider this possibility here, though.}  The transformation assignments of the fermions under the three $Z_2$ symmetries then dictate the Yukawa structure of the model according to the five physically-distinct ``types'' given in Tab.~\ref{tab:types}.

\begin{table}
	\centering
	\begin{tabular}{c c c c}
		\hline \hline
		Model & $\Phi_1$ & $\Phi_2$ & $\Phi_3$ \\
		\hline
Type-I & -- & $u, d, \ell$ & -- \\
Type-II & $d, \ell$ & $u$ & -- \\
Type-X or Lepton-specific & $\ell$ & $u,d$ & -- \\
Type-Y or Flipped & $d$ & $u, \ell$ & -- \\
Type-Z or Democratic & $d$ & $u$ & $\ell$ \\
	\hline \hline
	\end{tabular}
	\caption{The five types of 3HDM subject to NFC.  The table indicates which Higgs doublet is responsible for generating the mass of each type of fermion, wherein $u(d)$ refers to an up(down)-type quark and $\ell$ to a (charged) lepton.}
	\label{tab:types}
\end{table}

The most general SU(2)$_L \times $U(1)$_Y$ invariant potential subject to these $Z_2$ symmetries is~\cite{Cree2011}
\bea
        V &=& m^2_{11}\Phi^\dagger_1\Phi_1 + m^2_{22}\Phi^\dagger_2\Phi_2 + m^2_{33}\Phi^\dagger_3\Phi_3 \nonumber\\ &-& [m^2_{12}\Phi^\dagger_1\Phi_2 + m^2_{13}\Phi^\dagger_1\Phi_3+ m^2_{23}\Phi^\dagger_2\Phi_3 + \text{h.c.}] \nonumber\\
         &+& \frac{1}{2}\lambda_{1}(\Phi^\dagger_1\Phi_1)^2 + \frac{1}{2}\lambda_{2}(\Phi^\dagger_2\Phi_2)^2 + \frac{1}{2}\lambda_{3}(\Phi^\dagger_3\Phi_3)^2 \nonumber\\
         &+& \lambda_{12}(\Phi^\dagger_1\Phi_1)(\Phi^\dagger_2\Phi_2) + \lambda_{13}(\Phi^\dagger_1\Phi_1)(\Phi^\dagger_3\Phi_3)+ \lambda_{23}(\Phi^\dagger_2\Phi_2)(\Phi^\dagger_3\Phi_3) \nonumber\\
         &+& \lambda'_{12}(\Phi^\dagger_1\Phi_2)(\Phi^\dagger_2\Phi_1) + \lambda'_{13}(\Phi^\dagger_1\Phi_3)(\Phi^\dagger_3\Phi_1) + \lambda'_{23}(\Phi^\dagger_2\Phi_3)(\Phi^\dagger_3\Phi_2) \nonumber\\
         &+& \frac{1}{2}[\lambda^{\prime\prime}_{12}(\Phi^\dagger_1\Phi_2)^2 + \lambda^{\prime\prime}_{13}(\Phi^\dagger_1\Phi_3)^2 + \lambda^{\prime\prime}_{23}(\Phi^\dagger_2\Phi_3)^2 + \text{h.c.}],
\eea
where we have retained the terms $m^2_{ij}$ that break the $Z_2$ symmetries softly. 

The potential contains six complex parameters: the three soft-breaking masses, $m^2_{12}$, $m^2_{13}$, and $ m^2_{23}$, and three quartic couplings, $\lambda^{\prime\prime}_{12}$, $\lambda^{\prime\prime}_{13}$, and $\lambda^{\prime\prime}_{23}$.  Only four of the six CP-violating phases are physical, as the other two can be eliminated by phase rotations of $\Phi_1$, $\Phi_2$, and $\Phi_3$.\footnote{Only relative phase rotations are physically meaningful.  A common overall phase rotation of all three doublets corresponds to the U(1)$_Y$ hypercharge symmetry and has no effect on the potential.  This overall phase rotation can be used to choose one of the VEVs to be real and positive; we apply this to $v_3$.}  Instead of removing the imaginary part of two of the six complex parameters, we  use this phase freedom to make all three VEVs real and positive with no loss of generality.  This choice requires that we fix the imaginary parts of $m_{13}^2$ and $m_{23}^2$ as follows~\cite{Cree2011}: 
\begin{subequations} 
\bea \label{EqAr:realvevs}
 {\text{Im}}(m^2_{13}) &=& -\frac{v_2}{v_3}{\text{Im}}(m^2_{12}) + \frac{v_1 v_2^2}{2 v_3}{\text{Im}}(\lambda^{\prime\prime}_{12}) + \frac{v_1v_3}{2}{\text{Im}}(\lambda^{\prime\prime}_{13}),\\
 {\text{Im}}(m^2_{23}) &=& \frac{v_1}{v_3}{\text{Im}}(m^2_{12}) - \frac{v_1^2v_2}{2v_3}{\text{Im}}(\lambda^{\prime\prime}_{12}) + \frac{v_2v_3}{2}{\text{Im}}(\lambda^{\prime\prime}_{23}).
\eea
\end{subequations}

The remaining four independent complex phases are responsible for the Higgs sector CP violation in the form of mixing between the two would-be CP-odd and the three would-be CP-even neutral Higgs states, as well as a complex phase in the charged Higgs mass matrix, which results in CP violation in the couplings of the charged Higgs mass eigenstates.  For our purposes in this paper, we specialize to a constrained version of the model in which we turn off  CP violation in the neutral scalar sector; this is achieved by imposing the following three relations: 
\begin{subequations} \label{EqAr:CPsplit}
   \bea 	
	{\text{Im}}(\lambda^{\prime\prime}_{13}) &=& -\frac{v^2_2}{v_3^2}{\text{Im}}(\lambda^{\prime\prime}_{12}), \\
	{\text{Im}}(\lambda^{\prime\prime}_{23}) &=& \frac{v^2_1}{v_3^2}{\text{Im}}(\lambda^{\prime\prime}_{12}), \\
	{\text{Im}}(m^2_{12}) &=& v_1v_2{\text{Im}}(\lambda^{\prime\prime}_{12}).
   \eea
\end{subequations}
This leaves only one independent CP-violating parameter, which can be taken as {\rm Im}($\lambda_{12}^{\prime\prime})$.  The remaining CP-violating phase appears in the charged Higgs mass matrix.

Finally, minimizing the potential also allows three real parameters to be eliminated in favor of the (real) VEVs~\cite{Cree2011}:
\begin{subequations}
\bea
 m^2_{11} &=& \frac{v_2}{v_1}{\text{Re}}(m^2_{12}) + \frac{v_3}{v_1}{\text{Re}}(m^2_{13}) - \frac{v_1^2}{2}\lambda_1 \nonumber \\
  &&- \frac{v_2^2}{2}[\lambda_{12} + \lambda'_{12} + {\text{Re}}(\lambda^{\prime\prime}_{12})] 
  - \frac{v_3^2}{2}[\lambda_{13} + \lambda'_{13} + {\text{Re}}(\lambda^{\prime\prime}_{13})], \\
 m^2_{22} &=& \frac{v_1}{v_2}{\text{Re}}(m^2_{12}) + \frac{v_3}{v_2}{\text{Re}}(m^2_{23}) - \frac{v_2^2}{2}\lambda_2 \nonumber \\
 &&- \frac{v_1^2}{2}[\lambda_{12} + \lambda'_{12} + {\text{Re}}(\lambda^{\prime\prime}_{12})] 
  - \frac{v_3^2}{2}[\lambda_{23} + \lambda'_{23} + {\text{Re}}(\lambda^{\prime\prime}_{23})], \\
 m^2_{33} &=& \frac{v_1}{v_3}{\text{Re}}(m^2_{13}) + \frac{v_2}{v_3}{\text{Re}}(m^2_{23}) - \frac{v_3^2}{2}\lambda_3 \nonumber \\
&&- \frac{v_1^2}{2}[\lambda_{13} + \lambda'_{13} + {\text{Re}}(\lambda^{\prime\prime}_{13})] 
  - \frac{v_2^2}{2}[\lambda_{23} + \lambda'_{23} + {\text{Re}}(\lambda^{\prime\prime}_{23})].
\eea
\end{subequations}

\subsection{Charged Higgs sector}

CP violation in the charged Higgs sector emerges from the mixing of the gauge eigenstates $\phi_i^+$ ($i=1,2,3$) to form the charged Higgs mass eigenstates.  Following the notation of Ref.~\cite{Cree2011}, we define a mixing matrix $U$ according to
\begin{eqnarray}
	\left( \begin{array}{c}
	\phi_1^+ \\ \phi_2^+ \\ \phi_3^+ \end{array} \right) 
	= U^{\dagger} 
	\left( \begin{array}{c}
	G^+ \\ H_2^+ \\ H_3^+ \end{array} \right),
	\label{eq:Udefinition}
\end{eqnarray}
where $G^+$ is the charged Goldstone boson, and $H_2^+$ and $H_3^+$ are the physical charged Higgs mass eigenstates.  Here, $U$ is obtained by diagonalizing the charged Higgs mass-squared matrix, $V \supset \phi_i^- (\mathcal{M}^2_{H^\pm})_{ij} \phi_j^+$, where~\cite{Cree2011}
\be
 \mathcal{M}^2_{H^\pm} = \left(
\begin{array}{ccc}
 \frac{v_2}{v_1}A_{12} + \frac{v_3}{v_1}A_{13} & -A_{12}+i B & -A_{13}-i\frac{v_2}{v_3} B \\
 -A_{12}-i B & \frac{v_1}{v_2}A_{12} + \frac{v_3}{v_2}A_{23} & -A_{23}+i\frac{v_1}{v_3} B \\
 -A_{13}+i\frac{v_2}{v_3} B & -A_{23}-i\frac{v_1}{v_3} B & \frac{v_1}{v_3}A_{13} + \frac{v_2}{v_3}A_{23} \\
\end{array}
\right),
\ee
with 
\bea
 A_{12} &=& {\text{Re}}(m^2_{12}) - \frac{v_1v_2}{2}[\lambda'_{12} + {\text{Re}}(\lambda^{\prime\prime} _{12})], \\ \nonumber
 A_{23} &=& {\text{Re}}(m^2_{23}) - \frac{v_2v_3}{2}[\lambda'_{23} + {\text{Re}}(\lambda^{\prime\prime}_{23})], \\ \nonumber
 A_{13} &=& {\text{Re}}(m^2_{13}) - \frac{v_1v_3}{2}[\lambda'_{13} + {\text{Re}}(\lambda^{\prime\prime}_{13})], \\ \nonumber
 B &=& -{\text{Im}}(m^2_{12}) + \frac{v_1v_2}{2}{\text{Im}}(\lambda^{\prime\prime}_{12}). 
\eea
Notice that  CP violation enters only via $B$. In the case that we turn off CP violation in the neutral Higgs  sector by imposing Eqs.~(\ref{EqAr:CPsplit}), $B$ becomes
 \[ B = -\frac{v_1v_2}{2}\rm{Im}(\lambda^{\prime\prime}_{12}). \]
 
We now diagonalize the charged Higgs mass matrix.  We perform the rotation in two stages, starting with rotating to the Higgs basis using the rotation matrix
\begin{eqnarray}
	U_1 &=& \left( \begin{array}{ccc}
		\sin\gamma & 0 & \cos\gamma \\
		0 & 1 & 0 \\
		-\cos\gamma & 0 & \sin\gamma \end{array} \right)
		\left( \begin{array}{ccc}
		\cos\beta & \sin\beta & 0 \\
		-\sin\beta & \cos\beta & 0 \\
		0 & 0 & 1 \end{array} \right),
\end{eqnarray}
where we define the angles $\beta$ and $\gamma$ in terms of the VEVs:
\be\label{tanbetatangamma}
   \tan \beta = \frac{v_2}{v_1}, \qquad \tan \gamma = \frac{\sqrt{v^2_1 + v^2_2}}{v_3}.
 \ee
This rotation isolates the charged Goldstone boson, yielding the following mass matrix:
\be
 \mathcal{M}^{\prime 2}_{H^\pm}
 = U_1\mathcal{M}^2_{H^\pm}U^\dagger_1
 =  \left(
  \begin{array}{ccc}
   0 & 0 & 0 \\
   0 & \mathcal{M}^{\prime 2}_{22} & \mathcal{M}^{\prime 2}_{23} \\
   0 & \mathcal{M}^{\prime 2*}_{23} & \mathcal{M}^{\prime 2}_{33} \\
  \end{array}
 \right), \label{Chmatrix}
\ee
where \cite{Cree2011}:\footnote{We correct two typos in the expression for $\mathcal{M}^2_{22}$ in Eq.~(A8) of Ref.~\cite{Cree2011}.}
\bea
 \mathcal{M}^{\prime 2}_{22} &=& \frac{v^2_{12}}{v_1v_2}A_{12} + \frac{v^2_2v_3}{v_1v^2_{12}}A_{13} + \frac{v_1^2v_3}{v_2v^2_{12}}A_{23}, \\
  \mathcal{M}^{\prime 2}_{33} &=& \frac{v_1v^2}{v_3v^2_{12}}A_{13} + \frac{v_2v^2}{v_3v_{12}^2}A_{23}, \\
  \mathcal{M}^{\prime 2}_{23} &=& \frac{v_2v}{v_{12}^2}A_{13} - \frac{v_1v}{v_{12}^2}A_{23} + i\frac{v}{v_3}B, 
\eea
and $v_{12}^2=v_1^2+v_2^2$. The next step is to diagonalize the matrix in Eq.~(\ref{Chmatrix}). We do it with the matrix $U_2$,
\be
 U_2 = \left(
  \begin{array}{ccc}
   1 & 0 & 0 \\
   0 & e^{-i\delta} & 0 \\
   0 & 0 & 1 \\
  \end{array} \right)
  \left(
  \begin{array}{ccc}
   1 & 0 & 0 \\
   0 & \cos\theta & \sin\theta e^{i\delta} \\
   0 & -\sin\theta e^{-i\delta} & \cos\theta \\
  \end{array} \right),
\ee
where the CP-violating phase $\delta$ is given by
\be
 \delta = \text{phase}(\mathcal{M}^{\prime 2}_{23}),
\ee
with $0 \leq \delta < 2\pi$.  For later convenience, we choose the mixing angle $\theta$ to lie in the range $-\pi/2 \leq \theta \leq 0$,\footnote{In the Democratic (or Type-Z) 3HDM, the coupling of $H_2^+$ to leptons goes to zero when $\theta = 0$, likewise the coupling of $H_3^+$ to leptons goes to zero when $\theta = -\pi/2$.} so that either $H_2^{\pm}$ or $H_3^{\pm}$ can be the lighter physical charged Higgs boson.  

The full rotation matrix in Eq.~(\ref{eq:Udefinition}) is then given explicitly by \cite{Cree2011}:
\begin{equation}
	U^{\dagger}
	= (U_2 U_1)^{\dagger}
	= \left( \begin{array}{ccc}
	s_\gamma c_\beta & -c_\theta s_\beta e^{i\delta} - s_\theta c_\gamma c_\beta  	&  s_\theta s_\beta e^{i\delta} - c_\theta c_\gamma c_\beta \\
	s_\gamma s_\beta 
		& c_\theta c_\beta e^{i\delta} - s_\theta c_\gamma s_\beta & -s_\theta c_\beta e^{i\delta} - c_\theta c_\gamma s_\beta  \\
c_\gamma	
		& s_\theta s_\gamma & c_\theta s_\gamma 
	\end{array} \right),
	\label{eq:Uexplicit}
\end{equation}
where $s_{\beta} = \sin\beta$, $c_{\beta} = \cos\beta$ and similarly for the other mixing angles.  We give the explicit form for $U^{\dagger}$ rather than $U$ for later convenience in writing the Yukawa couplings.

{For simplicity, we assume that the masses of all the extra neutral scalars, that is, $H_{2,3}$ and $A_{2,3}$, are larger than those of the charged Higgs bosons, and we take the alignment limit so that the tree-level couplings of the 125~GeV Higgs boson $h$ are identical to those of the SM Higgs boson. In other words, we focus on the physics related to the charged Higgs sector so that our input parameters are the following six:}
\[ M_{H^\pm_2}, M_{H^\pm_3}, \tan\beta, \tan\gamma, \theta, \delta. \]
{Notice that our definitions of $\tan\beta$ and $\tan\gamma$ differ from those in Ref. \cite{Akeroyd2017}, where a similar analysis of the 3HDM was performed.}

\subsection{The Yukawa Lagrangian}
 
In what follows, we will focus on the Democratic 3HDM,\footnote{Hereafter, we 
adopt this nomenclature in preference to Type-Z, as the former was 
introduced in Ref.~\cite{Cree2011} prior to the latter in 
Ref.~\cite{Akeroyd2017}.} in which $\Phi_1$ gives mass 
to down-type quarks, $\Phi_2$ gives mass to up-type quarks, and $\Phi_3$ 
gives mass to charged leptons.  This version of the model gives rise to the 
most interesting EDM phenomenology arising from CP violation in the charged 
Higgs mixing matrix.  We will comment on the EDMs in the other versions of 
the 3HDM in Sec.~\ref{sec:hiding}.

The Yukawa Lagrangian takes the form
\be\label{YukLag}
	\mathcal{L}_\text{Yukawa} = - \{ \bar{Q}_L\Phi_1\mathcal{G}_dd_R + \bar{Q}_L\tilde{\Phi}_2\mathcal{G}_uu_R + \bar{L}_L\Phi_3\mathcal{G}_l l_R + \text{h.c} \},
\ee
where $\tilde \Phi$ is the conjugate doublet given by $i \sigma^2 \Phi^*$.
Here, $\mathcal{G}_f$ are the Yukawa matrices, which are determined in terms of the fermion mass matrices $\mathcal{M}_f$ by $\mathcal{M}_f = \mathcal{G}_f v_i /\sqrt{2}$.

The Yukawa couplings of the charged Higgs bosons are given by~\cite{Grossman1994}:
\bea 
 \mathcal{L}^\text{charged}_\text{Yukawa} &=& -\frac{\sqrt{2}}{v} \left\{ [X_2\bar{u}_LV \mathcal{M}_dd_R + Y_2\bar{u}_R\mathcal{M}_uVd_L + Z_2\bar{\nu}_L \mathcal{M}_ll_R]H^+_2 \right. \\ \nonumber
 	&+& \left. [X_3\bar{u}_LV \mathcal{M}_dd_R + Y_3\bar{u}_R\mathcal{M}_uVd_L + Z_3\bar{\nu}_L\mathcal{M}_ll_R]H^+_3 + \text{h.c.} \right\},
\eea
where $V$ is the CKM matrix and the coupling coefficients $X_i$, $Y_i$ and $Z_i$ are given in terms of the elements of the charged Higgs mixing matrix $U^{\dagger}$ in Eq.~(\ref{eq:Uexplicit}) by
\be
	X_i = \frac{U^\dagger_{1i}}{U^\dagger_{11}}, \qquad Y_i = -\frac{U^\dagger_{2i}}{U^\dagger_{21}}, \qquad Z_i = \frac{U^\dagger_{3i}}{U^\dagger_{31}},
	\label{XYZ}
\ee
where $i = 2, 3$.  Note that these expressions are for the Democratic 3HDM.  The coupling coefficients for the other types of 3HDM are collected in Tab.~\ref{tab:couplingfactors}.\footnote{In all types of 3HDM except the Democratic one, taking the limit $\tan\gamma \to \infty$ (i.e., $v_3 \to 0$) recovers the corresponding 2HDM plus a third, inert, doublet.}

\begin{table}
	\centering
	\begin{tabular}{c ccc}
		\hline \hline
		Model & $X_i$ & $Y_i$ & $Z_i$ \\
		\hline
Type-I & $\frac{U^\dagger_{2i}}{U^\dagger_{21}}$ & $-\frac{U^\dagger_{2i}}{U^\dagger_{21}}$ & $\frac{U^\dagger_{2i}}{U^\dagger_{21}}$ \\
Type-II & $\frac{U^\dagger_{1i}}{U^\dagger_{11}}$ & $-\frac{U^\dagger_{2i}}{U^\dagger_{21}}$ & $\frac{U^\dagger_{1i}}{U^\dagger_{11}}$ \\
Type-X or Lepton-specific & $\frac{U^\dagger_{2i}}{U^\dagger_{21}}$ & $-\frac{U^\dagger_{2i}}{U^\dagger_{21}}$ & $\frac{U^\dagger_{1i}}{U^\dagger_{11}}$ \\
Type-Y or Flipped & $\frac{U^\dagger_{1i}}{U^\dagger_{11}}$ & $-\frac{U^\dagger_{2i}}{U^\dagger_{21}}$ & $\frac{U^\dagger_{2i}}{U^\dagger_{21}}$ \\
Type-Z or Democratic & $\frac{U^\dagger_{1i}}{U^\dagger_{11}}$ & $-\frac{U^\dagger_{2i}}{U^\dagger_{21}}$ & $\frac{U^\dagger_{3i}}{U^\dagger_{31}}$ \\
	\hline \hline
	\end{tabular}
	\caption{Coefficients $X_i$, $Y_i$, and $Z_i$ appearing in the Yukawa Lagrangian for the $H_i^+$ couplings to down-type quarks, up-type quarks, and (charged) leptons, respectively, with $i = 2,3$.  The matrix $U^{\dagger}$ is defined in Eq.~(\ref{eq:Uexplicit}).}
	\label{tab:couplingfactors}
\end{table}

{In Figs. \ref{Fig:HpmBRs1}--\ref{Fig:HpmBRs2}, we show the branching ratios (BRs) of $H^\pm_2$ (upper panels) and $H^\pm_3$ (lower panels) as a function of $\tan\beta$, in the 3HDM Type-II, -X, -Y and the Democratic model.\footnote{We have used {\tt CalcHEP} \cite{Belyaev2013} to produce these plots. We will use it again to calculate widths and cross sections to compare against experimental constraints.} 
In Fig.~\ref{Fig:HpmBRs1} we take $M_{H_2^\pm}=100$ GeV and  $M_{H_3^\pm}=150$ GeV,
while in Fig.~\ref{Fig:HpmBRs2}
we take $M_{H_2^\pm}=200$ GeV and $M_{H_3^\pm}=250$ GeV, with $\theta = -\pi/4$ and $\delta = 0$ in both. The solid and dotted curves show the case for $\tan\beta = 2$ and $5$, respectively. We can see that a light charged Higgs boson (with $M_{H^\pm_i}<m_t$) predominantly decays to $\tau\nu$, although $cs$ is more dominant for some types  in specific $\tan\beta$ regions. Furthermore, the decay into $cb$ becomes relevant for higher $\tan\beta$ in the Type-Y and  Democratic models.
For a heavy  charged Higgs boson  (with $M_{H^\pm_i}>m_t$), the vastly dominant decay is into $tb$ except  for Type-X at large $\tan\beta$, where $\tau\nu$ dominates instead. Instead, for the Democratic model, $\tau\nu$ dominates for large values of $\tan\gamma$.
(Notice that, here, the 3HDM
 parameter values are chosen so we can directly compare with  Figs.~1 and 2 of \cite{Akeroyd2017}, where the parametrization of $\tan\beta$ and $\tan\gamma$ is, however, chosen differently from our work.\footnote{Furthermore, we use here the labeling $H_{2,3}^\pm$ in place of $H_{1,2}^\pm$ in Ref.~\cite{Akeroyd2017}, respectively.})
{We do not show the BRs for the charged Higgs bosons in the Type-I 3HDM because they are independent of $\tan\beta$, are the same for both of the charged Higgs bosons, and depend very little on the charged Higgs boson mass for $M_{H_i^\pm}<m_t$.  The most important BRs for the masses shown are to $cs$ (close to 70\%) and $\tau \nu$ (a little less than 30\%), with sub-dominant decays to $cb$ (just over 1\%) and $\mu \nu$ (around 0.1\%).  When the charged Higgs masses are above the top one - in the Type-I 3HDM, decays to $t b$ become overwhelmingly dominant, e.g.,  for the parameter choices of Fig.~\ref{Fig:HpmBRs2}, all of the decay BRs to fermion pairs other than $tb$ are below 0.2\%.}

\begin{figure}
	\centering
	\includegraphics[scale=0.4]{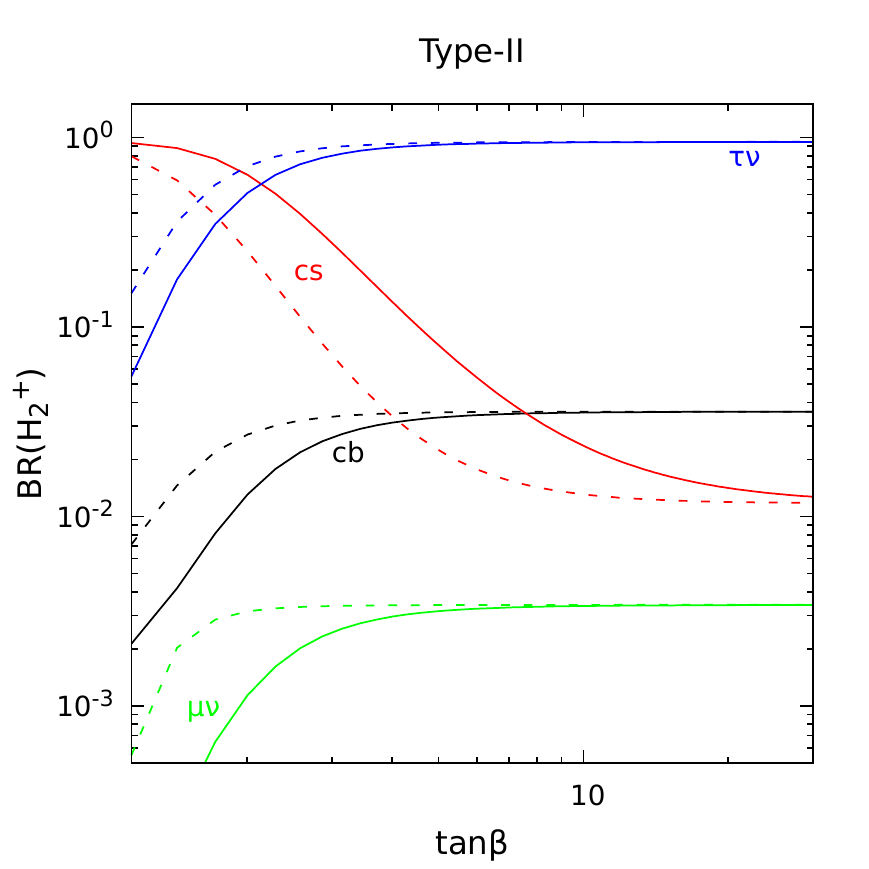}	\includegraphics[scale=0.4]{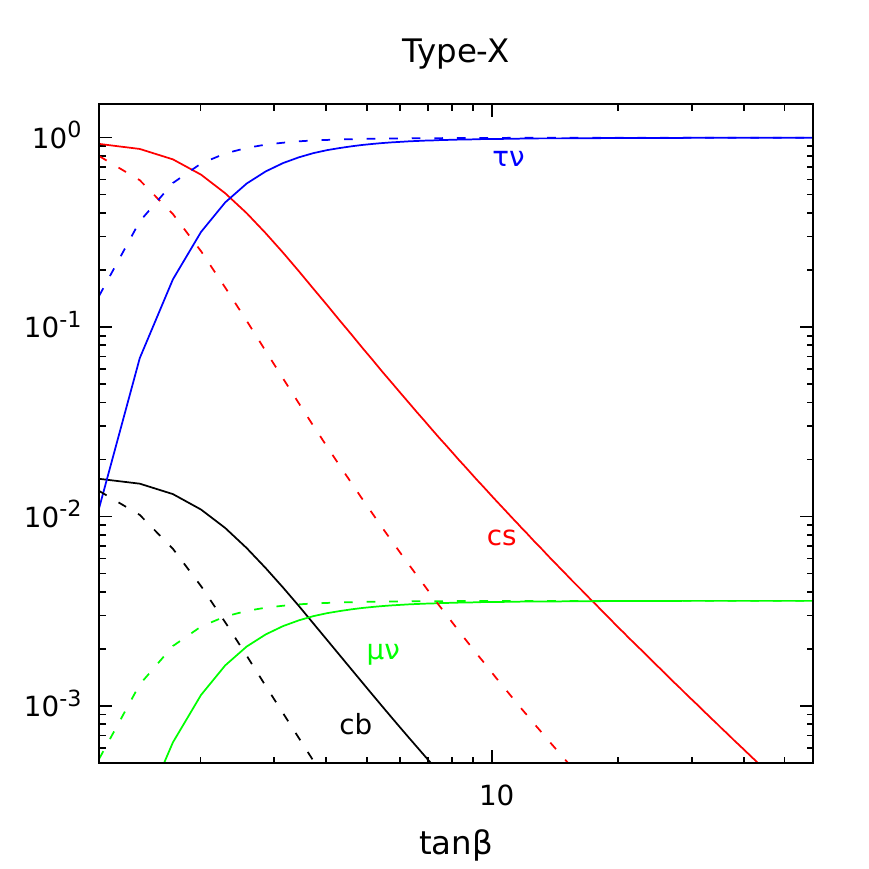}	\includegraphics[scale=0.4]{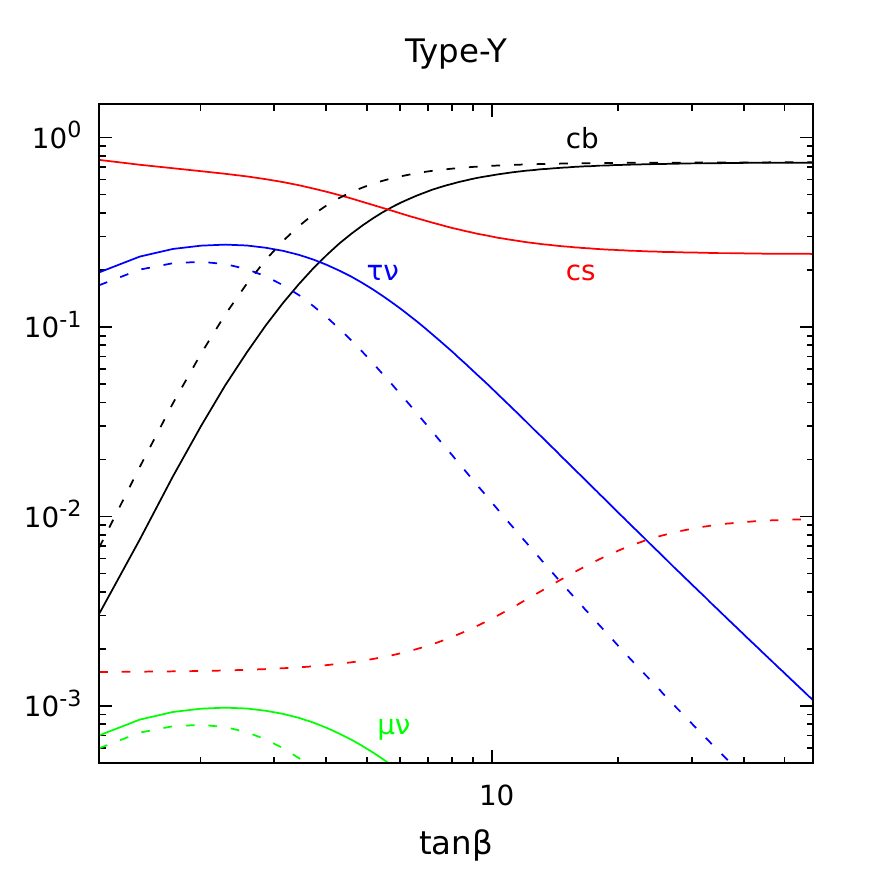}	\includegraphics[scale=0.4]{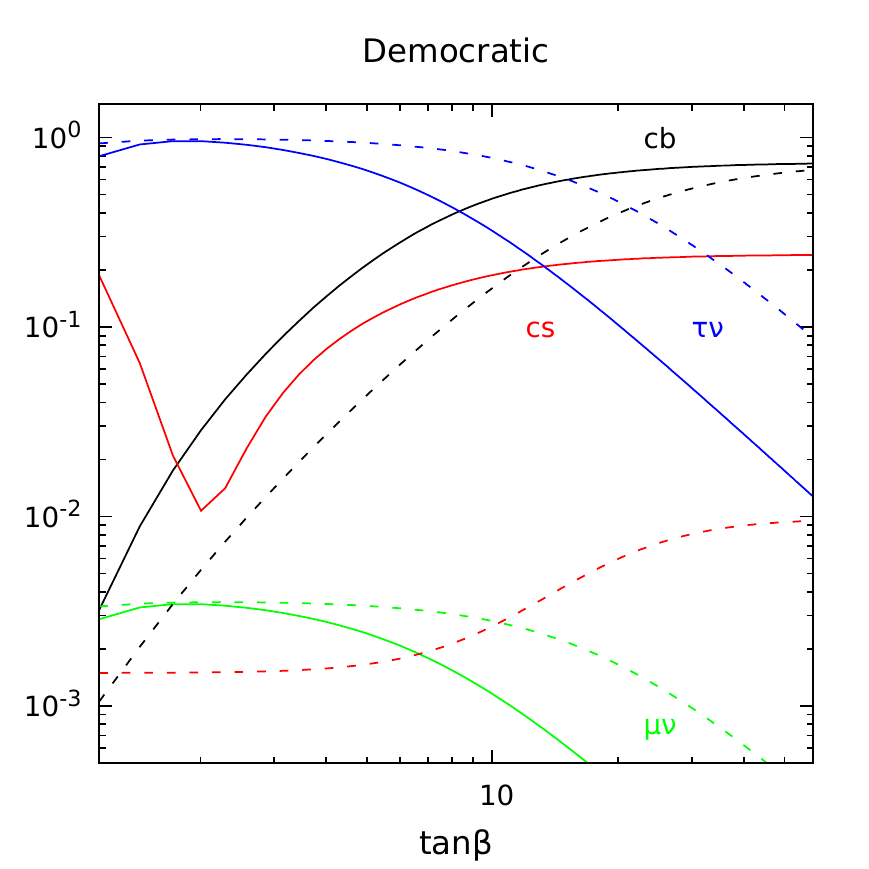}\\
		\includegraphics[scale=0.4]{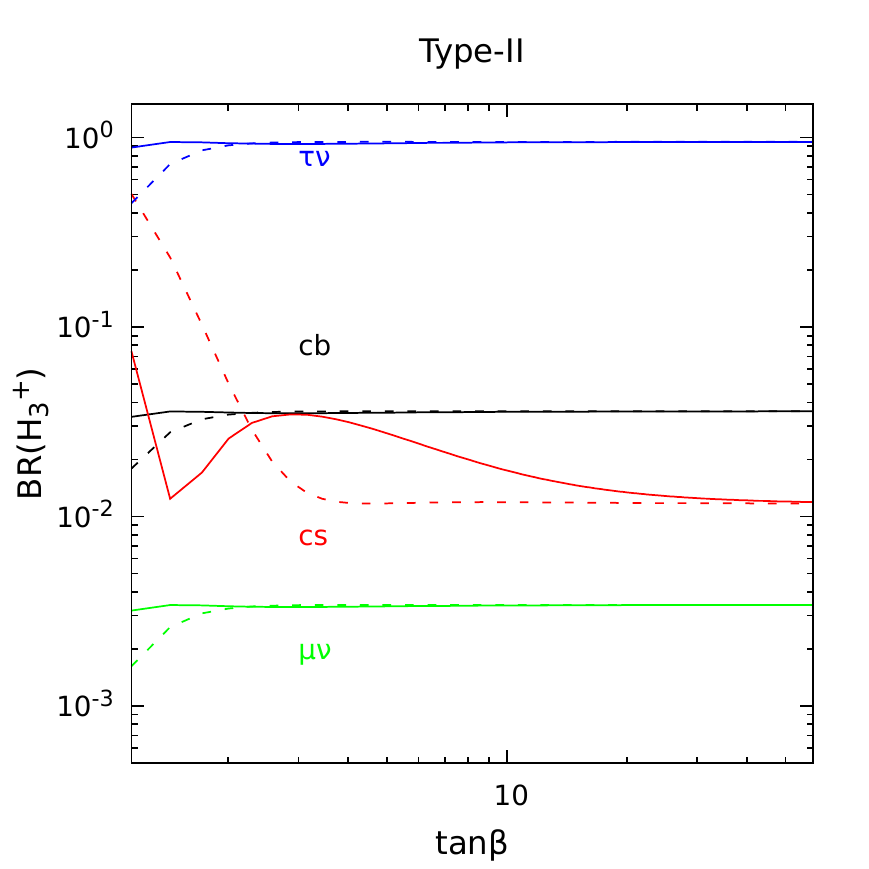}	\includegraphics[scale=0.4]{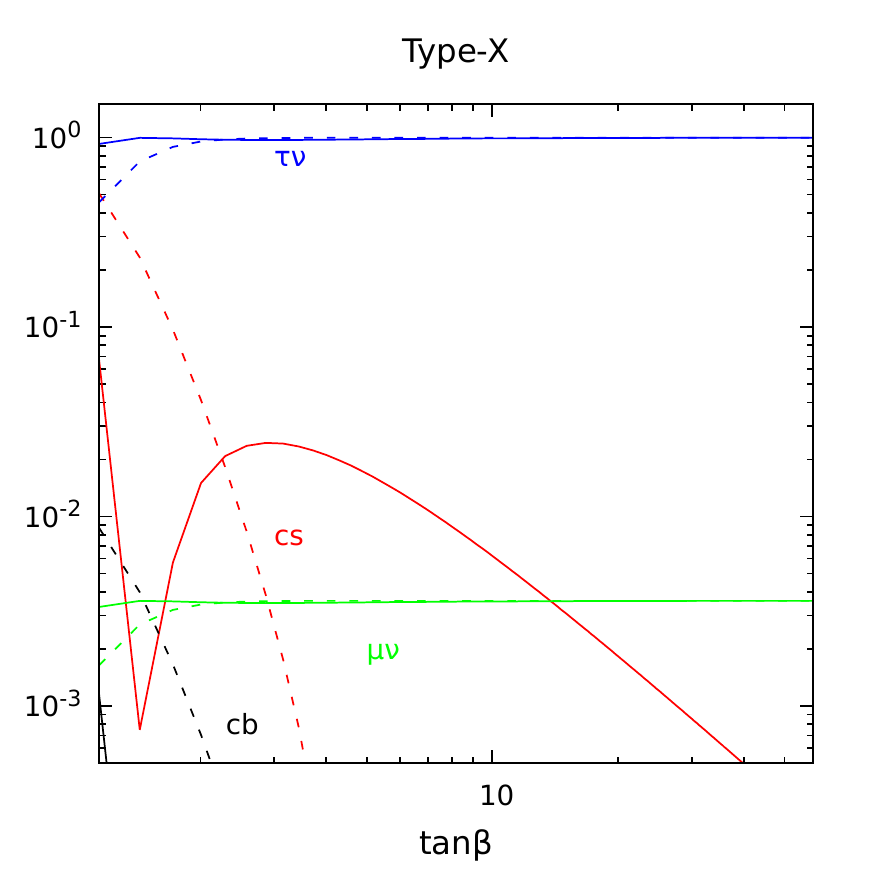}	\includegraphics[scale=0.4]{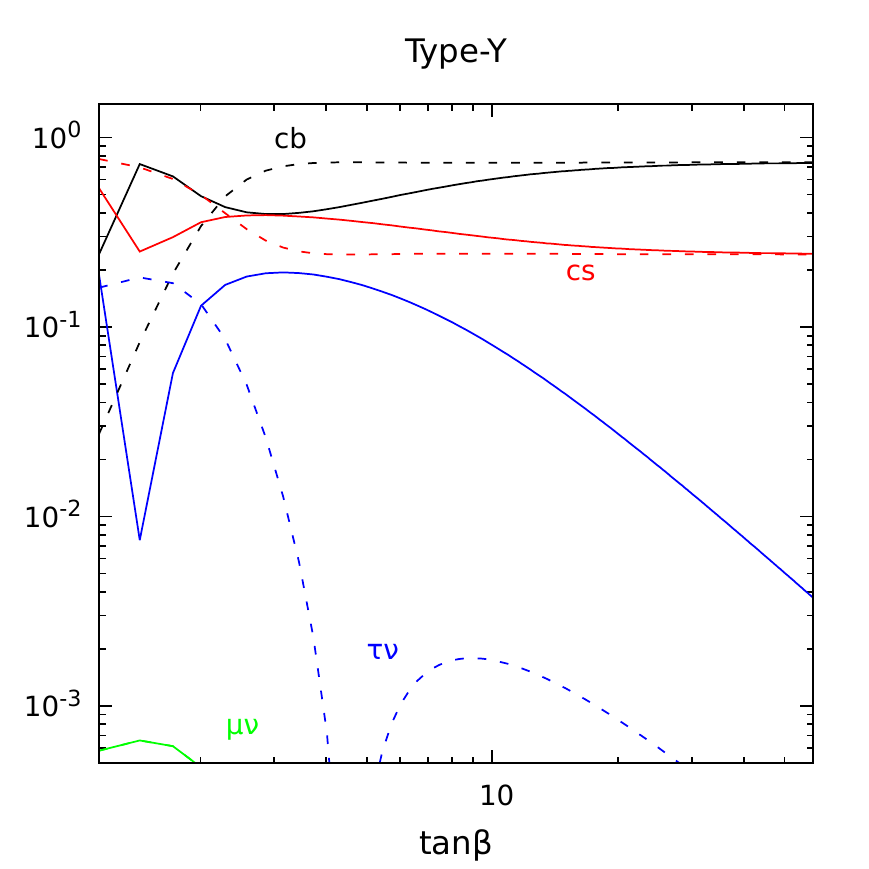}	\includegraphics[scale=0.4]{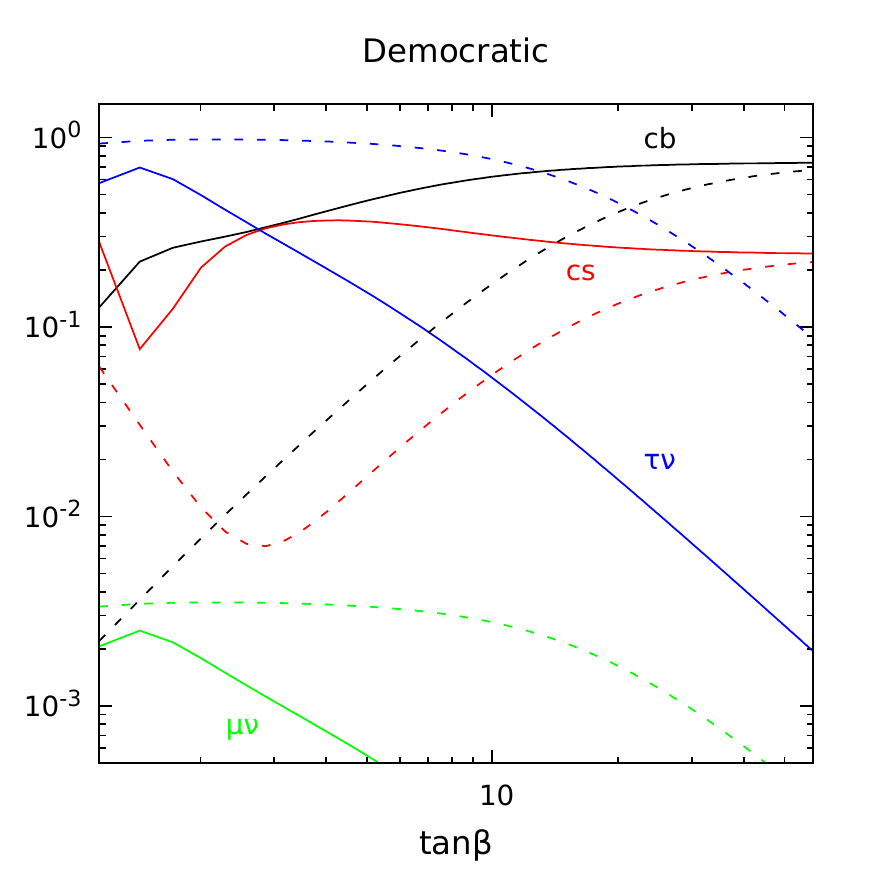}
	\caption{BRs of $H_2^+$ (upper panels) and $H_3^+$ (lower panels) as a function of $\tan\beta$ in, from left to right, the Type-II, -X, -Y, and Democratic 3HDMs. We take $M_{H_2^+}=100$ GeV, $M_{H_3^+}=150$ GeV, $\theta=-\pi/4$ and $\delta = 0$. The value of $\tan\gamma$ is 2 (5) for the solid (dotted) curves.}
	\label{Fig:HpmBRs1}
\end{figure}

\begin{figure}
	\centering
\includegraphics[scale=0.4]{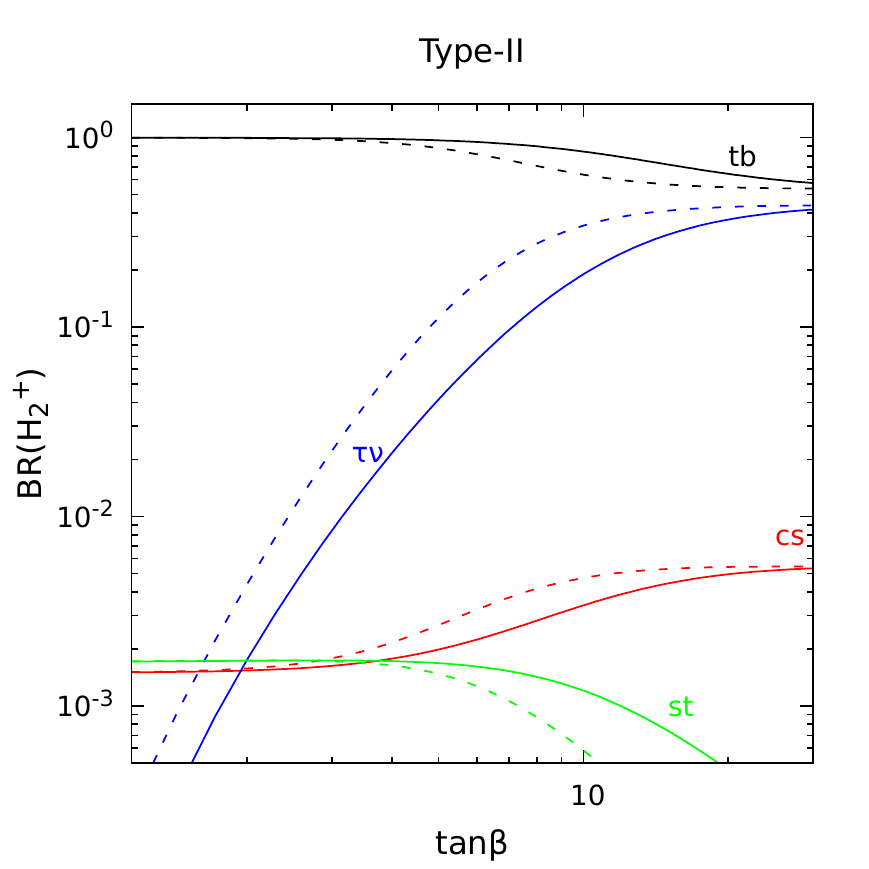}	\includegraphics[scale=0.4]{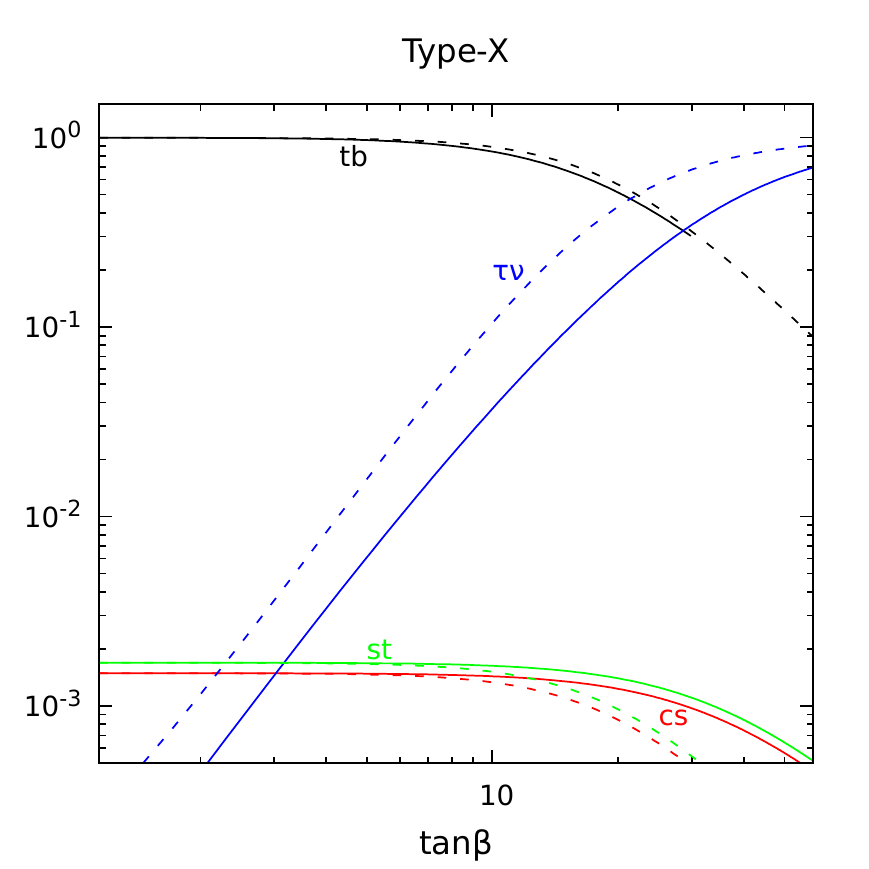}	\includegraphics[scale=0.4]{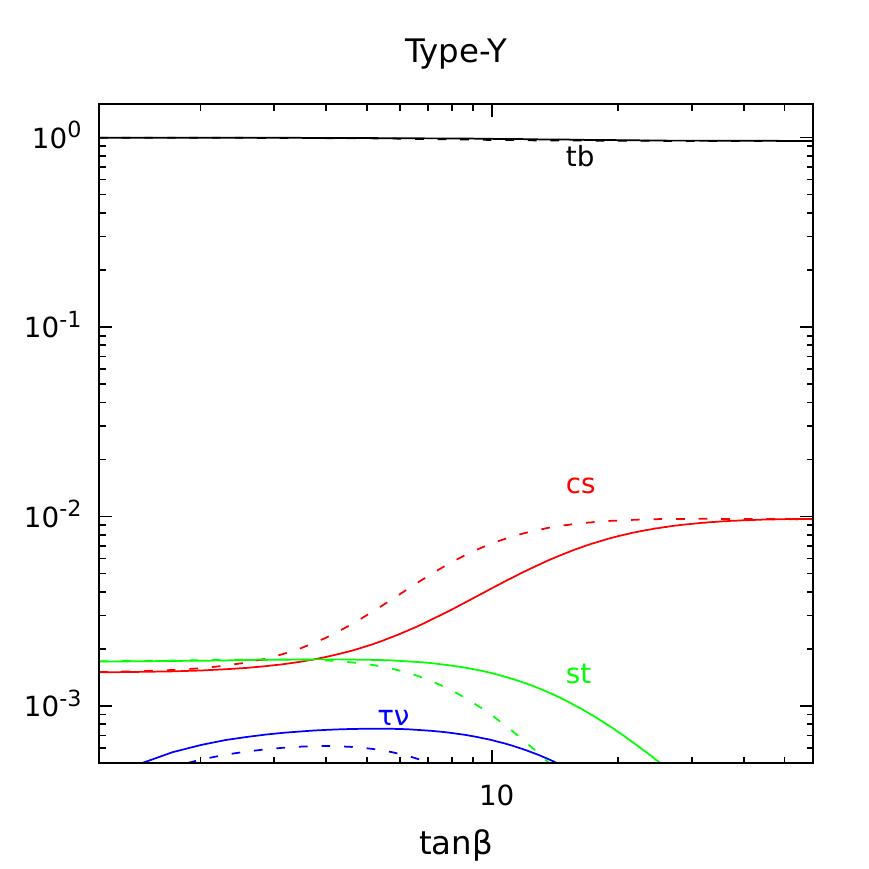}	\includegraphics[scale=0.4]{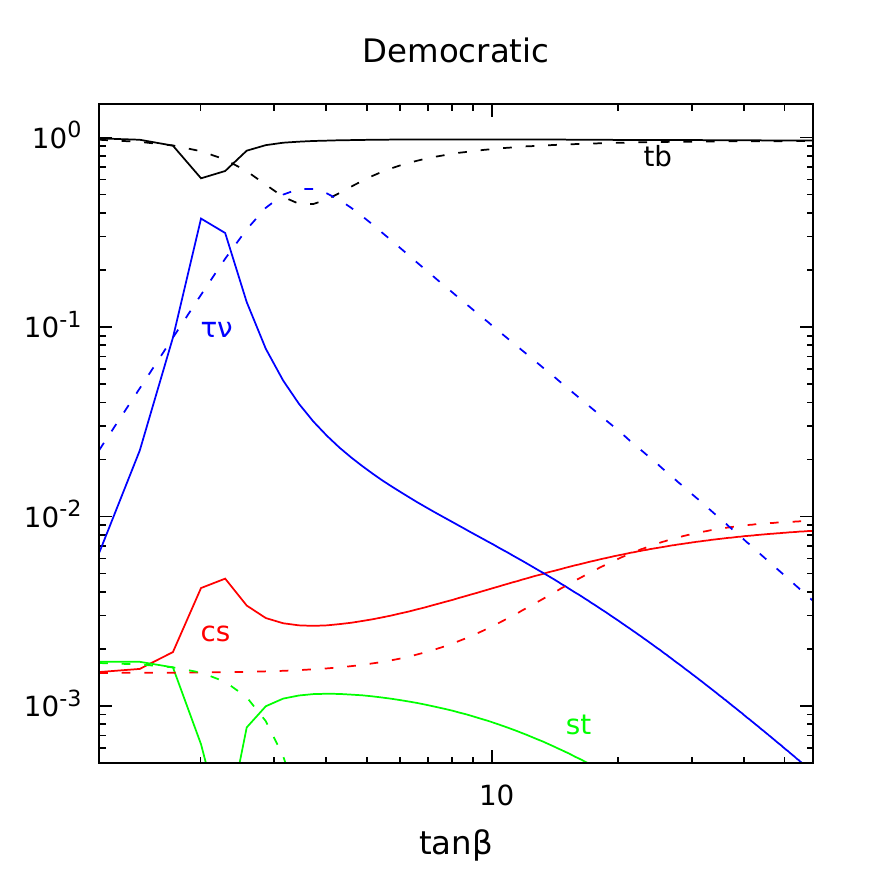}\\
\includegraphics[scale=0.4]{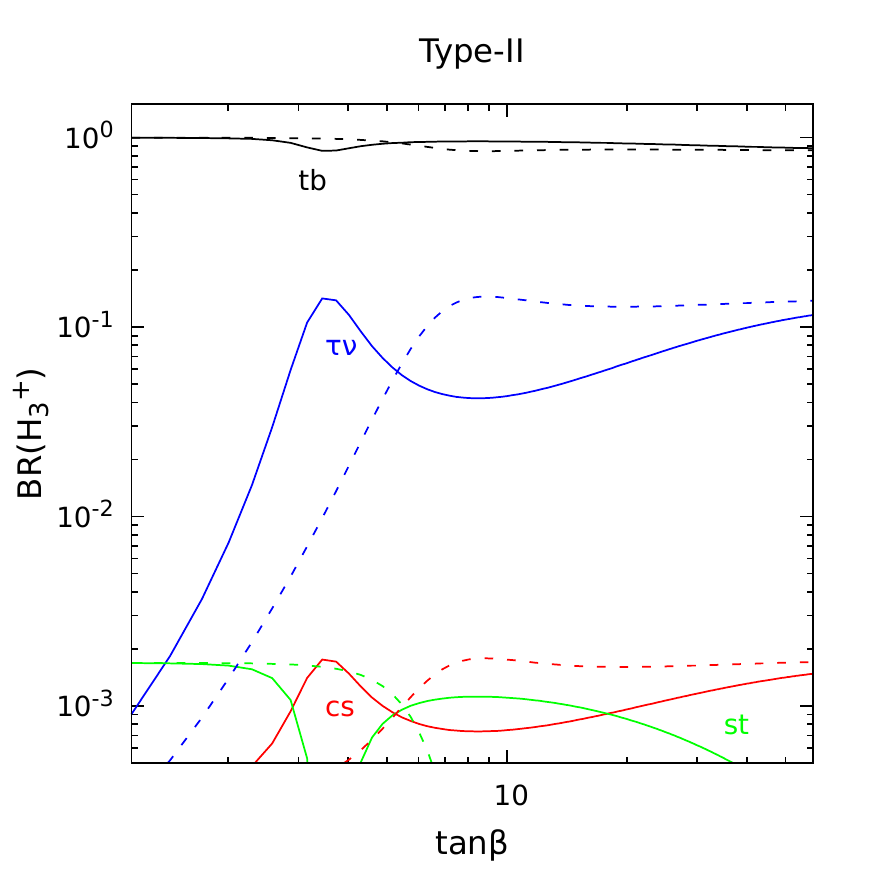}	\includegraphics[scale=0.4]{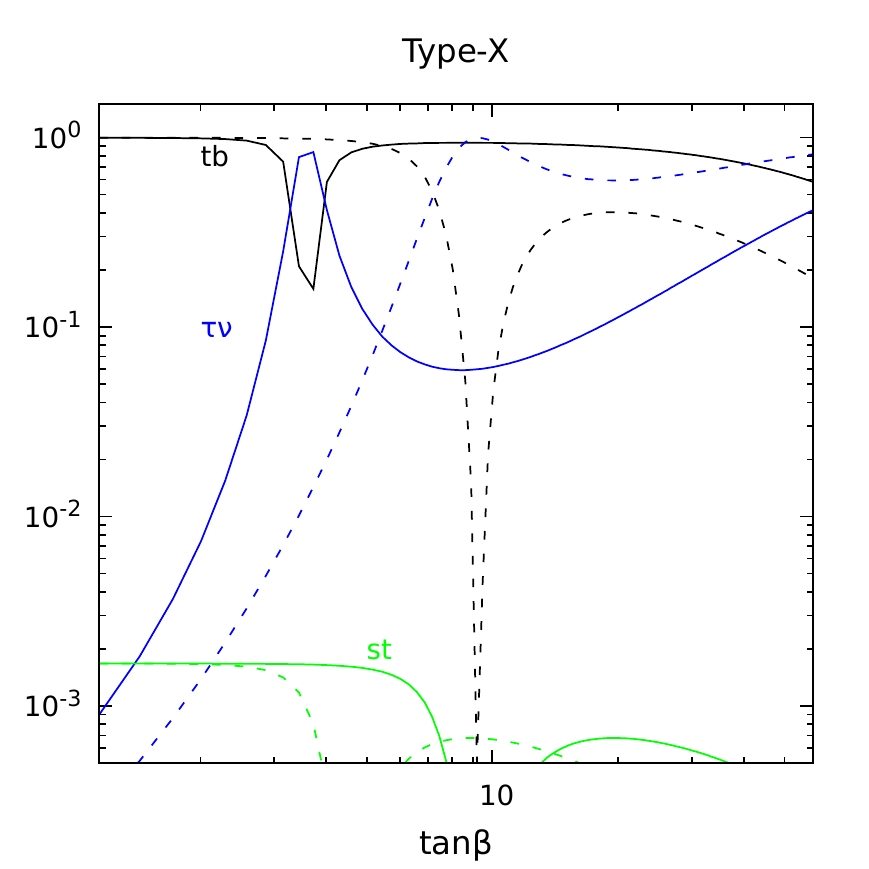}	\includegraphics[scale=0.4]{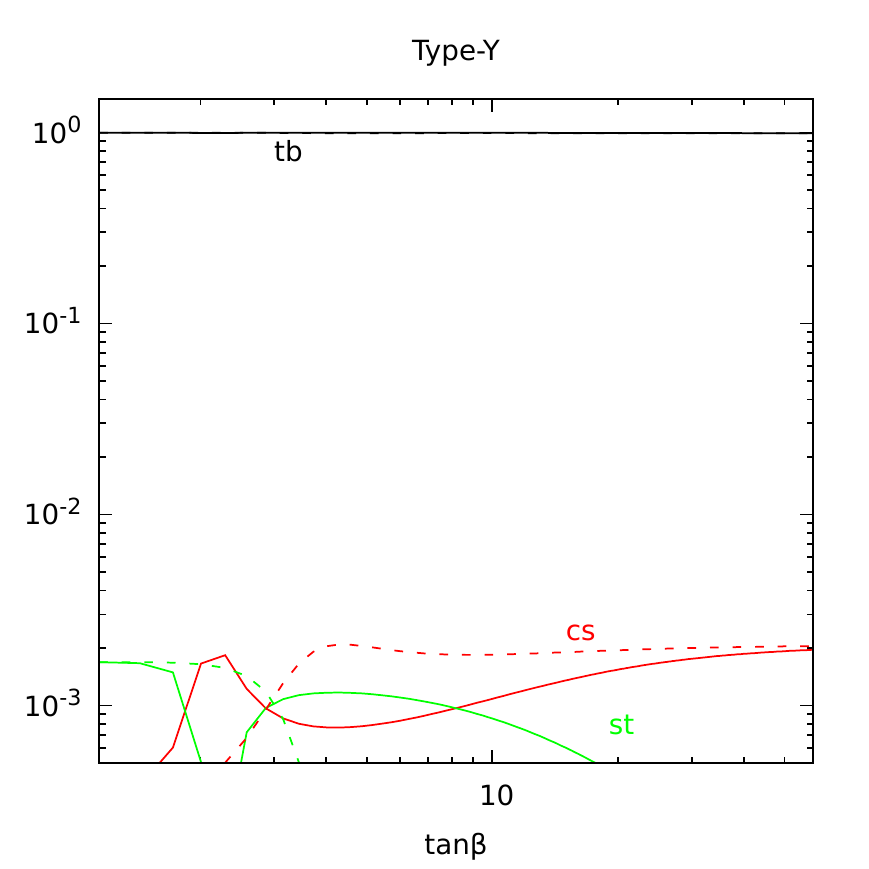}	\includegraphics[scale=0.4]{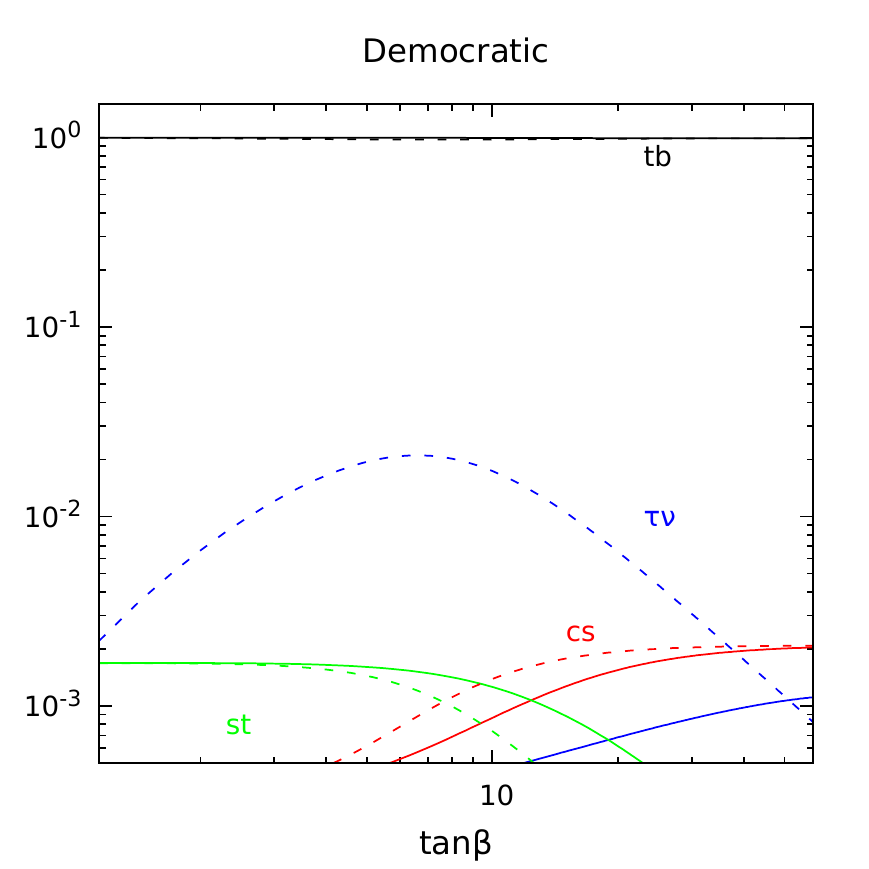}
	\caption{As in Fig.~\ref{Fig:HpmBRs1} but for $M_{H_2^+}=200$ GeV and $M_{H_3^+}=250$ GeV.}
	\label{Fig:HpmBRs2}
\end{figure}

\subsection{Collider constraints}

Charged Higgs boson production in hadronic collisions can be described by the subprocesses $gg,q\bar q\to t\bar b H^-$ + c.c. for both light ($M_{H^\pm_i}<m_t$)
and heavy ($M_{H^\pm_i}>m_t$) states \cite{Guchait2002,Assamagan2004}, as in the former case the dominant channel is $gg,q\bar q\to t\bar t\to t \bar b H^-$ + c.c. (i.e., $t$-quark pair production and decay) while in the latter case, it is $bg\to tH^-$ + c.c. (i.e., Higgs-strahlung off $b$-quarks).\footnote{Recall that $b$-(anti)quarks are produced inside protons from a gluon splitting.} Since the Higgs-strahlung cross section is much smaller than the one for  top-antitop quark production, a light charged Higgs boson is severely constrained while direct searches for a heavy one leave it largely unconstrained. However, when $M_{H^\pm_i}\approx M_{W^\pm}\approx 80$~GeV, the $t\to bW^+$ background overwhelms the $t\to b H^+$ signal, so that, even at the current  Large Hadron Collider (LHC), this  mass region is still allowed for a charged Higgs state in a 3HDM, no matter its decay mode \cite{Akeroyd2018,Akeroyd2020a}. Of relevance to our analysis are the constraints coming from  $H^\pm\to\tau\nu$ \cite{Sirunyan2019}, $cb$ \cite{Sirunyan2018} and $cs$ \cite{Aad2013} searches at the LHC (with the first channel generally being more constraining than the second and third ones), which have been performed by both ATLAS and CMS. 

In Fig.~\ref{Fig:munuconstraints}, we fix the values of $M_{H_2^\pm}=80$ GeV, $M_{H_3^\pm}=170$~(200)~GeV in the upper (lower) panels and $\tan\beta = 20$.\footnote{Comparing the upper and lower left panels of Fig.~\ref{Fig:munuconstraints} shows that the cross section times BR of $H_2^{\pm}$ is essentially unaffected by the mass of the heavier $H_3^{\pm}$, once it is at least comparable to the top quark mass.} 
We tested the region $-0.6<\theta <0$, $0.4<\tan\gamma<2.6$ against CMS searches for $H^\pm\to\tau\nu$~\cite{Sirunyan2019}.\footnote{{Although values of $\tan\gamma >2.6$ are allowed, we have chosen this region to better show the tension between the excluded areas for $H_2^\pm$ and $H_3^\pm$, respectively.}  }
In the case of $H_2^\pm$, it is preferable to take values of $\theta$ closer to zero, which is in tension with the cross section for $H_3^\pm$, which prefers $\theta\lesssim -0.4$. 
However, we can quench this tension if we choose $\tan\gamma\lesssim 2$, as the BR of both charged Higgs states to $\tau\nu$ are smaller (see Fig.~\ref{Fig:HpmBRs1}).
We can also notice that lower values for $M_{H_3^\pm}$ increase the cross section of $H^\pm_3\to\tau\nu$, thus making it harder to agree with collider limits. 
For example, this is very manifest  for the case of $M_{H^\pm_3}=150$ GeV, shown in Fig.~\ref{Fig:80150collider}, a scenario that is excluded by $H^\pm_3\to\tau\nu$ results.
For this value of $M_{H^\pm_3}$, we should also compare to the collider limits for $H^\pm_3\to cb$ and $cs$.
However, these are less constraining than the case of $\tau\nu$.
In the case when $m_t < M_{H^\pm_2} < M_{H^\pm_3}$, the BR of $H^\pm_2$ to $\tau\nu$ only dominates over the BR to $tb$ for small values of $\tan\beta$, as can be seen in Fig.~\ref{Fig:HpmBRs2}.
Later in this work, when we consider the masses of the charged Higgs bosons to be larger than the top-quark one, we take $\tan\beta>10$, and then this region readily satisfies collider limits.
Overall, notice that there is no significant  interference between $H^\pm_2$ and $H^\pm_3$, unless their mass difference is comparable to either of their widths, which is never the case for the benchmark points that we will study.\footnote{{When both charged Higgs boson masses are lower than $m_t$, their  widths become very small, so that very strong fine-tuning of their masses would be needed to achieve overlap of the lineshapes and thus interference in top quark decays involving on-shell $H^\pm_2$ and $H^\pm_3$. For example, if we take the parameter values of Fig.~\ref{Fig:lowmass-summary} (lower panels), the width of $H_2^\pm$ is around 5 MeV and the width of $H_3^\pm$ is 0.9 MeV (0.74 MeV) for $\delta = 0.8\pi$ $(0.95\pi)$.}}

\begin{figure}
	\centering
	\includegraphics[scale=0.45]{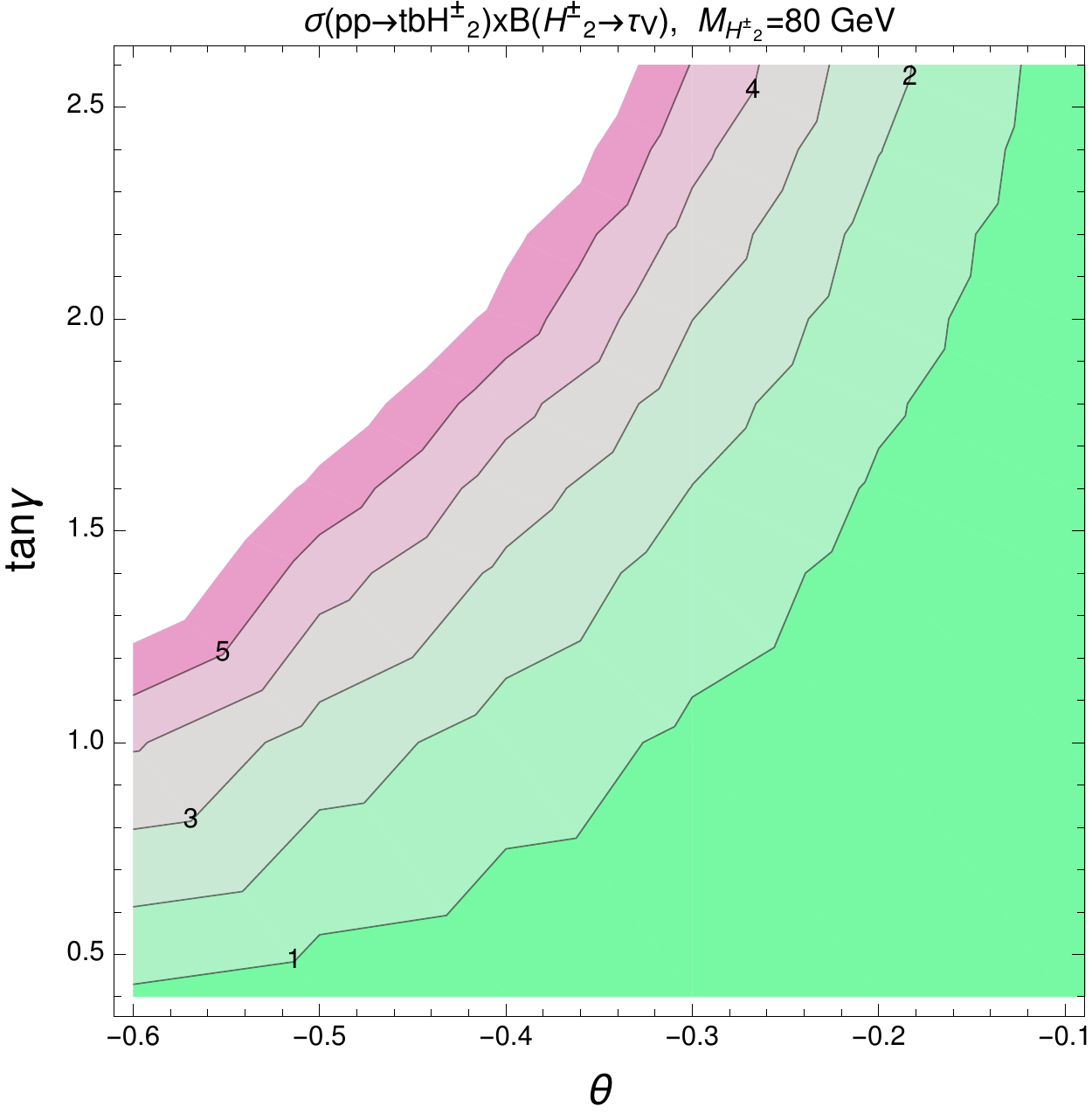}~
	\includegraphics[scale=0.45]{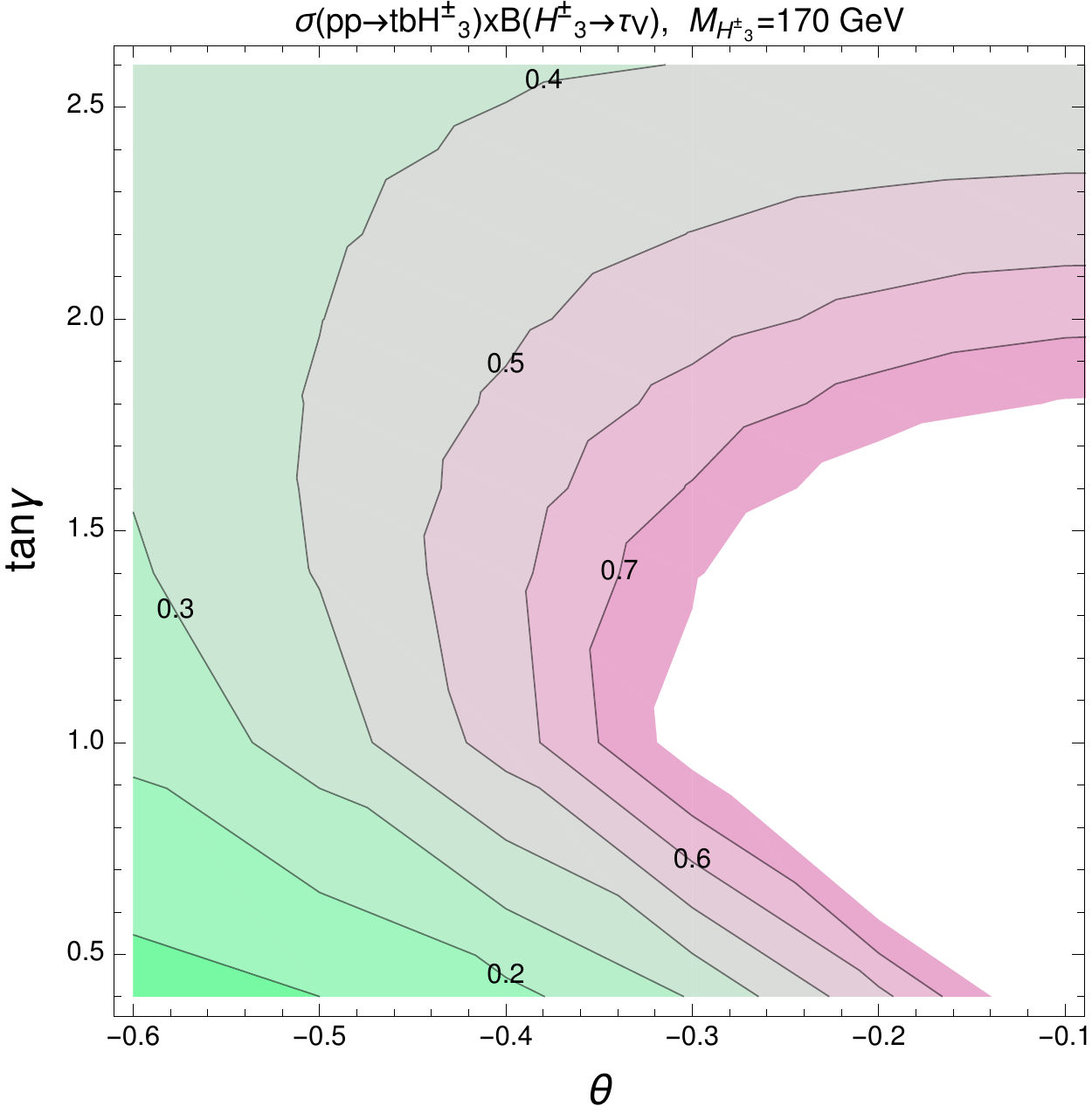}\\
	\includegraphics[scale=0.45]{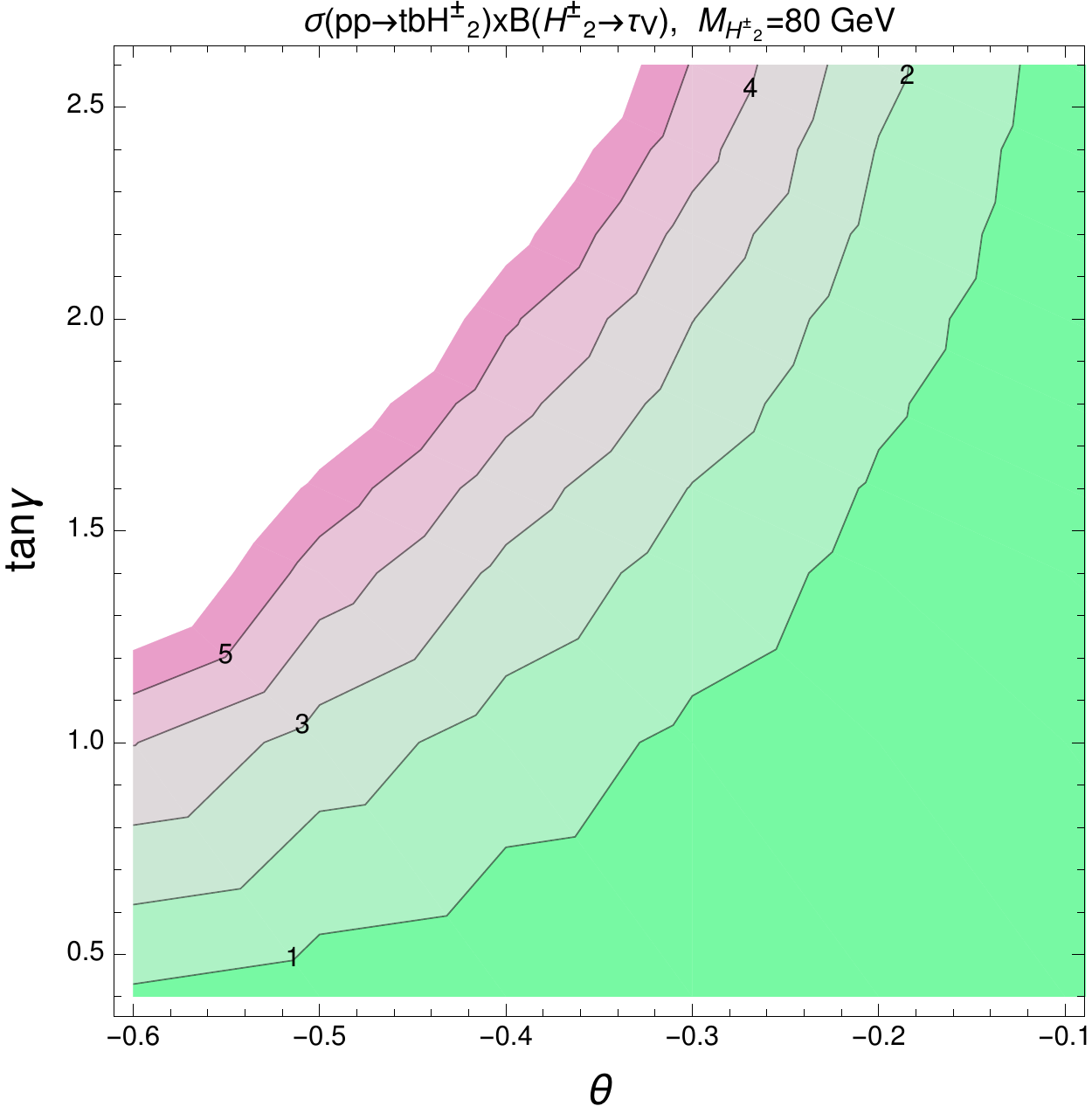}~
	\includegraphics[scale=0.45]{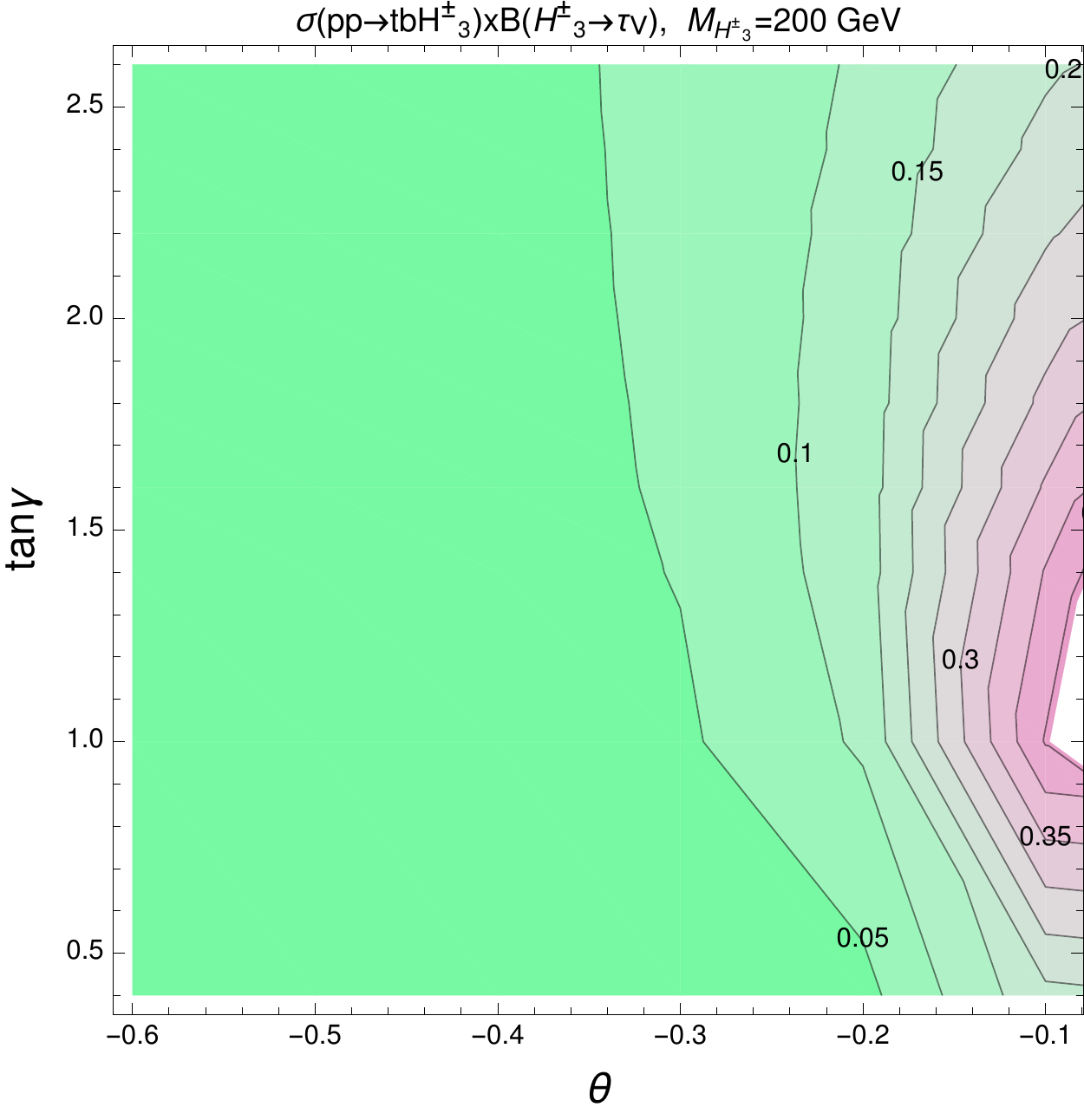}\\
	\caption{{Contour plots of $\sigma(pp\to tbH_i^\pm)\times {\rm BR}(H_i^\pm\to\tau\nu)$ for $H_2^{\pm}$ (left panels) and $H_3^{\pm}$ (right panels) on the ($\theta,\tan\gamma$) plane, for $\tan\beta = 20$, $\delta = 0.9\pi$, $M_{H_2^\pm} = 80$~GeV, and $M_{H_3^\pm}=170$~GeV (upper) and 200~GeV (lower). In the case of $H_2^\pm$ ($H_3^\pm$), the resulting rate is higher for lower (higher) values of $\theta$. The area in white is excluded by CMS~\cite{Sirunyan2019}. The lower the mass of $H_3^\pm$, the more challenging it is to find viable parameter space satisfying the direct experimental search bounds for both charged scalars.} }
	\label{Fig:munuconstraints}
\end{figure}

\begin{figure}
	\centering
\hspace*{-0.5truecm}
	\includegraphics[scale=0.4]{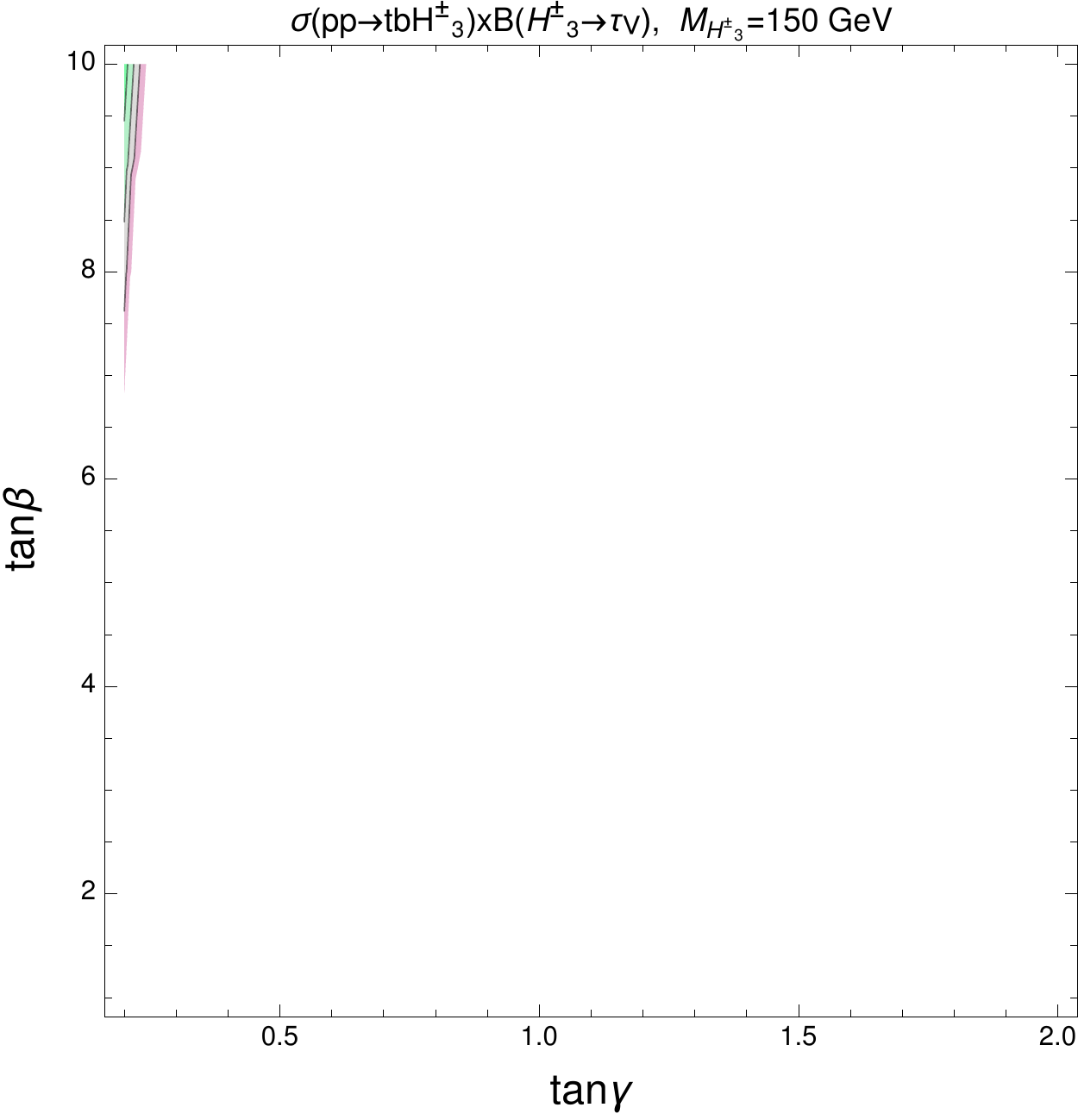}~
	\includegraphics[scale=0.4]{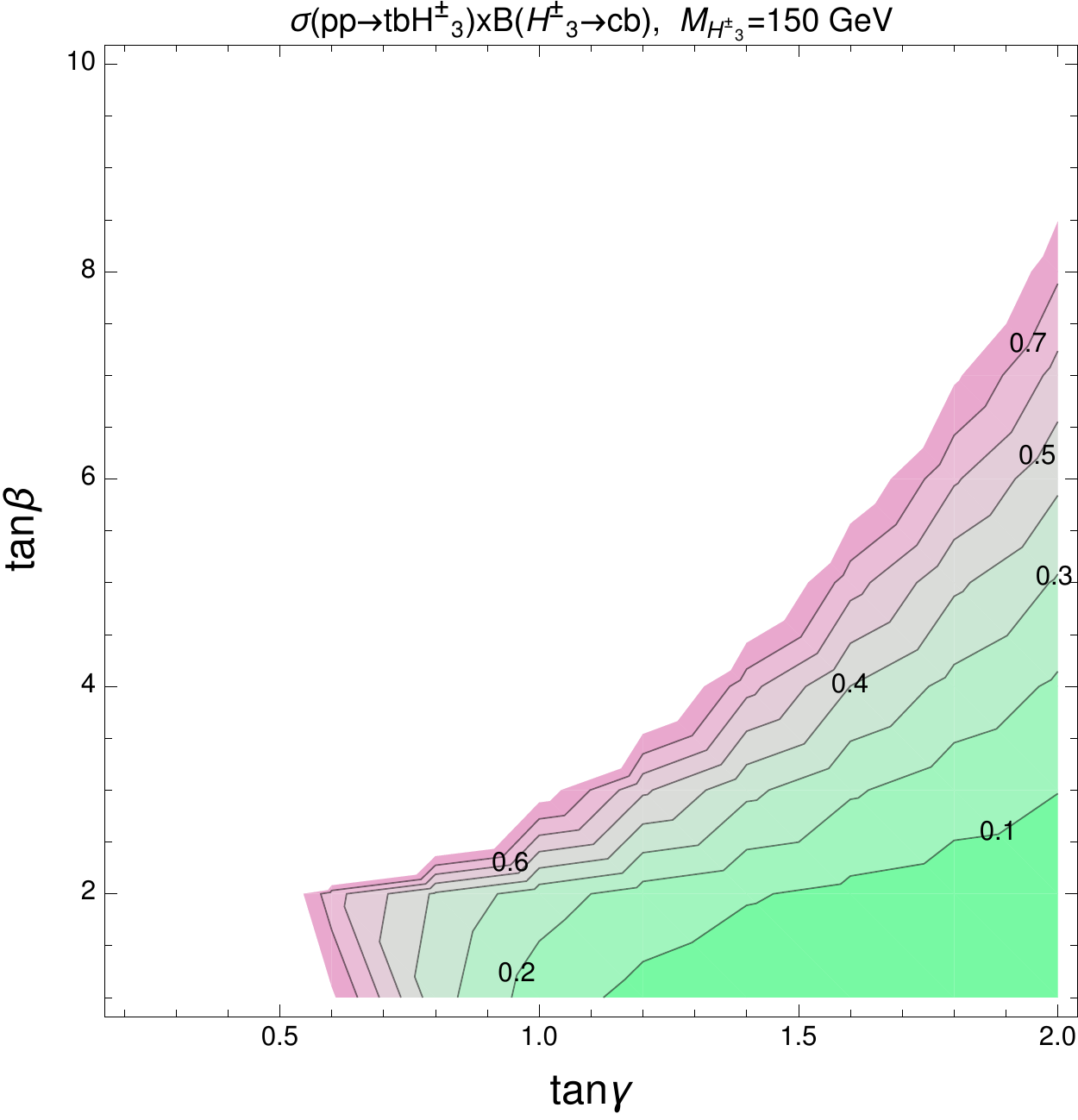}~
	\includegraphics[scale=0.4]{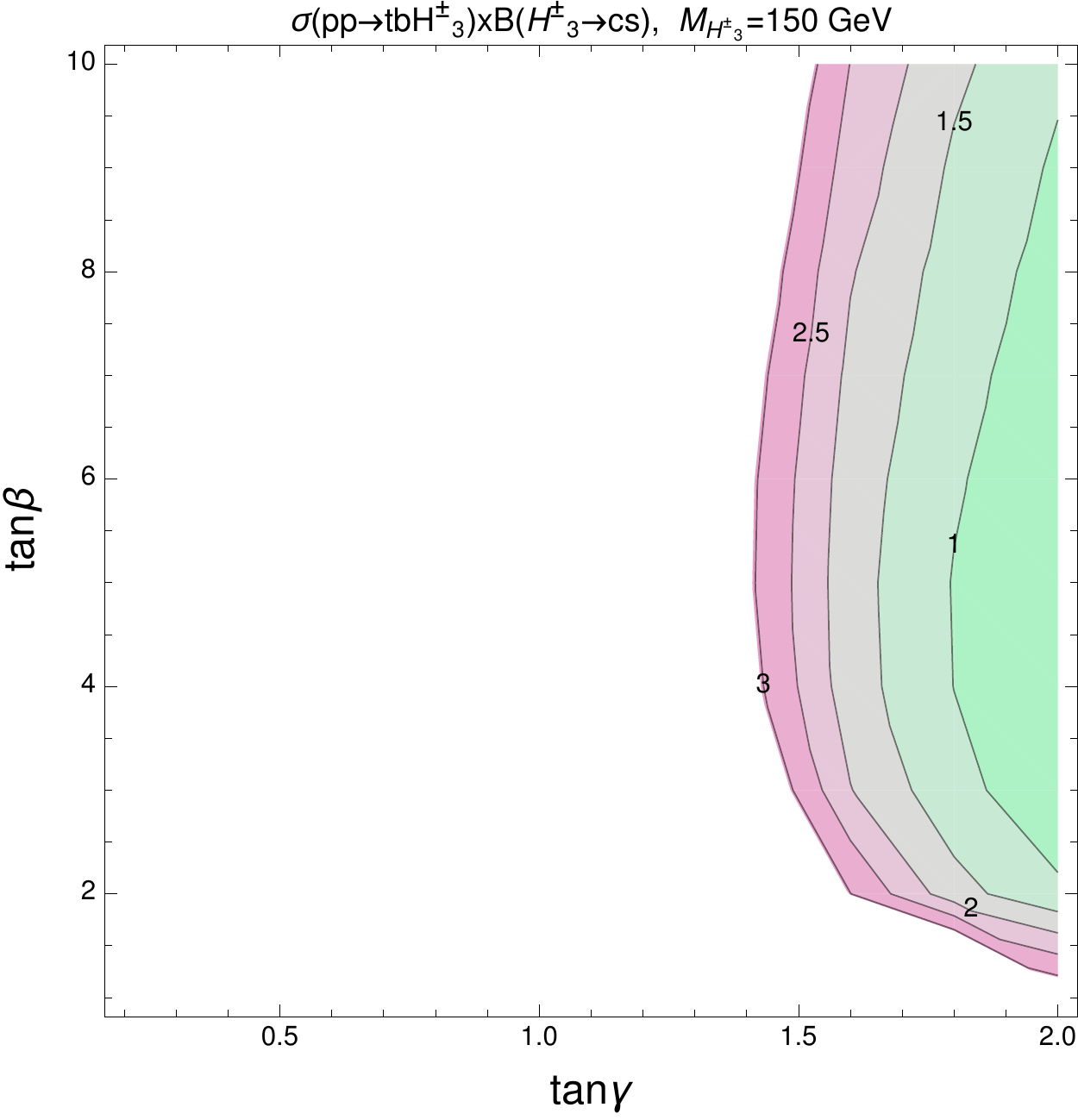}
	\caption{Production cross sections times BR for a 150~GeV heavier charged Higgs $H^\pm_3$ decaying to $\tau\nu$ (left), $cb$ (middle) and $cs$ (right) in the ($\tan\gamma,\tan\beta$) plane, for $M_{H^\pm_2} = 80$~GeV, $\theta=-0.5$, and $\delta = 0.95\pi$.  The white area represents the upper limits from LHC searches in Refs.~\cite{Sirunyan2019} (CMS), \cite{Sirunyan2018} (CMS), and \cite{Aad2013} (ATLAS), respectively.  Notice that the $H^\pm_3\to\tau\nu$ limits strongly exclude almost all of this scenario. The $H^\pm_2$ signal on the other hand is well below the collider limits.
}
\label{Fig:80150collider} 
\end{figure}


Charged Higgs boson parameters can also be constrained indirectly via measurements of the top-quark width, $\Gamma_t$, whenever $M_{H_i^\pm}<m_t$. 
{We add to the SM top quark width~\cite{Zyla2020} the partial width from the decays $t\to H^\pm_ib$, where \cite{Akeroyd2018}
\begin{equation}
	\Gamma(t\to H^\pm_ib) = \frac{G_Fm_t}{8\sqrt{2}\pi}\left[ m_t^2|Y_i|^2 + m_b^2|X_i|^2\right] \left(1 - M_{H_i^\pm}^2/m_t^2\right)^2,
\end{equation}
with $X_i,Y_i$ given in Eq. \eqref{XYZ}.}
This can be done by measuring $\Gamma_t$ from the top-quark visible decay products reconstructing its Breit-Wigner (BW) resonance.\footnote{Notice that, for our analysis, constraints obtained from measuring the single-top cross section are inapplicable, as these assume that $t\to bW^+$ is the only possible top-quark decay channel.}
According to Refs.~\cite{Zyla2020,ATLAS2019}, the most precise measurement to date is $\Gamma_t=(1.9\pm 0.5)$~GeV.
As can be seen from Fig.~\ref{Fig:topwidth},  to prevent the top-quark width from becoming too large, we need to select lower values of $\tan\beta$. 
Low values of $\tan\gamma$, in general, make the value of $\Gamma_t$ blow up. 
However, in a scenario  where the masses of the two charged Higgs bosons are close to the top-quark mass, we can still find very low values of $\tan\gamma$ that give an allowed $\Gamma_t$ value.
This will be relevant to find parameter space that can satisfy all constraints: from the top-quark width to collider searches for  $H^\pm_i$ states, EDMs, and $\bar{B}\to X_s\gamma$.

\begin{figure}
	\centering
	\includegraphics[scale=0.6]{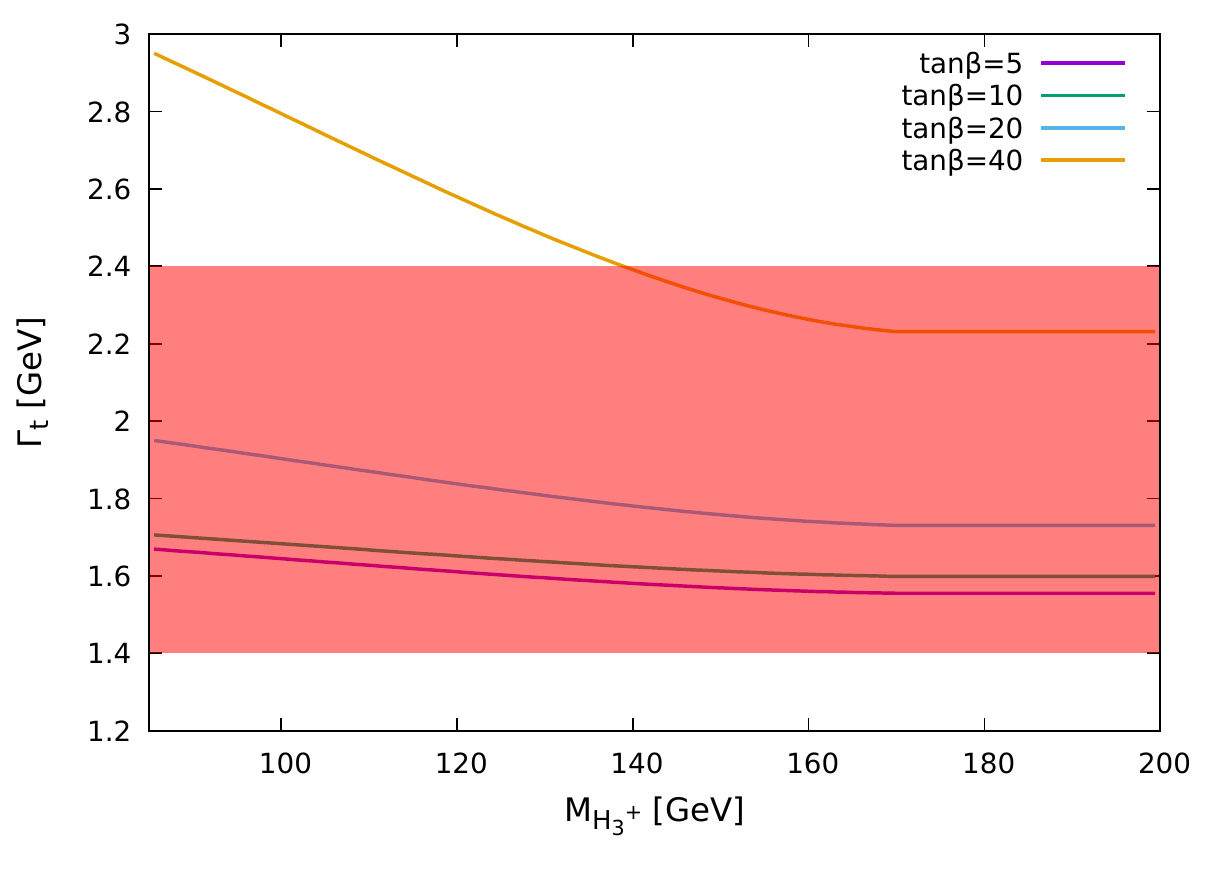} \\
	\includegraphics[scale=0.6]{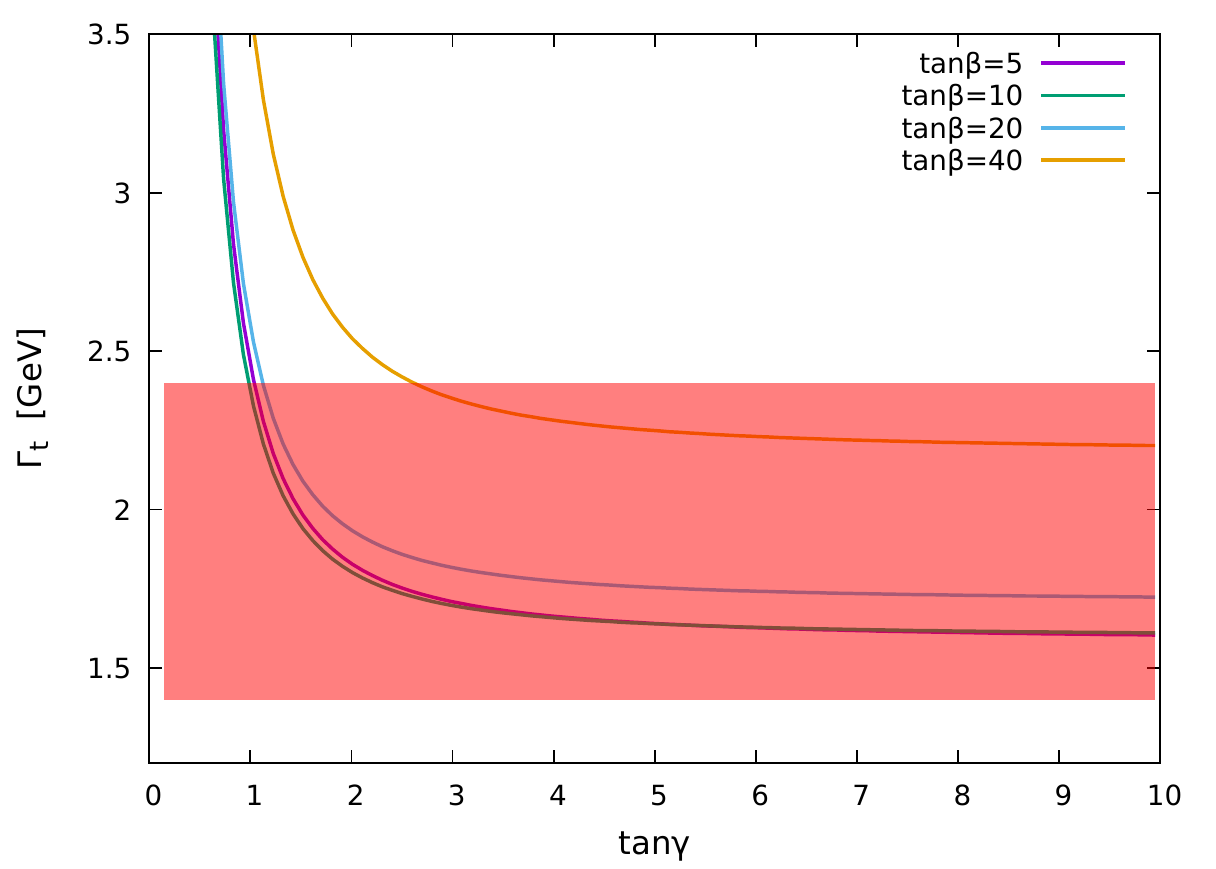}~
	\includegraphics[scale=0.6]{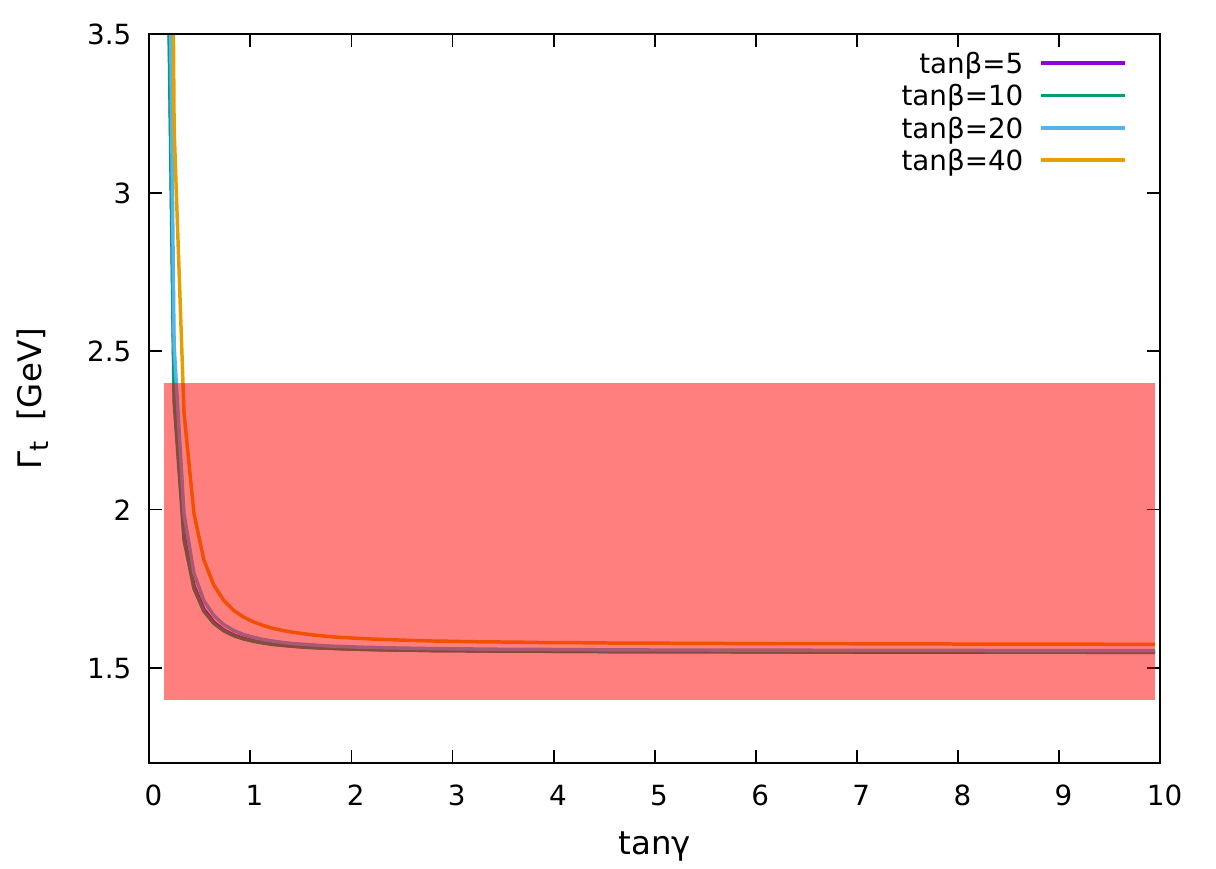}
	\caption{Predicted top quark width $\Gamma_t$ as a function of $M_{H_3^+}$ (upper) and $\tan\gamma$ (lower) in the Democratic 3HDM, for $\theta = -\pi/4$ and various values of $\tan\beta$.  
	In the upper panel $M_{H_2^+} = 85$~GeV, $\tan\gamma = 4$, and $\delta = 0.85 \pi$.
	The lower left panel has the same values of $M_{H_2^+}$ and $\delta$ but sets $M_{H_3^+} = 500$~GeV.
	In the lower right panel $M_{H_2^+} = 160$~GeV, $M_{H_3^+} = 170$~GeV, and $\delta = 0.9\pi$.
	The allowed range of top quark widths from Refs.~\cite{Zyla2020,ATLAS2019} lies in shaded red area.
}
	\label{Fig:topwidth}
\end{figure}

\subsection{Perturbativity constraints}

The ranges of $\tan\beta$ and $\tan\gamma$ can be constrained by requiring that the Yukawa couplings in Eq.~(\ref{YukLag}) remain sufficiently perturbative.  We adopt the approach introduced for the 2HDM in Ref.~\cite{Barger1990}, which required  the decay width $\Gamma_{H^+}$ of the charged Higgs boson into $t \bar b$ computed above the kinematic threshold to be no larger than $M_{H^+}/2$.  For example, at low $\tan\beta$, this leads to a constraint in the 2HDM Type-I of the form~\cite{Barger1990}
\begin{equation}
	\Gamma(H^+ \to t \bar b) \simeq \frac{3 G_F m_t^2}{4 \sqrt{2} \pi \tan^2\beta} M_{H^+} 
		< \frac{1}{2} M_{H^+}, \qquad {\rm or} \qquad \tan\beta \gtrsim 0.34,
\end{equation}
where we have used $m_t = 173$~GeV.  In the 2HDM Type-II, we can use the same approach to find an upper bound on $\tan\beta$, where at large $\tan\beta$ the bottom quark Yukawa dominates, and we have
\begin{equation}
	\Gamma(H^+ \to t \bar b) \simeq \frac{3 G_F m_b^2 \tan^2\beta}{4 \sqrt{2} \pi} M_{H^+}
		< \frac{1}{2} M_{H^+}, \qquad {\rm or} \qquad \tan\beta \lesssim 125,
\end{equation}
where we used $m_b \approx 4$~GeV (using the running bottom quark mass at the weak scale would yield an even higher upper bound on $\tan\beta$).  These bounds are generally loose compared to the ranges of $\tan\beta$ usually adopted in collider searches.  Nevertheless, we will adapt them to the 3HDM with this in mind.
However, 
the presence of two charged Higgs bosons in the 3HDM makes a direct adaptation of the above analysis rather opaque.  Instead, we interpret the constraints as upper bounds on the Yukawa couplings themselves, so that, applied to the 2HDM equivalent of Eq.~(\ref{YukLag}), these bounds on $\tan\beta$ are equivalent to imposing $\mathcal{G}_t \lesssim 3.07$ and $\mathcal{G}_b \lesssim 2.90$.

For uniformity we impose $\mathcal{G}_f \lesssim 3$ and derive constraints on $v_1 = v \cos\beta \sin\gamma$, $v_2 = v \sin\beta \sin\gamma$ and $v_3 = v \cos\gamma$ in the Democratic 3HDM using the $m_t$ and $m_b$ values quoted above (plus $m_{\tau} = 1.78$~GeV).  We find
\begin{equation}
	\sin\beta \sin\gamma \gtrsim 0.33, \qquad 
	\cos\beta \sin\gamma \gtrsim 0.0077, \qquad
	\tan\gamma \lesssim 290.
\end{equation}
The first two constraints yield an absolute lower bound on $\tan\gamma$,
\begin{equation}
	\tan\gamma \gtrsim 0.35.
\end{equation}

Later in this paper, we will show plots for $\tan\gamma = 1$ and 2.  For $\tan\gamma = 1$, the perturbativity analysis above requires $0.53 \lesssim \tan\beta \lesssim 92$ and the allowed $\tan\beta$ range expands as $\tan\gamma$ increases.

\section{Calculation of EDMs in the 3HDM}
\label{sec:edms}

In this section, we compute the dominant contributions to the electron and neutron EDMs from CP violation in the charged Higgs sector of the 3HDM.  All our results are obtained by a straightforward generalization of the charged Higgs contributions to EDMs that can arise in the 2HDM and are already available in the literature.

\subsection{Electron EDM from charged Higgs bosons in the 3HDM} 

Experimental sensitivity to the eEDM has improved by more than an order of magnitude in recent years, with a current upper bound from the ACME collaboration of~\cite{Andreev2018a}:
\be \label{de}
	|d_e| \leq 1.1 \times 10^{-29} \, e \, \text{cm}\, \, (90\% \,\, \text{C.L.}).
\ee

The charged Higgs bosons in the 3HDM give rise to contributions to the eEDM via the CP violation in their couplings to fermion pairs.  The one-loop contribution involving a charged Higgs loop is subdominant due to suppression by the tiny electron Yukawa coupling.  The dominant contribution comes from the two-loop Barr-Zee type diagrams as shown in Fig.~\ref{fig:barrzeetb}, first calculated in Ref.~\cite{BowserChao1997} in the 2HDM (see also Ref.~\cite{Jung2014}).

\begin{figure}[t]
\begin{center}\resizebox{0.5\textwidth}{!}{\includegraphics{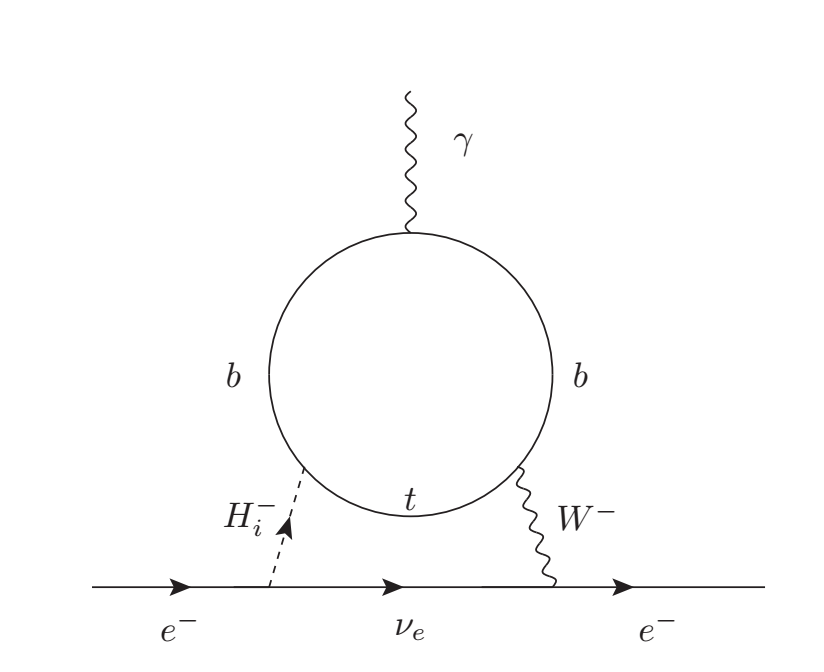}}\end{center}
\caption{One of the Barr-Zee type diagrams that give the dominant charged Higgs boson contribution to the eEDM in the 3HDM.}
\label{fig:barrzeetb}
\end{figure}

The charged Higgs sector also appears in the Barr-Zee type diagrams of Fig.~\ref{fig:barrzeeHp}, where $\phi^0$ is any of the neutral scalars in the model.  It was pointed out in Ref.~\cite{Kanemura2020}, in the context of the Aligned 2HDM, that these diagrams can contribute significantly and lead to interesting cancellations with the diagrams of Fig.~\ref{fig:barrzeetb}.  In the 3HDM scenario that we consider here, where CP violation is present in the charged Higgs sector but not in the neutral Higgs sector (which we have integrated out), these diagrams do not contribute to the eEDM because the $\phi^0 ee$ and $\phi^0 H_i^+ H_i^-$ couplings contain no CP phase.\footnote{It can be seen that, in the absence of neutral (pseudo)scalar sector CP violation, the latter coupling cannot contain a CP-violating phase because this term is Hermitian by itself and hence must have a real coefficient in the Lagrangian.}  The couplings $\phi^0 H_2^+ H_3^-$ do contain non-trivial CP phases, but these couplings do not appear in the diagrams of Fig.~\ref{fig:barrzeeHp} because the photon coupling to the charged Higgs boson is diagonal.

\begin{figure}[t]
\resizebox{\textwidth}{!}{\includegraphics{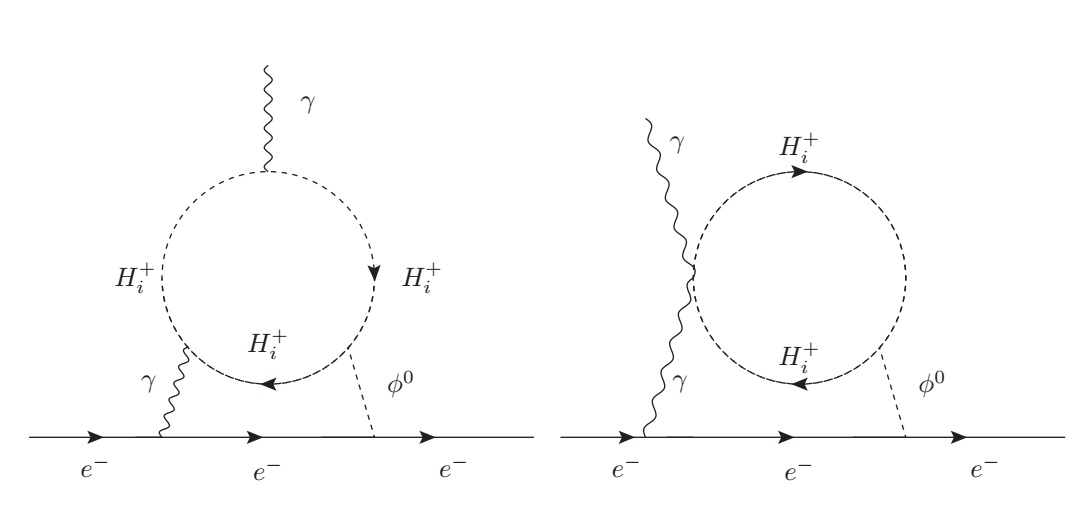}}
\caption{Two of the Barr-Zee type diagrams for the eEDM involving a charged Higgs boson in the loop.  These do not contribute in the 3HDM when CP violation is turned off in the neutral Higgs sector, as we assume in this paper.}
\label{fig:barrzeeHp}
\end{figure}

Under our assumption that the neutral Higgs sector is CP-conserving and that CP violation appears only in the charged Higgs sector, the dominant Barr-Zee type contribution of the charged Higgs to the eEDM in the 2HDM~\cite{BowserChao1997,Jung2014} can be generalized to the 3HDM as follows:
\begin{eqnarray}
\frac{d_e(M_{H^{\pm}_2},M_{H^{\pm}_3})_{BZ}}{2} &=& - m_e \frac{12 G^2_F M^2_W}{(4\pi)^4} |V_{tb}|^2   \nonumber  \\
&\times & \bigg[{\text{Im}} (- Y_2^*Z_2) \left(q_t F_t(\textit{z}_{H^{\pm}_2},\textit{z}_W) + q_b F_b(\textit{z}_{H^{\pm}_2},\textit{z}_W)\right)   \nonumber \\
&+& {\text{Im}} (- Y_3^*Z_3) \left(q_t F_t(\textit{z}_{H^{\pm}_3},\textit{z}_W) + q_b F_b(\textit{z}_{H^{\pm}_3},\textit{z}_W)\right) \bigg],
\label{eq: deforH}
\end{eqnarray}
where $q_t = 2/3$ and $q_b = -1/3$ are quark electric charges, $z_{a} =M_a^2/m_t^2$ and~\cite{BowserChao1997,Jung2014} 
\begin{eqnarray}
F_q(\textit{z}_{H^{\pm}_i},\textit{z}_W) &=& \frac{T_q(\textit{z}_{H^{\pm}_i}) - T_q(\textit{z}_W) }{\textit{z}_{H^{\pm}_i} - \textit{z}_W},   \nonumber \\
T_t(\textit{z}) &=& \frac{1- 3\textit{z}}{\textit{z}^2} \frac{\pi^2}{6} + \bigg(\frac{1}{\textit{z}} - \frac{5}{2}\bigg) \text{log}\textit{z} -\frac{1}{\textit{z}} - \bigg(2 - \frac{1}{\textit{z}}\bigg) \bigg(1 - \frac{1}{\textit{z}}\bigg) \text{Li}_2 (1 -\textit{z}),  \nonumber \\
T_b(\textit{z}) &=& \frac{2\textit{z} - 1}{\textit{z}^2} \frac{\pi^2}{6} + \bigg(\frac{3}{2} - \frac{1}{\textit{z}}\bigg) \text{log}\textit{z} + \frac{1}{\textit{z}} - \frac{1}{\textit{z}} \bigg(2 - \frac{1}{\textit{z}}\bigg) \text{Li}_2 (1 -\textit{z}).
\end{eqnarray}

Note that the original calculation of Ref.~\cite{BowserChao1997} was done setting $m_b = 0$ so that only the contribution involving the top-quark Yukawa couplings $m_t Y_i/v$ appears.  Keeping the non-zero bottom mass would introduce additional contributions proportional to $m_b X_i/v$, which could become important at large values of $\tan\beta$. Finally, all other eEDM contributions at the loop level that are purely fermionic or induced by gauge bosons 
\cite{Pospelov1991,Chang1991} remain identical to those in the SM and are negligible compared to the current experimental bound.

\subsection{Neutron EDM from charged Higgs bosons in the 3HDM}

The current measurement of the nEDM at the Paul Scherrer Institute with ultra-cold neutrons (UCN) provided an upper limit as follows~\cite{Abel2020}:\footnote{{If the (unrealistic) assumption is made that the nEDM is the sole
contribution to the atomic EDM of mercury, the most recent measurement of
the  latter yields a comparable limit, $|d_n| < 1.8 \times 10^{-26} e$~cm at 90\% C.L. \cite{Graner2016ses}.}}
\be\label{dn}
|d_n| \leq 1.8 \times 10^{-26} \, e \, \text{cm} \, \, (90\% \,\, \text{C.L.}).
\ee

CP violation from charged Higgs boson exchange enters this observable through a variety of effective operators.  Jung and Pich~\cite{Jung2014} point out three types of effective operators through which the charged Higgs boson contributes to the nEDM in the 2HDM. These are four-fermion operators involving the up- and down-type quarks which are induced by CP-violating Higgs exchange, the Weinberg operator (the CP-violating three-gluon operator) which is neither suppressed by quark masses nor CKM matrix elements, and the Barr-Zee type two-loop diagrams contributing to the EDMs and chromo-electric dipole moments (CEDMs) of the up- and down-type quarks. The light quark masses suppress the contributions of the four fermion operators and the up- and down-type quark (C)EDMs.  

\begin{figure}[t]
\resizebox{0.5\textwidth}{!}{\includegraphics{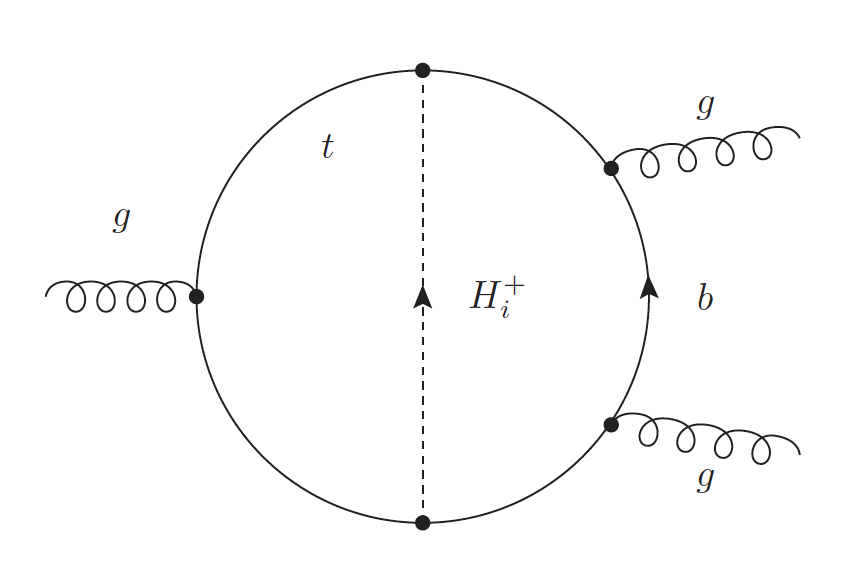}}
\resizebox{0.5\textwidth}{!}{\includegraphics{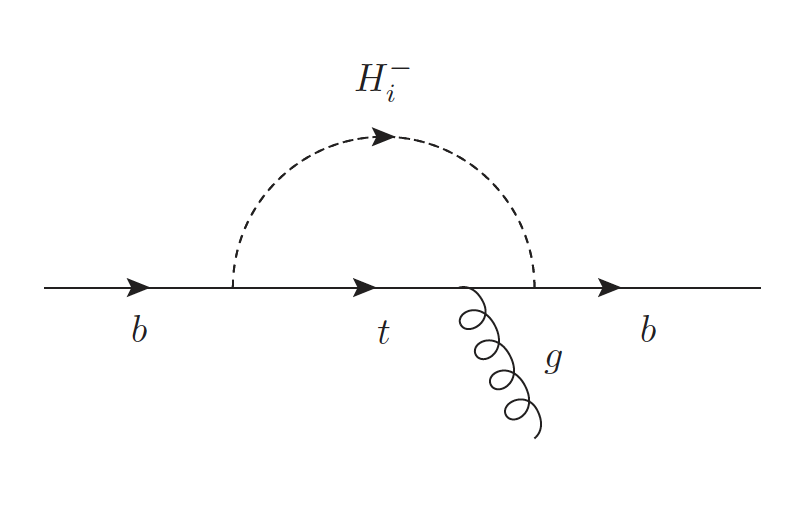}}
\caption{Left panel: Two-loop charged Higgs boson contribution to the Weinberg operator. 
Right panel: One-loop charged Higgs boson contribution to the bottom quark CEDM.  }
\label{fig:bCEDM}
\end{figure}

This leaves the Weinberg operator, the charged Higgs contribution to which is shown in the left panel of Fig.~\ref{fig:bCEDM}.  Following Ref.~\cite{Jung2014}, we compute this using an effective field theory approach~\cite{Braaten1990}, which amounts to computing only the one-loop short-distance piece at the high scale $\mu_{tH} = m_t$, which is the bottom quark CEDM shown in the right panel of Fig.~\ref{fig:bCEDM}.  

The contribution of the Weinberg operator to the nEDM is~\cite{Jung2014}:
\begin{equation}\label{formula1}
|d_n(C_W)/e|  = \left[1.0  {+1.0 \atop -0.5} \right] \times 20\,  \text{MeV} \,  C_W(\mu_h),
\end{equation}
where the sign is unknown, and the theoretical uncertainty on the magnitude is a factor of two.  In our numerical results, we follow Ref.~\cite{Jung2014} and use the central theoretical value.  
The Wilson coefficient $C_W$ evaluated at the hadronic scale $\mu_h \sim 1$~GeV is expressed as
\begin{equation}\label{formula2}
C_W(\mu_h) = \eta^{\kappa_W}_{c-h}  \eta^{\kappa_W}_{b-c} \bigg( \eta^{\kappa_W}_{t-b} C_W(\mu_{tH}) +  \eta^{\kappa_C}_{t-b} \frac{g^3_s(\mu_{b})}{8\pi^2 m_b} \frac{d^C_b(\mu_{tH})}{2} \bigg),
\end{equation}
where $C_W(\mu_{tH}) = 0$ because there is no short-distance contribution to the Weinberg operator involving the charged Higgs boson at the scale $m_t$.
$d_b^C(\mu_{tH})$ is the short-distance contribution to the bottom quark CEDM, given below.  The running of these short-distance contributions down to the scale $\mu_b = m_b$ is accomplished by the factors of $\eta_{t-b} = \alpha_s(\mu_{tH})/\alpha_s(\mu_b)$ raised to the appropriate power $\kappa_i = \gamma_i/(2 \beta_0$), where $\gamma_W = N_C + 2 n_f$ and $\gamma_C = 10 C_F - 4 N_C$ are the leading order (LO) anomalous dimensions of the Weinberg  and  $b$-quark CEDM operator, respectively, and $\beta_0 = (11 N_C - 2 n_f)/3$ is the one-loop beta function of QCD.  Here, $N_C = 3$, $C_F = 4/3$, and $n_f$ is the number of active quark flavors involved in the QCD running at the relevant scale (e.g.,  between the top and bottom masses, $n_f = 5$).  At the scale $\mu_b$, the bottom quark is integrated out and the operators matched, then the remaining Weinberg operator is run down to the hadronic scale $\mu_h$ in two steps (integrating out the charm quark at $\mu_c = m_c$), giving rise to two more factors, $\eta_{b-c}^{\kappa_W}$ and $\eta_{c-h}^{\kappa_W}$, in which the running of $\alpha_s$ and the exponent are evaluated with the appropriate value of $n_f$. At LO, $\alpha_s(\mu)$ is given by:
\begin{equation}
\alpha_s(\mu) = \frac{\alpha_s(M_Z)}{v(\mu)},
\end{equation}
with 
\begin{equation}
v(\mu) = 1 - \beta_0 \frac{\alpha_s(M_Z)}{2\pi} \log \bigg(\frac{M_Z}{\mu} \bigg).
\end{equation}

Finally, the high-scale one-loop charged Higgs boson contribution to the bottom quark CEDM in the right panel of Fig.~\ref{fig:bCEDM} has been calculated in the 2HDM in Ref.~\cite{Jung2014} (see also references therein). By adapting this to the 3HDM, one obtains  
\begin{eqnarray}
\frac{d^C_b(\mu_{tH})}{2} &=& - \frac{G_F}{\sqrt{2}} \frac{1}{16\pi^2}  |V_{tb}|^2 m_b(\mu_{tH}) 
	\left[{\text{Im}}(-X_2Y_2^*) x_{tH_2} \bigg(\frac{\log(x_{tH_2})}{(x_{tH_2} - 1)^3} + \frac{(x_{tH_2} - 3)}{2(x_{tH_2} - 1)^2}\bigg) \right. \nonumber \\
	&& \qquad \qquad \qquad \left. + {\text{Im}}(-X_3Y_3^*) x_{tH_3} \bigg(\frac{\log(x_{tH_3})}{(x_{tH_3} - 1)^3} + \frac{(x_{tH_3} - 3)}{2(x_{tH_3} - 1)^2}\bigg) \right],
\label{formula7}
\end{eqnarray}
where $x_{tH_i} = m_t^2 / M_{H^{\pm}_i}^2$.  Again, purely fermionic and gauge contributions \cite{Jung2014} remain identical to those in the SM and are negligible compared to the current experimental bound.

\section{Cancellation in the charged Higgs contributions to the EDMs}
\label{sec:hiding}

The CP-violating phase in the charged Higgs mixing matrix is responsible for generating CP-violating observables in this model.  The effects of this CP-violating phase in processes involving virtual charged Higgs boson exchange can be arbitrarily suppressed by making the two physical charged Higgs bosons sufficiently degenerate in mass, thereby avoiding constraints from EDMs.  This can be understood as a consequence of an analogue of the GIM mechanism~\cite{Glashow1970}, in particular, when $H^\pm_2$ and $H^\pm_3$ become degenerate, both the mixing angle $\theta$ and the CP-violating phase $\delta$ in their mixing matrix become non-physical.

Any internal charged Higgs propagator that begins and ends on a fermion line brings with it one factor of $X_i^*$, $Y_i^*$ or $Z_i^*$ and one factor of $X_i$, $Y_i$ or $Z_i$.  The combinations $X_iX_i^*$, $Y_iY_i^*$, and $Z_iZ_i^*$ are purely real and cannot contribute to CP-odd observables, leaving only the combinations $X_iY_i^*$, $X_iZ_i^*$ and $Y_iZ_i^*$ (or their complex conjugates) which can have an imaginary part.  Consider, for example, $X_iY_i^*$, which is given in the Democratic 3HDM in terms of the unitary rotation matrix in Eq.~(\ref{eq:Uexplicit}) by 
\begin{equation}
	X_i Y_i^* = - \frac{U^{\dagger}_{1i} U_{i2}}{U^{\dagger}_{11} U_{12}},
\end{equation}
where $i = 2$ or 3.  The denominator is real by construction since $U_{1j} = v_j/v$.  Computation of CP-odd observables in this context always involves a sum over the two charged Higgs bosons that can appear in the contributing diagrams, yielding
\begin{equation}
	\sum_{i = 2}^3 {\rm Im} (X_i Y_i^*) f(M_{H^+_i})
	= - \frac{1}{U_{11}^{\dagger} U_{12}} \left[ {\rm Im} (U_{12}^{\dagger} U_{22}) f(M_{H^+_2})
		+ {\rm Im} (U_{13}^{\dagger} U_{32}) f(M_{H^+_3}) \right],
	\label{eq:gimmech1}
\end{equation}
where $f(M_{H^+_i})$ represents the dependence of the diagram on the charged Higgs boson mass.  We can trivially add zero in the form of ${\rm Im} (U_{11}^{\dagger} U_{12}) f(m)$ inside the square brackets.  Then, in the limit $M_{H^\pm_2} = M_{H^\pm_3} \equiv m$, Eq.~(\ref{eq:gimmech1}) becomes
\begin{equation}
	\sum_{i = 2}^3 {\rm Im} (X_i Y_i^*) f(m) 
	=  - \frac{1}{U_{11}^{\dagger} U_{12}} {\rm Im} \left[ \sum_{i = 1}^3 U_{1i}^{\dagger} U_{i2} \right] f(m) 
	= - \frac{1}{U_{11}^{\dagger} U_{12}} {\rm Im} (\delta_{12}) f(m) = 0,
\end{equation}
where $\delta_{12}$ is the $(1,2)$ element of the Kronecker delta.
This also shows that ${\rm Im}(X_2 Y_2^*) = - {\rm Im}(X_3 Y_3^*)$, due to the unitarity of the charged Higgs mixing matrix, and similarly for the imaginary parts of $X_iZ_i^*$ and $Y_iZ_i^*$.
The form of Eq.~(\ref{eq:gimmech1}) also implies that, for small non-zero mass splitting $\Delta M_{H^\pm} \ll M_{H^\pm}$, CP-violating amplitudes must be linear in $\Delta M_{H^\pm} /M_{H^\pm}$, where $\Delta M_{H^\pm} \equiv M_{H^\pm_3} - M_{H^\pm_2}$ and $M_{H^\pm} \equiv (M_{H^\pm_3} + M_{H^\pm_2})/2$.\footnote{{The degeneracy of the charged Higgs boson masses favored by the avoidance of EDM constraints raises the possibility of interesting interference effects in direct collider production of on-shell charged Higgs bosons, if their mass splitting is comparable to or smaller than the decay widths of the two charged Higgs bosons so that the BW lineshapes of their decay products overlap in phase space. 
	 Unfortunately, for the case when both charged Higgs boson masses are below $m_t$, not only are their decay widths extremely narrow (as illustrated already), but it is also very difficult (maybe impossible) to find a viable set of model parameters that are not already ruled out by collider searches for which such a degeneracy can be achieved.  
	 For charged Higgs boson masses above $m_t$, however, collider constraints are much less stringent and the decay widths are larger, so that such a lineshape overlap could offer interesting future possibilities for experimental exploration. 
	 For example, for the set of values in Fig.~17, the width of $H_3^\pm$ is of the order of 1 GeV when $M_{H^\pm_2}=M_{H^\pm_3}=200$ GeV and of the order of $10^{2}$ GeV for $M_{H^\pm_2}=M_{H^\pm_3}=800$ GeV whereas the width of $H_2^\pm$ is of the order of $10^{-3}$ GeV when $M_{H^\pm_2}=M_{H^\pm_3}=200$ GeV and of the order of 0.1 GeV for $M_{H^\pm_2}=M_{H^\pm_3}=800$ GeV.	} 	}

In this paper, we focus on the Democratic 3HDM because CP violation in the charged Higgs sector gives rise to interesting contributions to the EDMs of both the electron and  neutron.  In the other types of 3HDM, the effects of charged Higgs CP violation are more limited because, in these models, at least two of $X_i$, $Y_i$, and $Z_i$ become identical (see Tab.~\ref{tab:couplingfactors}).  In particular, the dominant charged Higgs contribution to the eEDM, proportional to ${\rm Im}(-Y_i^* Z_i)$, is zero in the Type-I and Type-Y (Flipped) 3HDMs because in those models $Y_i = Z_i$.  Similarly, the dominant charged Higgs contribution to the nEDM, proportional to ${\rm Im}(-X_i Y_i^*)$, is zero in the Type-I and Type-X (Lepton-specific) 3HDMs because in those models $X_i = Y_i$.  In the Type-II 3HDM, $X_i = Z_i$,  so that this model also leads to CP-violating charged Higgs boson contributions to both the electron and neutron EDMs.

\section{Numerical results}
\label{sec:numerics}

We now present our results for the Democratic 3HDM as a function of the relevant coupling parameters ($\theta$, $\tan\beta$, $\tan\gamma$, and $\delta$)
and masses ($M_{H^{\pm}_{2}}$ and $M_{H^{\pm}_{3}}$)  against the eEDM and nEDM constraints. 
We will also impose the constraints from direct searches for charged Higgs bosons, as well as from the measurement of BR($\bar B \to X_s \gamma$), which provides the most stringent indirect constraint on the charged Higgs masses.  Details of our implementation of the $\bar B \to X_s \gamma$ constraint are given in Appendix~\ref{sec:bsg}.

To start with, 
it is instructive to compare the 3HDM results with those available in the literature for the analogous case in a 2HDM, which we do
by presenting the nEDM and eEDM constraints against the Yukawa coupling combinations Im$(X_iY_i^*)$ and Im$(Y_i^*Z_i)$ ($i=2$).  In  Fig.~\ref{2hdmnedm}, we show the Aligned 2HDM results in the plane of the charged Higgs mass and the imaginary part of the relevant combination of Yukawa coupling factors, to be compared to  Figs.~3, 4, and 5 of Ref.~\cite{Jung2014},\footnote{Herein, there is no subscript 2 for the couplings and masses of the 2HDM, as only one charged Higgs state is present in the model.} updated using the latest nEDM and eEDM experimental limits as given in Eqs.~(\ref{dn}) and (\ref{de}), respectively.  The shaded areas in Fig.~\ref{2hdmnedm} represent the viable parameter regions in both cases. The newest bounds from both nEDM and eEDM  induce a strong suppression on  the allowed parameter space corresponding to the  imaginary contributions of the couplings $X_2Y_2^*$  and $Y_2^*Z_2$. In Figs. \ref{3hdmnedm300} and \ref{3hdmnedm500}, we show the 3HDM cases as a function of $M_{H_2^{\pm}}$ with $M_{H^{\pm}_3} = 85$ and 300 GeV, respectively. We can then see that the parameter space is generally enlarged in the Democratic 3HDM with respect to the Aligned 2HDM, particularly in the $M_{H_2^\pm}=M_{H_3^\pm}$ limit, clearly  illustrating the aforementioned cancellation mechanism between the two charged Higgs states of the 3HDM. It is worth noticing here that, while in the exact mass degeneracy case there is virtually no constraint applicable to the Democratic 3HDM from either nEDM or eEDM, even when the $M_{H_2^\pm}=M_{H_3^\pm}$ condition is lifted,   there are substantial differences in the values allowed for the Yukawa couplings between the two scenarios at both small and large values of the lightest charged Higgs boson mass.

Next, we consider the effect of the coupling parameters $\theta$, $\tan\beta$, $\tan\gamma$, and $\delta$ for various scenarios for the charged Higgs masses $M_{H^{\pm}_2}$ and $M_{H^{\pm}_3}$ within the Democratic 3HDM.  We consider two classes of mass scenarios: the first in which either or both $H_i^\pm$ masses are  lighter than $ m_t$  (in Sec.~\ref{sec: lightmass}) and the second in which they are both heavier than $m_t$ (in Sec.~\ref{sec: heavymass}).  Explicit expressions for the parameter combinations ${\rm Im}(-X_2 Y_2^*)$ and ${\rm Im}(-Y_2^* Z_2)$ that enter the calculations of the EDMs are given in Appendix~\ref{sec:appB}; in particular we note that these quantities are proportional to $\sin\delta$ and to the product $\sin\theta \cos\theta$, so that the CP-violating effects are largest when $\delta = \pi/2$ or $3\pi/2$ and $\theta = -\pi/4$.

\begin{figure}[t!]
    \centering
     \begin{subfigure}
    {
        \includegraphics[scale=.45]{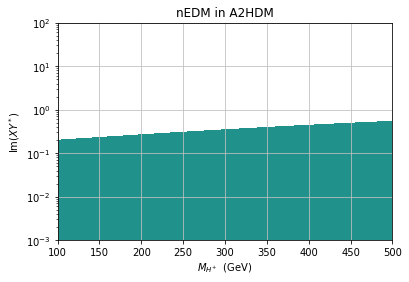}
    }
    \hfill
     {
        \includegraphics[scale=.45]{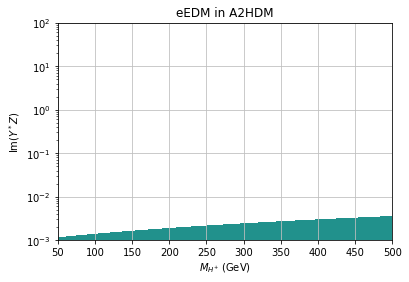}
    }
    \caption{Constraint from the nEDM (left) and the eEDM (right) on $| {\rm Im}(X Y^*)|$ and $| {\rm Im}(Y^*Z)|$, respectively, in the Aligned 2HDM as a function of the charged Higgs mass ($M_{H^+}$).  The blue shaded region is allowed.
    }  \label{2hdmnedm}
    \vspace*{0.75cm}
     \subfigure
    {
        \includegraphics[scale=.45]{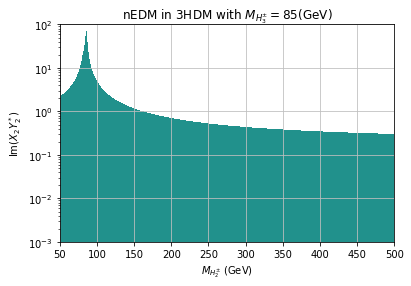}
    }
     \hfill
    {
        \includegraphics[scale=.45]{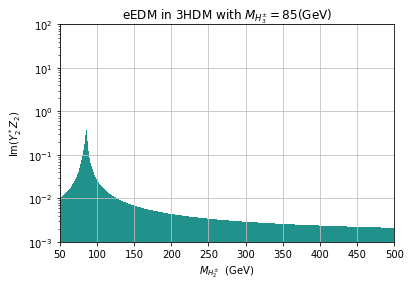}
    }
    \caption
    {Constraint from the nEDM (left) and the eEDM (right) on $|{\rm Im}(X_2 Y_2^*)|$ and $|{\rm Im}(Y_2^*Z_2)|$ in the 3HDM as a function of the mass of $H_2^+$.  $M_{H_3^+}$ is fixed to be 85~GeV.  The structure of the model forces Im$(X_3Y_3^{*}) = - {\rm Im}(X_2Y_2^{*})$ and Im$(Y_3^*Z_3) = -{\rm Im}(Y_2^*Z_2)$.
}  \label{3hdmnedm300}
    \vspace*{0.75cm}
    \subfigure
    {
        \includegraphics[scale=.45]{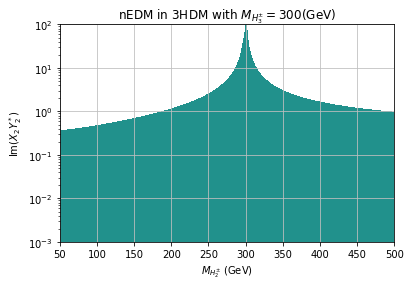}
    }
      \hfill
    {
        \includegraphics[scale=.45]{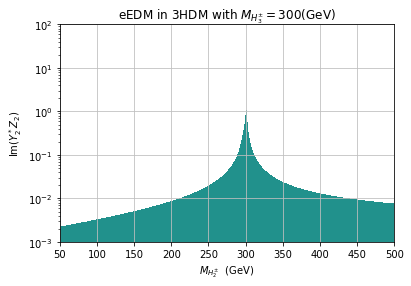}
    }
    \end{subfigure}
    \caption
    {Same as in Fig.~\ref{3hdmnedm300} but for $M_{H^{\pm}_3} = 300$ GeV.
    } 
    \label{3hdmnedm500}
\end{figure}

\subsection{Light charged Higgses} \label{sec: lightmass}

\subsubsection{The $M_{H_2^\pm}<m_t<M_{H_3^\pm}$ case}

In Fig.~\ref{Fig:80,200EDM}, we show the constraints from $\bar{B}\to X_s\gamma$, eEDM and nEDM on the $[\delta,\theta]$ plane, for $M_{H_2^+} = 80$~GeV, $M_{H_3^+} = 200$~GeV, and small values of $\tan\beta$ and $\tan\gamma$ so as to be compliant with collider limits, as seen previously.
 Notice that the $\bar{B}\to X_s\gamma$ constraint is satisfied within the green and grey shaded areas while 
the two EDM constraints are satisfied outside the corresponding closed curves.  (Details of our calculation of the $\bar{B}\to X_s\gamma$ constraint are given in Appendix~\ref{sec:bsg}.) {The shaded areas correspond to the $\pm 2 \sigma$ allowed region of BR($\bar B \to X_s \gamma$), with the green (grey) area corresponding to values below (above) the experimental central value.} From these plots, we learn that we need $\delta$ to be very close to $\delta = n \pi$  to satisfy all three constraints at once.
That is, we are forced to find solutions very close to the CP-conserving limit; furthermore, the constraint from $\bar B \to X_s \gamma$ furthermore tends to favour $\delta \simeq \pi$.

In Fig.~\ref{fig:deltatheta20}, we show the effect of varying $\tan\gamma$ and increasing the mass of the heavier charged Higgs state while keeping $M_{H_2^\pm} = 80$~GeV and fixing $\tan\beta = 20$.
As can be seen, increasing $M_{H_3^\pm}$ from 200 to 500 GeV makes it more difficult to find regions that can survive all constraints, in line with the requirements of the aforementioned cancellation mechanism.  Comparing with Fig.~\ref{Fig:80,200EDM} we also see that larger values of $\tan\beta$ lead to tighter constraints from the nEDM while larger values of $\tan\gamma$ lead to tighter constraints from the eEDM.

In Fig.~\ref{Fig:80,200constraints}, we show the same constraints on the $[\tan\gamma,\tan\beta]$ plane instead, {for $\theta = -0.3$ and two characteristic values of $\delta$ chosen to be very close to $\pi$, i.e., $\delta = 0.975 \pi$ and $0.985\pi$.
We have also added here the constraints from the top-quark width and perturbativity of the $H_i^+ b\bar t$ vertex. The allowed region is the portion of the green and grey shaded areas that lies to the right of the black dotted line and above the blue curve. For all the parameter regions shown, the collider limits are satisfied.
We can see that, for $\tan\gamma > 1.5$ and $\tan\beta > 8$, we can satisfy all other constraints for these values of $\delta$.}

\begin{figure}[h!]
	\centering
\includegraphics[scale=0.5]{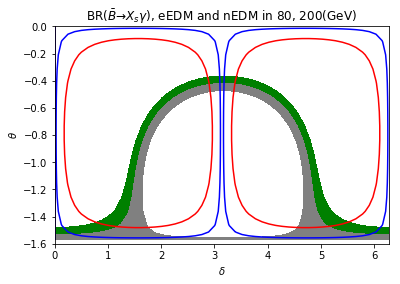}\includegraphics[scale=0.5]{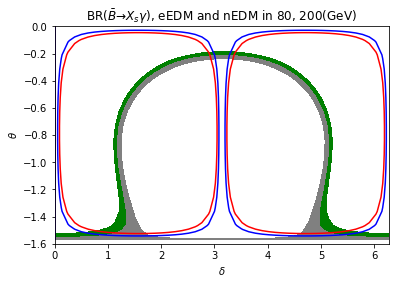}
	\caption{The allowed regions from $\bar{B} \to X_s \gamma$ (within the green and grey shaded areas), eEDM (outside the blue curves), and nEDM (outside the red curves) in the [$\delta, \theta$] plane, with $M_{H_2^+} = 80$~GeV, $M_{H_3^+} = 200$~GeV, $\tan\gamma = 1$, and $\tan\beta = 5$ (left) or 10 (right).  {Here, the shaded areas correspond to the $\pm 2 \sigma$ allowed region of BR($\bar B \to X_s \gamma$), with the green (grey) area corresponding to values below (above) the experimental central value.} }
	\label{Fig:80,200EDM}
\end{figure}

\begin{figure}[h!]
\centering
\includegraphics[scale=0.5]{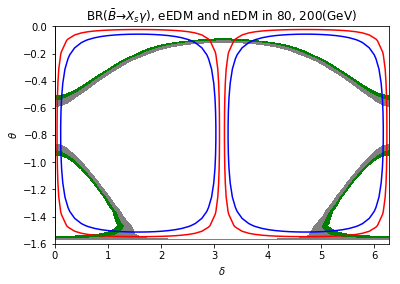}
\includegraphics[scale=0.5]{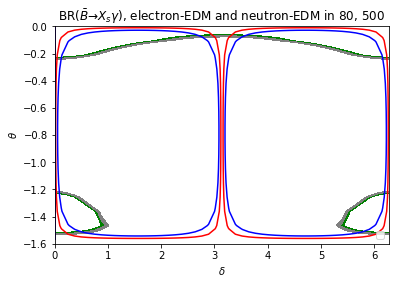}\\
\includegraphics[scale=0.5]{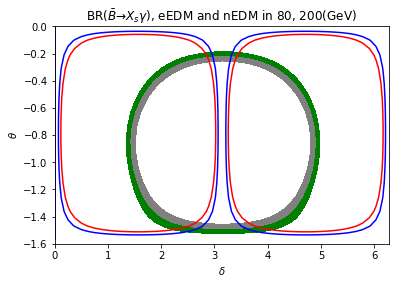}
\includegraphics[scale=0.5]{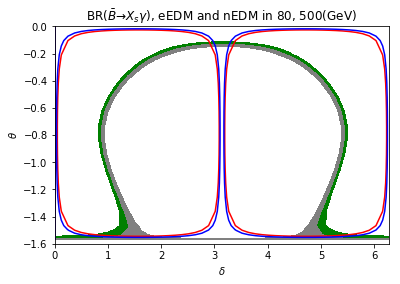}
\caption{The allowed regions from $\bar{B} \to X_s \gamma$ (within the green and grey shaded areas), eEDM (outside the blue curves), and nEDM (outside the red curves) in the [$\delta, \theta$] plane, with $M_{H_2^+} = 80$~GeV and $\tan\beta = 20$.  $M_{H_3^+} = 200$~GeV in the left panels and 500~GeV in the right panels. Here, $\tan\gamma = 1$ in the upper panels and 2 in the lower panels. }
\label{fig:deltatheta20}
\end{figure}  

\begin{figure}[h!]
\centering
\includegraphics[scale=0.522]{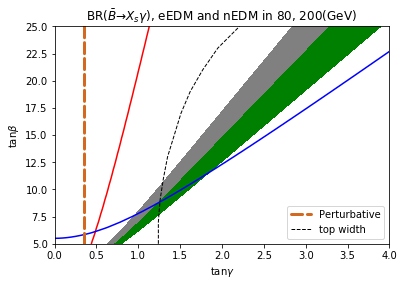}
\includegraphics[scale=0.522]{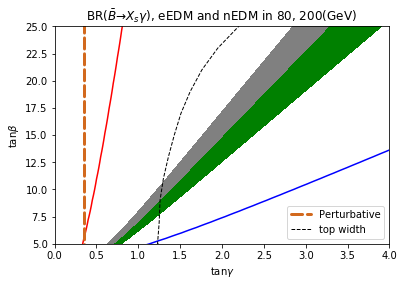}
	\caption{The allowed regions from $\bar{B} \to X_s \gamma$ (within the green and grey shaded areas), eEDM (above the blue line) and nEDM (to the right of the red line) in the [$\tan \gamma,\tan \beta$] plane, with $M_{H_2^+} = 80$~GeV, $M_{H_3^+} = 200$~GeV, $\theta = -0.3$, and $\delta = 0.975\pi$ (left) or $0.985\pi$ (right).   We also show constraints from the top-quark width (black dotted line) and perturbativity (orange dashed line), wherein the region to the right of the respective curves is allowed.}
	\label{Fig:80,200constraints}
\end{figure}

\subsubsection{The $M_{H^\pm_2}<M_{H^\pm_3}<m_t$ case}

Similarly to the previous case, also here we need low values of $\tan\beta$ to satisfy the top-quark width measurements.
However, this is in tension with the region of parameter space that satisfies simultaneously the constraints from $\bar{B}\to X_s\gamma$, eEDM, and nEDM,
despite which, as can be seen in Fig.~\ref{Fig:edmlow}, we could have a somewhat wider interval of $\delta$ around  $\pi$ for large values of $\tan\beta$ and $\tan\gamma$.
There also seems to be a broader band satisfying the $\bar{B}\to X_s\gamma$ constraint for lower values of $M_{H^\pm_3}$, while keeping $M_{H^\pm_2}=80$ GeV.
This, again, is in tension with the aforementioned experimental constraints.
However, in this case, it  is the collider limit on $H^\pm\to\tau\nu$ that becomes too restrictive on the $H^\pm_3$ properties as we decrease its mass.
But we can prevent this from happening if we  keep $M_{H^\pm_3}=170$ GeV and increase instead the mass of $M_{H^\pm_2}$, which is what we do in Fig. \ref{Fig:lowmass-summary}.
In the upper panel of this figure, we show the case $M_{H^\pm_2}=80$~GeV and  $M_{H^\pm_3}=170$~GeV.
In this case, the top-quark width measurement is very constraining, and very low values of $\tan\gamma$ are ruled out.
In the lower panel of this figure, we show the case $M_{H^\pm_2}=160$~GeV and $M_{H^\pm_3}=170$~GeV.
Here, the top-quark width measurement is not that constraining, and very low values of $\tan\gamma$ are allowed.  With the two charged Higgs masses closer to being degenerate, a larger range of the CP-violating phase $\delta$ also becomes allowed.

\begin{figure}
	\centering
\includegraphics[scale=0.5]{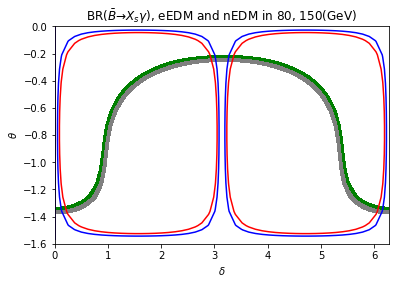}\includegraphics[scale=0.5]{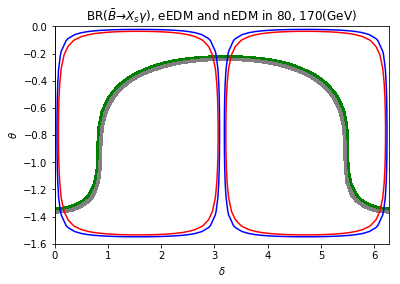}
\includegraphics[scale=0.5]{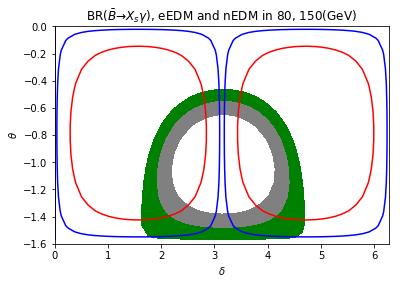}\includegraphics[scale=0.5]{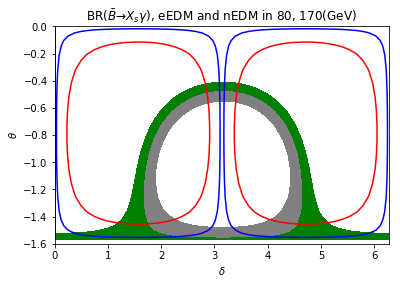}
\includegraphics[scale=0.5]{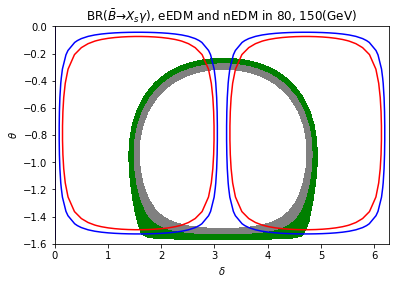}\includegraphics[scale=0.5]{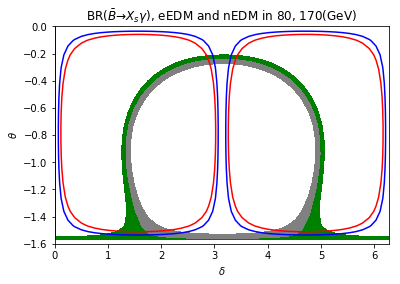}
	\caption{The allowed regions from $\bar{B} \to X_s \gamma$ (within the green and grey shaded areas), eEDM (outside the blue curves), and nEDM (outside the red curves) in the [$\delta, \theta$] plane, with $M_{H_2^+} = 80$~GeV and $M_{H_3^+} = 150$ (left) or 170 (right)~GeV.  From top to bottom, $(\tan\beta, \tan \gamma) = (5, 0.5)$; $(5, 1)$; and $(10, 1)$.
	}
	\label{Fig:edmlow}
\end{figure}

\begin{figure}
\centering
\includegraphics[scale=0.522]{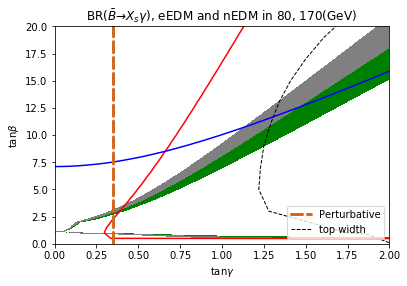}
\includegraphics[scale=0.522]{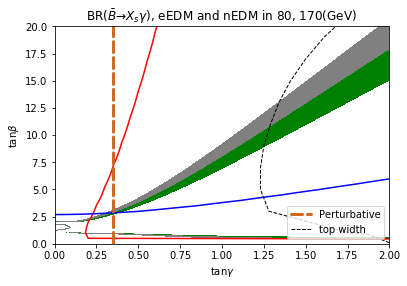}\\
\includegraphics[scale=0.522]{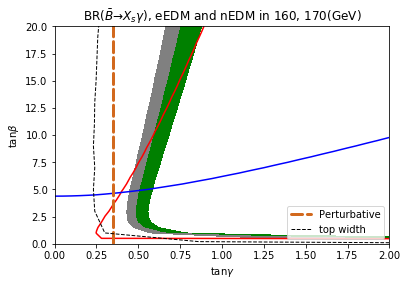}
\includegraphics[scale=0.522]{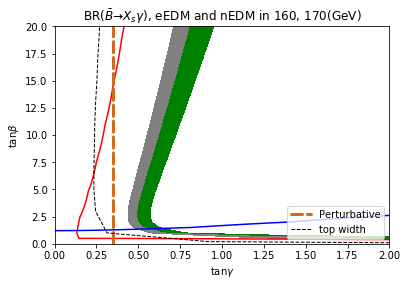}
	\caption{The allowed regions from $\bar{B} \to X_s \gamma$ (within the green and grey shaded areas), eEDM (above the blue line), and nEDM (to the right of the red line) in the [$\tan \gamma,\tan \beta$] plane, with $M_{H_3^+} = 170$~GeV.  In the upper panels $M_{H_2^+} = 80$~GeV, $\theta = -0.3$, and $\delta = 0.96\pi$ (left) or $0.985\pi$ (right).  In the lower panels $M_{H_2^+} = 160$~GeV, $\theta = -0.5$, and $\delta = 0.8\pi$ (left) or $0.95\pi$ (right).  We also show constraints from the top-quark width (black dotted line) and perturbativity (orange dashed line), wherein the region to the right of the respective curves is allowed.
	}
	\label{Fig:lowmass-summary}
\end{figure}

\subsection{Heavy charged Higgses}\label{sec: heavymass}

In the case that both the $H_2^\pm$ and $H_3^\pm$ masses are heavier than the top-quark mass, collider searches no longer significantly  limit the parameter space, so we present the $\bar{B} \to X_s \gamma$, eEDM and nEDM  constraints on the [$M_{H^{\pm}_2},M_{H^{\pm}_3}$] plane with different choices for the mixing parameters ($\tan\beta$, $\tan \gamma$, $\theta$, and $\delta$). We choose the parameters $\theta = - 0.476\pi$ $(-\pi/4)$,  $\tan\beta = 20\; (40)$ and $\tan \gamma =1\; (2)$ to plot from Fig.~\ref{fig:theta2105} to Fig.~\ref{fig:theta49}. Specifically,  Figs.~\ref{fig:theta2105}--\ref{fig:theta219} are plotted for three different $\delta$ values for the same $\theta=-0.476\pi$, where $\delta = 0.5\pi$ (maximum CP-violating scenario), 0.85$\pi$, and 0.9$\pi$ (two choices closer to the CP-conserving limit). In Fig.~\ref{fig:theta2105}, the  two bottom panels clearly show that the most constraining limit comes from the nEDM when $\tan\beta = 40$. For the choice of $\tan \beta = 20$ and $\tan\gamma = 2$, the top right panel shows instead that the eEDM constraint is the one limiting most of the parameter space.  In Figs.~\ref{fig:theta2185} and~\ref{fig:theta219}, a large expanse of  parameter space is allowed by both the eEDM and nEDM constraints. In fact, here, EDM constraints no longer strictly limit the parameter space so that $\bar{B} \to X_s \gamma$ becomes the essential constraint, especially as $\delta$ gets close to $\pi$. The typical funnel shape of the allowed region along the mass diagonal for the EDM constraints illustrates again the impact
of the  GIM-like cancellation mechanism driven by the charged Higgs mass degeneracy, the more so the smaller their absolute values.  Such a cancellation is not present in the $\bar{B} \to X_s \gamma$ constraint, since this observable receives  both real and imaginary contributions from $X_iY_i^*$ terms, with the real components of $X_2Y^*_2$ and  $X_3Y^*_3$ not being strongly correlated as their imaginary parts are; the corresponding shape thus departs from the funnel one and depends more on a judicious choice of $\theta$ for given values of $\tan\beta$ and $\tan\gamma$. 

In the case of $\theta = -\pi/4$,  three similar figures,  Figs.~\ref{fig:theta405}, \ref{fig:theta485}, and \ref{fig:theta49}, are presented for $\delta = 0.5\pi$, $0.85\pi$ and $0.9\pi$, respectively. For this $\theta$ value, it is  intriguing to note that even the exact degeneracy case between $H_2^\pm$ and $H_3^\pm$ fails the $\bar B\to X_s\gamma$ constraint for the smallest $\delta$ choice.  In contrast, for the other $\delta$ values, the main effect is a significant restriction of the parameter space allowed by $\bar B\to X_s\gamma$ along the $M_{H_2^\pm}=M_{H_3^\pm}$ diagonal while, conversely, the EDM constraints are less invasive. This is a generalized feature quite irrespectively of the value of $\tan\beta$, so long as $\tan\gamma$ remains small.

\begin{figure}
	\centering
	\includegraphics[scale=0.5]{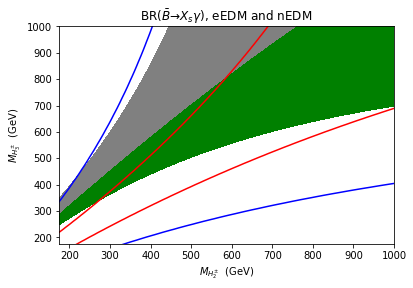}	\includegraphics[scale=0.5]{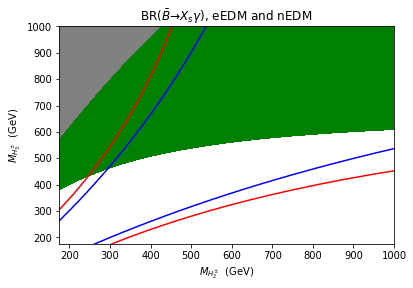}\\	\includegraphics[scale=0.5]{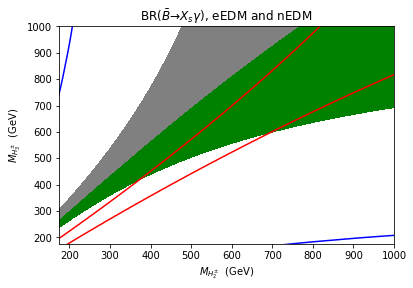}	\includegraphics[scale=0.5]{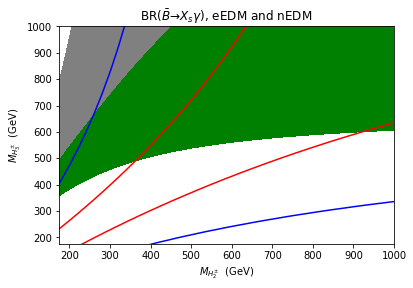}
	\caption{The allowed regions from $\bar{B} \to X_s  \gamma$ (within the green and grey shaded areas),  eEDM (between the blue lines), and  nEDM (between the red lines) in the [$M_{H^{\pm}_2} , M_{H^{\pm}_3} $] plane, for 
 $\theta = -0.476\pi$ and $\delta = 0.5\pi$ (i.e., maximal CP violation), with $\tan\beta = 20$ (upper panels) or 40 (lower panels) and $\tan\gamma = 1$ (left panels) or 2 (right panels). 
}
\label{fig:theta2105}
\end{figure}

\begin{figure}
	\centering
	\includegraphics[scale=0.5]{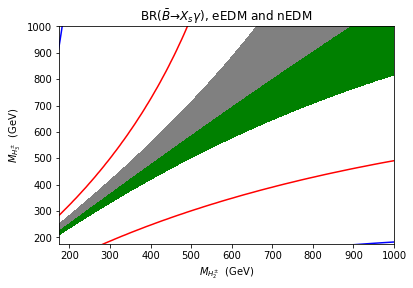}	\includegraphics[scale=0.5]{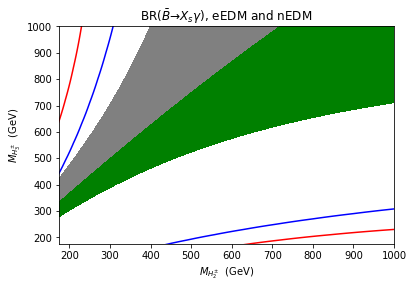}\\	\includegraphics[scale=0.5]{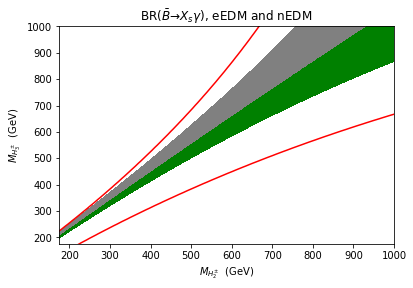}	\includegraphics[scale=0.5]{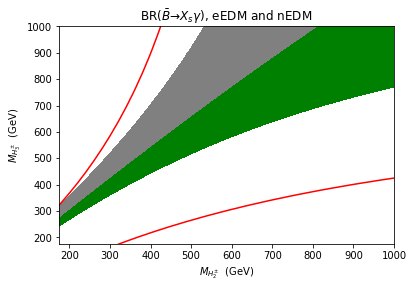}
	\caption{Same as Fig.~\ref{fig:theta2105} but with $\delta=0.85\pi$.
}
\label{fig:theta2185}
\end{figure}

\begin{figure}
	\centering
	\includegraphics[scale=0.5]{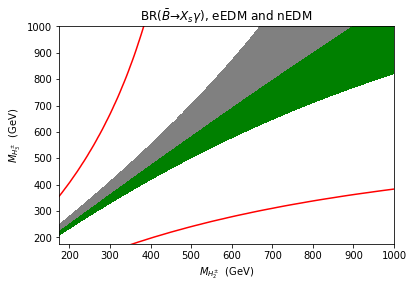}	\includegraphics[scale=0.5]{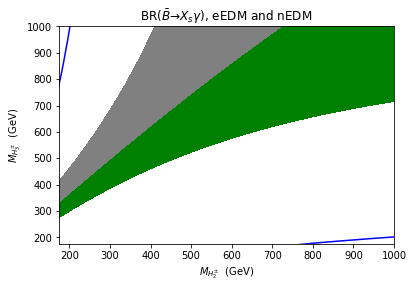}\\	\includegraphics[scale=0.5]{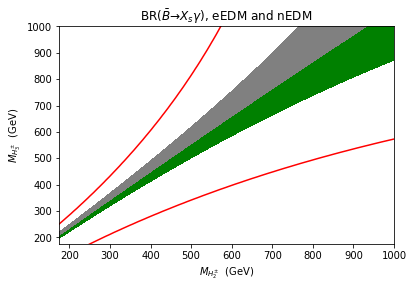}	\includegraphics[scale=0.5]{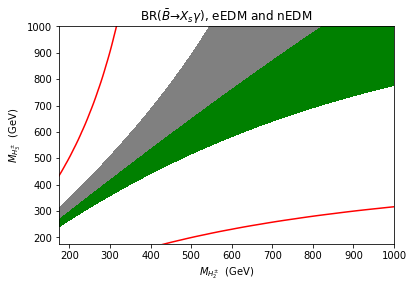}
	\caption{Same as Fig.~\ref{fig:theta2105} but with $\delta=0.9\pi$.
}
\label{fig:theta219}
\end{figure}

\begin{figure}
	\centering
	\includegraphics[scale=0.5]{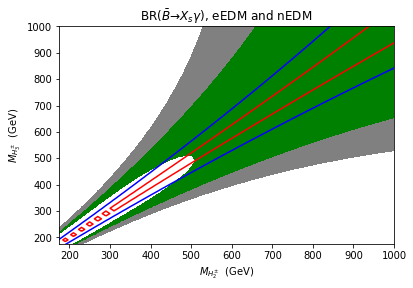}	\includegraphics[scale=0.5]{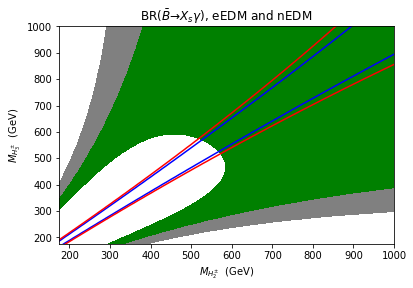}\\	\includegraphics[scale=0.5]{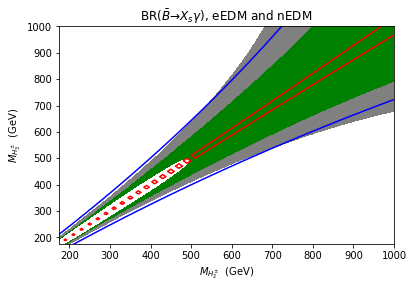}	\includegraphics[scale=0.5]{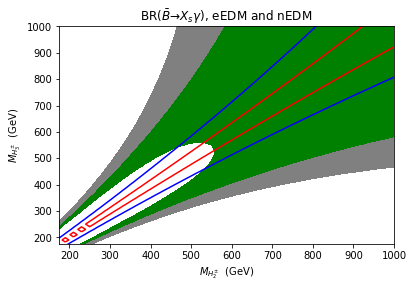}
	\caption{Same as Fig.~\ref{fig:theta2105} but with $\theta=-\pi/4$.
}
\label{fig:theta405}
\end{figure}

\begin{figure}
	\centering
	\includegraphics[scale=0.5]{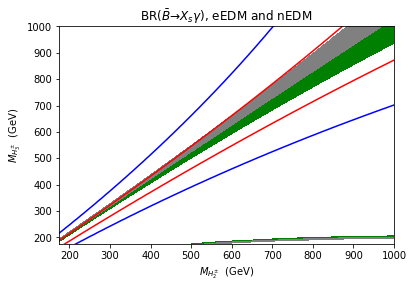}	\includegraphics[scale=0.5]{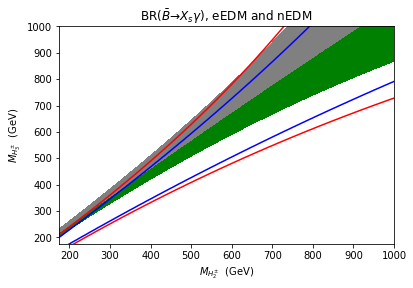}\\	\includegraphics[scale=0.5]{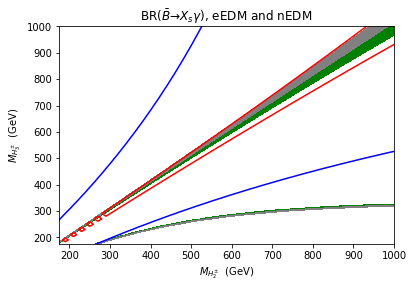}	\includegraphics[scale=0.5]{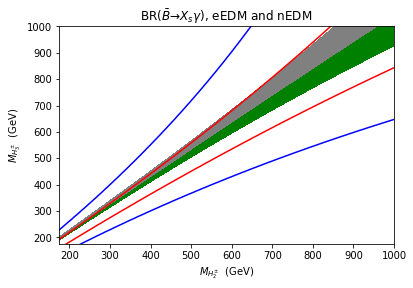}
	\caption{Same as Fig.~\ref{fig:theta2185} but with $\theta=-\pi/4$.
}
\label{fig:theta485}
\end{figure}

\begin{figure}
	\centering
	\includegraphics[scale=0.5]{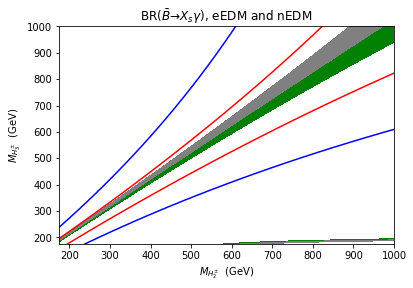}	\includegraphics[scale=0.5]{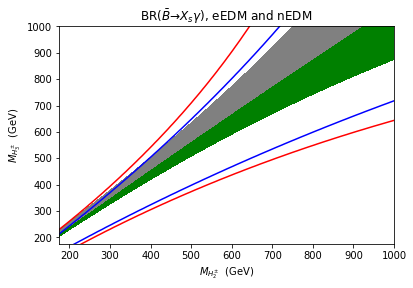}\\	\includegraphics[scale=0.5]{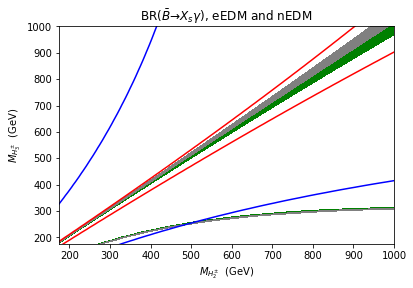}	\includegraphics[scale=0.5]{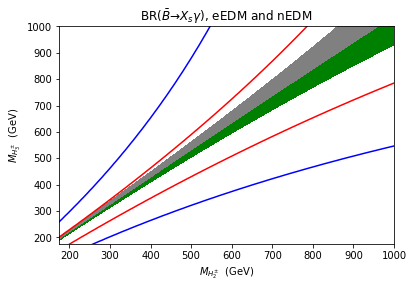}
	\caption{Same as Fig.~\ref{fig:theta219} but with $\theta=-\pi/4$.
}
\label{fig:theta49}
\end{figure}

\section{Conclusions}
\label{sec:conclusions}

In this paper, we have studied a version of the 3HDM, called Democratic, wherein each amongst the down-type  quarks, up-type quarks, and charged leptons gain their mass from one only of the three VEVs of the Higgs doublets, in the presence of explicit CP violation in the charged Higgs sector, which  consists of two physical states, each with mass varying from 80~GeV to the TeV scale. While an enlarged neutral Higgs sector also exists in this framework, consisting, in addition to the SM-like Higgs state already discovered, of four other neutral Higgs states, two CP-even and two CP-odd, these have been assumed to be sufficiently heavy compared to the charged Higgs bosons so as to not significantly affect the low energy phenomenology of the 3HDM.  In particular, we showed that it is possible to isolate the effects of CP violation to the charged Higgs sector only, and derived the conditions on the complex parameters of the scalar potential required to achieve this.  We have studied the charged Higgs sector in terms of the following experimental observables, all very sensitive to new CP-violating effects emerging alongside those contained in the CKM matrix: BR$(\bar B\to X_s\gamma)$, eEDM, and nEDM.  

We have tested the parameter space of the 3HDM, mapped in terms of the two charged Higgs boson masses $M_{H_2^\pm}$ and  $M_{H_3^\pm}$ and four parameters entering their Yukawa couplings, $\tan\beta$, $\tan\gamma$, $\theta$, and the CP violating phase $\delta$, against experimental measurements of these three observables. In doing so, we have discovered a sort of GIM-like cancellation mechanism between the two charged Higgs contributions to eEDM and nEDM driven by the unitarity of the charged Higgs mixing matrix.\footnote{An equivalent cancellation occurs in the CP-odd asymmetries in $\bar B \to X_s \gamma$, considered in Ref.~\cite{Akeroyd2020}.} Such a cancellation becomes exact when $M_{H_2^\pm}$ = $M_{H_3^\pm}$. As a consequence, it is then possible to evade the experimental constraints enforced through the aforementioned CP-odd observables whichever the values of $\tan\beta$, $\tan\gamma$, $\theta$, and $ \delta$.     

Even in less fine-tuned conditions, when we lift the charged Higgs boson mass degeneracy, interesting phenomenology emerges. Specifically,  light 
${H_2^\pm}$ and/or  ${H_3^\pm}$ states, with mass below $m_t$, are still allowed not only by the BR$(\bar B\to X_s\gamma)$, eEDM and nEDM constraints but also by those induced by the experimental measurements of the top quark decay width and direct searches for top quark decays to charged Higgs bosons with subsequent charged Higgs decays to $\tau^+\nu$, $c\bar s$, and $c\bar b$ at the LHC, as well as the theoretical requirement of perturbativity of the Yukawa couplings. While this is really only possible near the CP-conserving limit and when the lightest of the two charged Higgs states has a mass close to $M_{W}$ (so as to be unconstrained by the LHC, owing to the overwhelming  irreducible $W^\pm$ background herein), also for small values of $\tan\beta$ and $\tan\gamma$, it nonetheless opens us the possibility of searching for the corresponding signals at the LHC, wherein one could attempt to isolate CP-violating asymmetries in the top-quark decay rates between the positively and negatively charged $H^\pm_i$ ($i=2,3$) channels. The region of viable 3HDM parameter space in which these two states are both heavier than the top quark is much larger in comparison, and the mass difference
$M_{H_2^\pm} - M_{H_3^\pm}$ can be up to 200~GeV or so, albeit for selected values of the other parameters so that the constraint from BR($\bar B \to X_s \gamma$) can be satisfied. In this case, while it may be difficult to access $H^\pm_i$ signals in direct searches at the LHC, and consequently the possible CP-violating nature of the 3HDM, the latter could well be established in
CP asymmetries of ${\bar B}\to X_s/X_d\gamma$ observables at $B$-factories  (e.g., Belle-II), as demonstrated in Ref.~\cite{Akeroyd2020}, which in fact can capture striking signals in the case of light charged Higgs bosons too.

\acknowledgments

This work was supported by the grant H2020-MSCA-RISE-2014 No.\ 645722 (NonMinimalHiggs). 
H.E.L.\ was also supported by the Natural Sciences and Engineering Research Council of Canada. 
S.M. is supported in part through the NExT Institute and the STFC Consolidated Grant No. ST/L000296/1.
D.R.-C. is supported by the Royal Society Newton International Fellowship NIF/R1/180813 and
by the National Science Centre (Poland) under the research Grant No. 2017/26/E/ST2/00470.
D.R.-C. and M.S. thank Carleton University for hospitality during the initial stages of this work.
The authors thank Shinya Kanemura and Kei Yagyu for useful conversations.



\newpage
\appendix

\section{Experimental constraints from $\bar{B} \to X_s  \gamma$} 
\label{sec:bsg}

\begin{table}
\hspace*{-0.6cm}
\begin{tabular}{|c|c|c|c|}
	\hline
	$m_c/ m_b$ = 0.29 & $m_b - m_c$ = 3.39 GeV & $m_t$ = 173 GeV & $G_{\text{F}}$ = 1.1663787 $\times 10^{-5}$ GeV$^{-2}$ \\	
	$\alpha_{\text{em}}$ = 1/ 130.3 & $M_Z$ = 91.1875 GeV & $M_{W^\pm}$ = 80.33 GeV &  $\alpha({M_Z})$ = 0.119  \\
	$\lambda$ = 0.22650 &$A$ = 0.790 & $\bar{\rho}$ = 0.141 &$\bar{\eta}$ = 0.357  \\
	BR$_{SL}$ = 0.1049  &  & &\\
	\hline
\end{tabular}
\caption{Input values for the SM parameters. The central value of $\bar{B}\to X_s \gamma$ used here is obtained from these. 
We refer to \cite{Borzumati1998} for the choice of fermion masses. 
The Wolfenstein parameters of the CKM matrix are taken from Ref. \cite{Zyla2020}.  }
\label{tab:bsg}
\end{table}

Current values for the average experimental measurement \cite{Amhis2019} and SM prediction \cite{Misiak2020}  for the BR($\bar{B} \to X_s \gamma$) are as follows:
\bea
{\rm BR}(\bar{B} \to X_s \gamma)^{\rm exp} = (3.32 \pm 0.15) \times 10^{-4} \,\, \text{with} \,\,\, E_{\gamma} > 1.6\, \text{GeV},
\label{eq:bsgexp} \\
{\rm BR}(\bar{B} \to X_s \gamma)^{\rm SM} = (3.40 \pm 0.17) \times 10^{-4} \,\, \text{with} \,\,\, E_{\gamma} > 1.6\, \text{GeV}.
\label{eq:SMbsg}
\eea
In our BR$(\bar{B} \to X_s \gamma)$ numerical evaluation, we used the explicit formulas computed for the 2HDM with leading order
(LO) and  next-to-LO (NLO) effective Wilson coefficients running from the $\mu_W$ scale to $\mu_b$ with the  scheme adopted  in \cite{Borzumati1998} and extrapolated it to 3HDM as in Ref. \cite{Akeroyd2020}. 

The two LO Wilson coefficients at the $\mu_W$ scale which are affected by charged Higgs contributions are $C^{0,\text{eff}}_7 (\mu_W)$ and $C^{0,\text{eff}}_8 (\mu_W)$, obtained as follows:
\bea
C^{0,\text{eff}}_7 (\mu_W) &=& C^{0}_{7,SM}  + |Y_2 |^2 C^{0}_{7,Y_2Y_2} + |Y_3 |^2 C^{0}_{7,Y_3Y_3}  \nonumber \\ && + (X_2Y_2^*) C^{0}_{7,X_2Y_2} + (X_3Y_3^*) C^{0}_{7,X_3Y_3}, \\
C^{0,\text{eff}}_8 (\mu_W) &=& C^{0}_{8,SM}  + |Y_2 |^2 C^{0}_{8,Y_2Y_2} + |Y_3 |^2 C^{0}_{8,Y_3Y_3}  \nonumber \\  && + (X_2Y_2^*) C^{0}_{8,X_2Y_2} + (X_3Y_3^*) C^{0}_{8,X_3Y_3} ,
\eea
where $|Y_2 |^2$, $|Y_3 |^2$, $(X_2Y_2^*)$, and $(X_3Y_3^*)$ are the contribution of charged Higgs mixing couplings.  The other LO Wilson coefficients are $C^{0,\text{eff}}_2 (\mu_W)= 1$ and $C^{0,\text{eff}}_i (\mu_W) = 0$ ($i =1,3,4,5,6$). The SM contributions are functions of $m^2_t / M^2_{W}$ while the charged Higgs contribution are functions of $m^2_t / M^2_{H^{\pm}_2}$ and $m^2_t / M^2_{H^{\pm}_3}$, entering the $C^{0}_{n,{\rm SM}}, C^{0}_{n,X_2Y_2}, C^{0}_{n,X_3Y_3} \,\, (n = 7,8)$ terms in Ref.~\cite{Borzumati1998}.

The NLO Wilson coefficients at the matching scale ($\mu_{W}$) are as follow:
\bea
C^{1,\text{eff}}_1 (\mu_W) &=&  15 + 6 \hspace{0.2cm}  \text{ln} \frac{\mu^2_{W}}{M^2_W} ,\\
C^{1,\text{eff}}_4 (\mu_W) &= & E_0 + \frac{2}{3}  \hspace{0.2cm}  \text{ln}  \frac{\mu^2_{W}}{M^2_W} + |Y_2 |^2 E_{H_2} + |Y_3 |^2 E_{H_3}, \\
C^{1,\text{eff}}_i (\mu_W) &=& 0 \quad (i = 2,3,5,6), \\
C^{1,\text{eff}}_7 (\mu_W) &=& C^{1,\text{eff}}_{7,SM}(\mu_W)  + |Y_2 |^2 C^{1,\text{eff}}_{7,Y_2Y_2}(\mu_W)  + |Y_3 |^2 C^{1,\text{eff}}_{7,Y_3Y_3}(\mu_W)  \nonumber \\ & & + (X_2Y_2^*) C^{1,\text{eff}}_{7,X_2Y_2}(\mu_W) + (X_3Y_3^*) C^{1,\text{eff}}_{7,X_3Y_3}(\mu_W), \\
C^{1,\text{eff}}_8 (\mu_W) &=& C^{1,\text{eff}}_{8,SM}(\mu_W)  + |Y_2 |^2 C^{1,\text{eff}}_{8,Y_2Y_2}(\mu_W)  + |Y_3 |^2 C^{1,\text{eff}}_{8,Y_3Y_3}(\mu_W) \nonumber \\ & & + (X_2Y_2^*) C^{1,\text{eff}}_{8,X_2Y_2}(\mu_W) + (X_3Y_3^*) C^{1,\text{eff}}_{8,X_3Y_3}(\mu_W).
\eea
Explicit forms for all functions are given in \cite{Borzumati1998}. Renormalization group running is then used to evaluate the Wilson coefficients at the scale $\mu=m_b$. The $\bar{B} \to X_{s} \gamma $ decay rates can then be evaluated through
\bea
\Gamma (\bar{B} \to X_{s} \gamma) &=& \frac{G^2_F}{32\pi^4} |V^{*}_{ts} V_{tb} |^2 \alpha_{em} m^5_b \nonumber   \\ &\times &
\Bigg \{ |\bar{D} |^2 + A + \frac{\delta^{NP}_{\gamma}}{m^2_b} |\text{C}^{0,\text{eff}}_{7} (\mu_b) |^2  \nonumber \\  &+& \frac{\delta^{NP}_{c}}{m^2_c} {\rm Re} \Bigg[  [\text{C}^{0,\text{eff}}_{7} (\mu_b)]^* \times \bigg ( \text{C}^{0,\text{eff}}_{2} (\mu_b) - \frac{1}{6} \text{C}^{0,\text{eff}}_{1} (\mu_b)\bigg) \Bigg] \Bigg\},\\
{\rm BR}(\bar{B} \to X_s \gamma) &=& \frac{\Gamma (\bar{B} \to X_s \gamma)}{\Gamma_{SL}} {\rm BR}_{SL}.  
\eea 
where the expressions for $ |\bar{D}|$ ($b \to s \gamma$), $A$ ($b \to s \gamma g$), and the semileptonic width $\Gamma_{SL}$ are taken from \cite{Borzumati1998}.
For our BR$(\bar{B} \to X_s \gamma)$ numerical evaluations, we took the Tab.~\ref{tab:bsg} input values as SM parameters. In such a case,  $m_c$, $m_b$ and  $m_t$ are pole masses of the charm-, bottom- and top-quark, respectively, and $A$, $\lambda$, $\bar{\rho}$ and $\bar{\eta}$ are the Wolfenstein parameters of the CKM matrix taken from Ref. \cite{Zyla2020}. 

Our implementation of the BR$(\bar{B} \to X_s \gamma)$ calculation yields a SM value of $3.39 \times 10^{-4}$, which is extremely close to the result of the state-of-the-art calculation given in Eq.~(\ref{eq:SMbsg}).  In the figures in this paper we use coloured bands to indicate the allowed range of $\pm 2\sigma$ about the experimental central value, where we have combined the experimental and theoretical uncertainties in quadrature.  In particular, we show values of BR$(\bar{B} \to X_s \gamma)$ in the range ($3.32$--$3.77) \times 10^{-4}$ in grey and values in the range ($2.87$--$3.32) \times 10^{-4}$ in green.\footnote{By coincidence, these ranges are equivalent to taking the $3\sigma$ allowed range using the experimental uncertainty only.}

\section{Charged Higgs Yukawa couplings}
\label{sec:appB}

In this section we collect the combinations of the Yukawa coupling coefficients $X_i$, $Y_i$, and $Z_i$   ($i = 2,3$) that appear in the various calculations in this paper, and give their expressions as a function of the four mixing parameters ($\theta$, $\tan\gamma$, $\tan\beta$ and $\delta$) in the Democratic 3HDM.  We use the shorthand notation $s_\theta$, $c_\theta$, $t_\theta$ for $\sin\theta$, 
$\cos\theta$, and $\tan\theta$, respectively, and analogously for the other angles.

Starting from Eqs.~(\ref{eq:Uexplicit}) and (\ref{XYZ}), the Yukawa coupling coefficients in our parameterization are:
\bea
X_2 &=&  \frac{U^\dagger_{12}}{U^\dagger_{11}} =   \frac{- c_\theta s_\beta (c_\delta + i s_\delta ) -  s_\theta c_\gamma c_\beta}{c_\beta s_\gamma},\\
Y_2  &=& - \frac{U^\dagger_{22}}{U^\dagger_{21}} =  \frac{- c_\theta c_\beta (c_\delta + i s_\delta) +  s_\theta c_\gamma s_\beta}{s_\beta s_\gamma},\\
Z_2 &=& \frac{U^{\dagger}_{32}}{U^\dagger_{31}} = \frac{s_\theta s_\gamma}{c_\gamma}, \\
X_3 &=& \frac{U^\dagger_{13}}{U^\dagger_{11}} = \frac{s_\theta s_\beta  (c_\delta + i s_\delta)  - c_\theta c_\gamma  c_\beta }{c_\beta s_\gamma} ,\\
Y_3  &=& - \frac{U^\dagger_{23}}{U^\dagger_{21}} =  \frac{   s_\theta c_\beta (c_\delta + i s_\delta) +  c_\theta c_\gamma s_\beta}{s_\beta s_\gamma}, \\
Z_3 &=& \frac{U^{\dagger}_{33}}{U^\dagger_{31}} = \frac{c_\theta s_\gamma}{c_\gamma}.
\eea

The combinations that appear in the EDM calculations are:
\bea
{\rm Im}(- X_2Y_2^* ) &=& \frac{s_\theta c_\theta s_\delta}{s_\beta c_\beta s_\gamma t_\gamma} 
	= - {\rm Im}(- X_3 Y_3^*) , \\
{\rm Im}(- Y_2^* Z_2) &=& -\frac{s_\theta c_\theta s_\delta}{t_\beta c_\gamma}
	= - {\rm Im}(- Y_3^* Z_3).
\eea

The following contribute to the calculation of BR($\bar B \to X_s \gamma$).
The real components of $X_iY_i^*$ ($i = 2,3$) are as follows:
\bea
{\rm Re}(X_2Y_2^*) 
&=&\frac{c_\theta^2}{s_\gamma^2} + \frac{c_\delta c_\theta s_\theta}{t_\beta t_\gamma s_\gamma} - \frac{c_\delta t_\beta c_\theta s_\theta }{t_\gamma s_\gamma} - \frac{s_\theta^2}{t_\gamma^2}, \\
{\rm Re}(X_3Y_3^*) 
&=&  \frac{s_\theta^2}{s_\gamma^2} + \frac{c_\delta t_\beta c_\theta s_\theta }{t_\gamma s_\gamma}  - \frac{c_\delta c_\theta s_\theta }{t_\beta t_\gamma s_\gamma} - \frac{c_\theta^2}{t_\gamma^2}.
\eea
Finally for $|Y_2^2|$ and $|Y_3^2|$ we have:
\bea
|Y_2^2| 
&=& \frac{c_\delta^2 c_\theta^2}{t_\beta^2 s_\gamma^2} - \frac{s_\delta^2 c_\theta^2}{t_\beta^2 s_\gamma^2} - \frac{2 c_\delta c_\theta s_\theta}{t_\beta t_\gamma s_\gamma} + \frac{s_\theta^2}{t_\gamma^2}, \\
|Y_3^2| 
&=& \frac{c_\delta^2 s_\theta^2}{t_\beta^2 s_\gamma^2}  - \frac{s_\delta^2 s_\theta^2}{t_\beta^2 s_\gamma^2} + \frac{2 c_\delta c_\theta s_\theta}{t_\beta t_\gamma s_\gamma} + \frac{c_\theta^2}{t_\gamma^2}. 
\eea

\bibliography{jabref.bib}   

\providecommand{\href}[2]{#2}\begingroup\raggedright\begin{thebibliography}{10}

\bibitem{Canetti2012}
L.~Canetti, M.~Drewes and M.~Shaposhnikov, \emph{Matter and antimatter in the
  universe}, \href{https://doi.org/10.1088/1367-2630/14/9/095012}{\emph{New J.
  Phys.} {\bfseries 14} (2012) 095012}
  [\href{https://arxiv.org/abs/1204.4186}{{\ttfamily 1204.4186}}].

\bibitem{Sakharov1991a}
A.~Sakharov, \emph{{Violation of CP Invariance, C asymmetry, and baryon
  asymmetry of the universe}},
  \href{https://doi.org/10.1070/PU1991v034n05ABEH002497}{\emph{Sov. Phys. Usp.}
  {\bfseries 34} (1991) 392}.

\bibitem{Abel2020}
{\scshape nEDM} collaboration, \emph{{Measurement of the permanent electric
  dipole moment of the neutron, EDM}},
  \href{https://doi.org/10.1103/PhysRevLett.124.081803}{\emph{Phys. Rev. Lett.}
  {\bfseries 124} (2020) 081803}
  [\href{https://arxiv.org/abs/2001.11966}{{\ttfamily 2001.11966}}].

\bibitem{Andreev2018a}
{\scshape ACME} collaboration, \emph{Improved limit on the electric dipole
  moment of the electron, edm},
  \href{https://doi.org/10.1038/s41586-018-0599-8}{\emph{Nature} {\bfseries
  562} (2018) 355}.

\bibitem{CorderoCid2016}
A.~Cordero-Cid, J.~Hern{\'{a}}ndez-S{\'{a}}nchez, V.~Keus, S.F.~King,
  S.~Moretti, D.~Rojas et~al., \emph{{CP violating scalar Dark Matter}},
  \href{https://doi.org/10.1007/jhep12(2016)014}{\emph{Journal of High Energy
  Physics} {\bfseries 2016} (2016) 014}
  [\href{https://arxiv.org/abs/1608.01673}{{\ttfamily 1608.01673}}].

\bibitem{Azevedo2018}
D.~Azevedo, P.M.~Ferreira, M.M.~Muhlleitner, S.~Patel, R.~Santos and
  J.~Wittbrodt, \emph{{CP} in the dark},
  \href{https://doi.org/10.1007/JHEP11(2018)091}{\emph{JHEP} {\bfseries 11}
  (2018) 091} [\href{https://arxiv.org/abs/1807.10322}{{\ttfamily
  1807.10322}}].

\bibitem{Carena2019}
M.~Carena, M.~Quir\'os and Y.~Zhang, \emph{{Electroweak Baryogenesis from
  Dark-Sector CP Violation}},
  \href{https://doi.org/10.1103/PhysRevLett.122.201802}{\emph{Phys. Rev. Lett.}
  {\bfseries 122} (2019) 201802}
  [\href{https://arxiv.org/abs/1811.09719}{{\ttfamily 1811.09719}}].

\bibitem{Okawa2019}
S.~Okawa, M.~Pospelov and A.~Ritz, \emph{{Electric Dipole Moments From Dark
  Sectors, EDM}},
  \href{https://doi.org/10.1103/PhysRevD.100.075017}{\emph{Phys. Rev. D}
  {\bfseries 100} (2019) 075017}
  [\href{https://arxiv.org/abs/1905.05219}{{\ttfamily 1905.05219}}].

\bibitem{Carena2020}
M.~Carena, M.~Quir\'os and Y.~Zhang, \emph{{Dark CP violation and gauged lepton
  or baryon number for electroweak baryogenesis}},
  \href{https://doi.org/10.1103/PhysRevD.101.055014}{\emph{Phys. Rev. D}
  {\bfseries 101} (2020) 055014}
  [\href{https://arxiv.org/abs/1908.04818}{{\ttfamily 1908.04818}}].

\bibitem{Keus2020}
V.~Keus, \emph{Dark origins of matter-antimatter asymmetry},
  \href{https://doi.org/10.22323/1.376.0059}{\emph{PoS} {\bfseries CORFU2019}
  (2020) 059} [\href{https://arxiv.org/abs/2003.02141}{{\ttfamily
  2003.02141}}].

\bibitem{CorderoCid2020}
A.~Cordero-Cid, J.~Hern{\'{a}}ndez-S{\'{a}}nchez, V.~Keus, S.~Moretti,
  D.~Rojas-Ciofalo and D.~Soko{\l}owska, \emph{Collider signatures of dark {CP}
  violation}, \href{https://doi.org/10.1103/physrevd.101.095023}{\emph{Physical
  Review D} {\bfseries 101} (2020) 095023}
  [\href{https://arxiv.org/abs/2002.04616}{{\ttfamily 2002.04616}}].

\bibitem{Kanemura2020}
S.~Kanemura, M.~Kubota and K.~Yagyu, \emph{{Aligned CP-violating Higgs sector
  canceling the electric dipole moment, EDM}},
  \href{https://doi.org/10.1007/JHEP08(2020)026}{\emph{JHEP} {\bfseries 08}
  (2020) 026} [\href{https://arxiv.org/abs/2004.03943}{{\ttfamily
  2004.03943}}].

\bibitem{Glashow1970}
S.~Glashow, J.~Iliopoulos and L.~Maiani, \emph{Weak interactions with
  lepton-hadron symmetry},
  \href{https://doi.org/10.1103/PhysRevD.2.1285}{\emph{Phys. Rev. D} {\bfseries
  2} (1970) 1285}.

\bibitem{Glashow1977}
S.L.~Glashow and S.~Weinberg, \emph{Natural conservation laws for neutral
  currents}, \href{https://doi.org/10.1103/PhysRevD.15.1958}{\emph{Phys. Rev.
  D} {\bfseries 15} (1977) 1958}.

\bibitem{Paschos}
E.A.~Paschos, \emph{Diagonal neutral currents},
  \href{https://doi.org/10.1103/physrevd.15.1966}{\emph{Physical Review D}
  {\bfseries 15} (1977) 1966}.

\bibitem{Jung2014}
M.~Jung and A.~Pich, \emph{{Electric Dipole Moments in Two-Higgs-Doublet
  Models, EDM}}, \href{https://doi.org/10.1007/JHEP04(2014)076}{\emph{JHEP}
  {\bfseries 04} (2014) 076} [\href{https://arxiv.org/abs/1308.6283}{{\ttfamily
  1308.6283}}].

\bibitem{Cree2011}
G.~Cree and H.E.~Logan, \emph{Yukawa alignment from natural flavor
  conservation}, \href{https://doi.org/10.1103/PhysRevD.84.055021}{\emph{Phys.
  Rev. D} {\bfseries 84} (2011) 055021}
  [\href{https://arxiv.org/abs/1106.4039}{{\ttfamily 1106.4039}}].

\bibitem{Akeroyd2017}
A.~Akeroyd, S.~Moretti, K.~Yagyu and E.~Yildirim, \emph{{Light charged Higgs
  boson scenario in 3-Higgs doublet models}},
  \href{https://doi.org/10.1142/S0217751X17501457}{\emph{Int. J. Mod. Phys. A}
  {\bfseries 32} (2017) 1750145}
  [\href{https://arxiv.org/abs/1605.05881}{{\ttfamily 1605.05881}}].

\bibitem{Grossman1994}
Y.~Grossman, \emph{{Phenomenology of models with more than two Higgs
  doublets}}, \href{https://doi.org/10.1016/0550-3213(94)90316-6}{\emph{Nucl.
  Phys. B} {\bfseries 426} (1994) 355}
  [\href{https://arxiv.org/abs/hep-ph/9401311}{{\ttfamily hep-ph/9401311}}].

\bibitem{Belyaev2013}
A.~Belyaev, N.D.~Christensen and A.~Pukhov, \emph{{CalcHEP} 3.4~for collider
  physics within and beyond the {Standard Model}},
  \href{https://doi.org/10.1016/j.cpc.2013.01.014}{\emph{Computer Physics
  Communications} {\bfseries 184} (2013) 1729}
  [\href{https://arxiv.org/abs/1207.6082}{{\ttfamily 1207.6082}}].

\bibitem{Guchait2002}
M.~Guchait and S.~Moretti, \emph{{Improving the discovery potential of charged
  Higgs bosons at Tevatron run II}},
  \href{https://doi.org/10.1088/1126-6708/2002/01/001}{\emph{JHEP} {\bfseries
  01} (2002) 001} [\href{https://arxiv.org/abs/hep-ph/0110020}{{\ttfamily
  hep-ph/0110020}}].

\bibitem{Assamagan2004}
K.~Assamagan, M.~Guchait and S.~Moretti, \emph{{Charged Higgs bosons in the
  transition region $M(H^{\pm}) \sim m(t)$ at the LHC}},  in \emph{{3rd Les
  Houches Workshop on Physics at TeV Colliders}}, 2, 2004
  [\href{https://arxiv.org/abs/hep-ph/0402057}{{\ttfamily hep-ph/0402057}}].

\bibitem{Akeroyd2018}
A.~Akeroyd, S.~Moretti and M.~Song, \emph{{Light charged Higgs boson with
  dominant decay to quarks and its search at the LHC and future colliders}},
  \href{https://doi.org/10.1103/PhysRevD.98.115024}{\emph{Phys. Rev. D}
  {\bfseries 98} (2018) 115024}
  [\href{https://arxiv.org/abs/1810.05403}{{\ttfamily 1810.05403}}].

\bibitem{Akeroyd2020a}
A.~Akeroyd, S.~Moretti and M.~Song, \emph{{Light charged Higgs boson with
  dominant decay to a charm quark and a bottom quark and its search at LEP2 and
  future $e^+e^-$ colliders}},
  \href{https://doi.org/10.1103/PhysRevD.101.035021}{\emph{Phys. Rev. D}
  {\bfseries 101} (2020) 035021}
  [\href{https://arxiv.org/abs/1908.00826}{{\ttfamily 1908.00826}}].

\bibitem{Sirunyan2019}
{\scshape CMS} collaboration, \emph{{Search for charged Higgs bosons in the
  H$^{\pm}$ $\to$ $\tau^{\pm}\nu_\tau$ decay channel in proton-proton
  collisions at $\sqrt{s} =$ 13 TeV}},
  \href{https://doi.org/10.1007/JHEP07(2019)142}{\emph{JHEP} {\bfseries 07}
  (2019) 142} [\href{https://arxiv.org/abs/1903.04560}{{\ttfamily
  1903.04560}}].

\bibitem{Sirunyan2018}
{\scshape CMS} collaboration, \emph{{Search for a charged Higgs boson decaying
  to charm and bottom quarks in proton-proton collisions at $ \sqrt{s}=8 $
  TeV}}, \href{https://doi.org/10.1007/JHEP11(2018)115}{\emph{JHEP} {\bfseries
  11} (2018) 115} [\href{https://arxiv.org/abs/1808.06575}{{\ttfamily
  1808.06575}}].

\bibitem{Aad2013}
{\scshape ATLAS} collaboration, \emph{{Search for a light charged Higgs boson
  in the decay channel $H^+ \to c\bar{s}$ in $t\bar{t}$ events using pp
  collisions at $\sqrt{s}$ = 7 TeV with the ATLAS detector}},
  \href{https://doi.org/10.1140/epjc/s10052-013-2465-z}{\emph{Eur. Phys. J. C}
  {\bfseries 73} (2013) 2465}
  [\href{https://arxiv.org/abs/1302.3694}{{\ttfamily 1302.3694}}].

\bibitem{Zyla2020}
{\scshape Particle Data Group} collaboration, \emph{Review of particle
  physics}, \href{https://doi.org/10.1093/ptep/ptaa104}{\emph{PTEP} {\bfseries
  2020} (2020) 083C01}.

\bibitem{ATLAS2019}
{\scshape ATLAS} collaboration, \emph{{Direct top-quark decay width measurement
  at $\sqrt{s}$ = 13 TeV with the ATLAS experiment}},
  \href{https://doi.org/10.22323/1.367.0089}{\emph{PoS} {\bfseries
  LeptonPhoton2019} (2019) 089}.

\bibitem{Barger1990}
V.D.~Barger, J.~Hewett and R.~Phillips, \emph{{New Constraints on the Charged
  Higgs Sector in Two Higgs Doublet Models}},
  \href{https://doi.org/10.1103/PhysRevD.41.3421}{\emph{Phys. Rev. D}
  {\bfseries 41} (1990) 3421}.

\bibitem{BowserChao1997}
D.~Bowser-Chao, D.~Chang and W.-Y.~Keung, \emph{{Electron electric dipole
  moment from {CP} violation in the charged Higgs sector, {EDM}}},
  \href{https://doi.org/10.1103/PhysRevLett.79.1988}{\emph{Phys. Rev. Lett.}
  {\bfseries 79} (1997) 1988}
  [\href{https://arxiv.org/abs/hep-ph/9703435}{{\ttfamily hep-ph/9703435}}].

\bibitem{Pospelov1991}
M.~Pospelov and I.~Khriplovich, \emph{{Electric dipole moment of the W boson
  and the electron in the Kobayashi-Maskawa model, EDM}}, {\emph{Sov. J. Nucl.
  Phys.} {\bfseries 53} (1991) 638}.

\bibitem{Chang1991}
D.~Chang, W.-Y.~Keung and T.~Yuan, \emph{{Two loop bosonic contribution to the
  electron electric dipole moment, EDM}},
  \href{https://doi.org/10.1103/PhysRevD.43.R14}{\emph{Phys. Rev. D} {\bfseries
  43} (1991) 14}.

\bibitem{Graner2016ses}
B.~Graner, Y.~Chen, E.G.~Lindahl and B.R.~Heckel, \emph{{Reduced Limit on the
  Permanent Electric Dipole Moment of Hg199}},
  \href{https://doi.org/10.1103/PhysRevLett.116.161601}{\emph{Phys. Rev. Lett.}
  {\bfseries 116} (2016) 161601}
  [\href{https://arxiv.org/abs/1601.04339}{{\ttfamily 1601.04339}}].

\bibitem{Braaten1990}
E.~Braaten, C.-S.~Li and T.-C.~Yuan, \emph{{The Evolution of Weinberg's Gluonic
  CP Violation Operator}},
  \href{https://doi.org/10.1103/PhysRevLett.64.1709}{\emph{Phys. Rev. Lett.}
  {\bfseries 64} (1990) 1709}.

\bibitem{Akeroyd2020}
A.~Akeroyd, S.~Moretti, T.~Shindou and M.~Song, \emph{{CP asymmetries of
  ${\overline B}\to X_s/X_d\gamma$ in models with three Higgs doublets}},
  \href{https://arxiv.org/abs/2009.05779}{{\ttfamily 2009.05779}}.

\bibitem{Borzumati1998}
F.~Borzumati and C.~Greub, \emph{{2HDMs predictions for ${\overline B} \to X_s
  \gamma$ in NLO QCD}},
  \href{https://doi.org/10.1103/PhysRevD.58.074004}{\emph{Phys. Rev. D}
  {\bfseries 58} (1998) 074004}
  [\href{https://arxiv.org/abs/hep-ph/9802391}{{\ttfamily hep-ph/9802391}}].

\bibitem{Amhis2019}
{\scshape HFLAV} collaboration, \emph{Averages of $b$-hadron, $c$-hadron, and
  $\tau$-lepton properties as of 2018},
  \href{https://arxiv.org/abs/1909.12524}{{\ttfamily 1909.12524}}.

\bibitem{Misiak2020}
M.~Misiak, A.~Rehman and M.~Steinhauser, \emph{{Towards $ \overline{B}\to
  {X}_s\gamma $ at the NNLO in QCD without interpolation in $m_c$}},
  \href{https://arxiv.org/abs/2002.01548v2}{{\ttfamily 2002.01548v2}}.

\end{thebibliography}\endgroup
\end{document}